\documentclass[12pt]{article}
\usepackage{graphicx}
\usepackage{lscape}
\usepackage{pdflscape}
\usepackage{afterpage}
\usepackage{pdfpages}
\usepackage{float}
\usepackage{booktabs}
\usepackage[utf8]{inputenc}
\usepackage{indentfirst}
\usepackage{arydshln}
\usepackage{arydshln}
\usepackage{tikz}
\usepackage{lipsum}
\usepackage{pgffor}
\usepackage{dsfont}
\usepackage{amsfonts}
\usepackage{amsmath}
\allowdisplaybreaks
\usepackage{mathtools}
\newcommand\independent{\protect\mathpalette{\protect\independenT}{\perp}}
\def\independenT#1#2{\mathrel{\rlap{$#1#2$}\mkern2mu{#1#2}}}

\DeclareMathOperator*{\arginf}{\arg\!\inf}
\usepackage{xfrac}
\usepackage{amssymb}
\usepackage{bigints}
\usepackage{graphicx}
\usepackage{url}				
\usepackage{caption}
\usepackage{subcaption} 
\usepackage{enumitem}
\usepackage{relsize,exscale}
\usepackage{multirow}
\usepackage{natbib}
\setcitestyle{authoryear, open={(},close={)}}
\usepackage{datetime}

\newdateformat{monthyeardate}{%
	\monthname[\THEMONTH] \THEYEAR}

\usepackage{titlesec}
\titleformat*{\section}{\large\bfseries}
\titleformat*{\subsection}{\large\bfseries}


\newcounter{parentnumber}
\usepackage{theorem}

\newtheorem{assumption}{Assumption}

\newtheorem{corollary}{Corollary}

\newtheorem{example}{Example}

\newtheorem{proposition}{Proposition}

\newtheorem{remark}{Remark}

\newenvironment{proof}[1][Proof]{\noindent \textbf{#1.} }{\  \rule{0.5em}{0.5em}}

\usepackage[margin=1in]{geometry}

\usepackage{hyperref} 
\hypersetup{
	colorlinks=true, breaklinks=true, bookmarks=true,bookmarksnumbered,
	urlcolor=blue, linkcolor=blue, citecolor=blue, 
	pdftitle={}, 
	pdfauthor={\textcopyright}, 
	pdfsubject={}, 
	pdfkeywords={}, 
	pdfcreator={pdfLaTeX}, 
	pdfproducer={LaTeX with hyperref and ClassicThesis} 
}

\begin{document}
	\begin{center}
		{\Large \textbf{Crime and Mismeasured Punishment: \\ Marginal Treatment Effect with Misclassification}}\footnote{I would like to thank the editor, Xiaoxia Shi, and four anonymous referees for their invaluable comments and suggestions. Moreover, I would like to thank Santiago Acerenza, Nathan J. Canen, Xiaohong Chen, Bruno Ferman, John Finlay, Paul Goldsmith-Pinkham, John Eric Humphries, Hugo Jales, Désiré Kédagni, Helena Laneuville, Julian Martinez-Iriarte, Marcela Mello, Michael Mueller-Smith, Yusuke Narita, Ahyan Panjwani, Takuya Ura and Edward Vytlacil, the institutional support of Associação Brasileira de Jurimetria, and seminar participants at Yale University, Indiana University, 2021 Latin American Meeting of the Econometric Society, 2021 Causal Data Science Meeting, 2022 Texas Economics of Crime Workshop, New York Camp Econometrics XVI, Federal University of Pernambuco, Universidad ORT Uruguay, PUC-Rio and Insper for helpful suggestions.}

			Vitor Possebom\footnote{\emph{Email:} vitor.possebom@fgv.br} (Sao Paulo School of Economics - FGV)

	\newsavebox{\tablebox} \newlength{\tableboxwidth}

\monthyeardate\today

	\end{center}

	\textbf{Abstract:} I partially identify the marginal treatment effect (MTE) when the treatment is misclassified. I explore two restrictions, allowing for dependence between the instrument and the misclassification decision. If the signs of the derivatives of the propensity scores are equal, I identify the MTE sign. If those derivatives are similar, I bound the MTE. To illustrate, I analyze the impact of alternative sentences (fines and community service v. no punishment) on recidivism in Brazil, where Appeals processes generate misclassification. The estimated misclassification bias may be as large as 10\% of the largest possible MTE, and the bounds contain the correctly estimated MTE.


	\newpage

\section{Introduction}\label{Sintro}

Evaluating a policy with a misclassified treatment variable is theoretically challenging \citep{Ura2018,Calvi2019,Acerenza2021}. However, this type of problem is widespread in empirical economics. For example, when analyzing the effect of incarceration or alternative sentences in Crime Economics, the treatment variable will be misclassified if the researcher has information only about the trial judge's ruling. In this context, measurement error is created by the appeal process because Appeals Court judges may reverse the trial judge's ruling \citep{Green2010}. Furthermore, education attainment \citep{Black2003} and welfare program participation \citep{Hernandez2007} are likely to be mismeasured. Last but not least, with the increasing availability of large data sets, prediction methods are being used to infer the treatment status in a variety of empirical questions \citep{ArellanoBover2020}. Since no prediction algorithm perfectly classifies the treatment status, the observed treatment variable is misclassified.

In this paper, I provide easy-to-compute bounds around the marginal treatment effect (MTE) function when the treatment variable is misclassified. To do so, I propose two partial identification strategies under increasingly restrictive sets of assumptions, extending the MTE framework \citep{Heckman2006} to scenarios with a misclassified treatment variable. Importantly, I allow the instrument to depend on the potential misclassified treatment variables and on the misclassification decision.

The MTE is a function that captures the effect of a treatment for the individual who is indifferent between taking the treatment or not. In the Crime Economics example, the MTE function captures the effect of being punished with an alternative sentence (fines or community service) on recidivism for the defendant who is at the margin of being punished or not conditional on her judge's leniency level. By analyzing this treatment effect parameter at different margins of judge's leniency, we may find that its heterogeneity is correlated with judge's leniency. Therefore, understanding the unobserved heterogeneity of alternative sentences' impact is key to understanding its benefits and costs. 

My partial identification strategy for the MTE function with a misclassified treatment variable offers a menu of estimates based on two sets of assumptions. These assumptions simultaneously address endogenous selection into treatment and non-classical measurement error. Since these sets gradually add stronger assumptions to tighten the bounds around the MTE function, the researcher can transparently analyze the informational content of each assumption \citep{Tamer2010}.

Four assumptions are common to all sets of assumptions used in this paper. First, the instrument is independent of the potential outcomes and the latent heterogeneity variable that defines the true treatment. Second, the instrument is relevant, impacting the correctly measured and mismeasured probabilities of receiving the treatment. Third, the latent heterogeneity variable that defines the treatment is continuous. Finally, the potential outcomes' first moments are finite. These assumptions are standard in the literature about instrumental variables \citep{Heckman2006} and address endogenous selection into treatment.

Under these four assumptions, I show that the Local Instrumental Variable (LIV) estimand is biased relative to the MTE function when the treatment variable is misclassified. This bias depends on the instrument's value and may move the LIV estimand in any direction. Consequently, even with expert knowledge, predicting the direction of the misclassification bias is challenging. Moreover, when using a misclassified treatment variable, the standard instrument validity tests (\citealp{Fradsen2019}, and \citealp{Heckman2006}) may fail, and the IV weights may not integrate to one even when all weights are positive.

To account for misclassification of the treatment variable, I add two increasingly strong assumptions to those four common assumptions.

First, I impose that the instrument's impacts on the correctly measured probability of treatment and on the mismeasured probability of treatment have the same sign. With this assumption, I identify the sign of the MTE function at any point in the instrument's support.

Second, I impose that the instrument's impacts on the mismeasured probability of treatment and on the correctly measured probability of treatment are similar. With this assumption, I uniformly bound the magnitude of the MTE function at any point and analyze the bounds' sensitivity to the degree of misclassification.

To estimate the bounds around the MTE function, I suggest a parametric model. I impose a polynomial model for the correctly measured propensity score and for the conditional expectation of the treatment effect as a function of the latent resistance to treatment, implying that the outcome equation's reduced-form model is a polynomial function too. By combining this object with a reduced-form polynomial model for the misclassified treatment variable, I estimate the mismeasured LIV estimand, and the MTE function's sign and bounds.

To exemplify the identification tools proposed in this work, I evaluate the effect of alternative sentences (fines and community service in comparison to no punishment) on recidivism in the State of São Paulo, Brazil, between 2010 and 2019.\footnote{São Paulo is the largest state in Brazil, with a population above 44.4 million people according to the Brazilian Census in 2022.} To do so, I observe trial judge's full sentences and divide them into two groups: (i) punished (treated group), containing defendants who were fined or sentenced to community services, and (ii) not punished (untreated group), containing defendants who were acquitted or whose cases were dismissed. To measure recidivism, I check whether the defendant's name appears in any criminal case within two years after the final sentence date.

This context illustrates my framework in two dimensions. First, we need an instrumental variable because the econometrician does not observe all the variables that influence the defendant's future criminal behavior and are used by the trial judge to decide the defendant's punishment. To address endogenous selection into punishment, I use the trial judge's leave-one-out rate of punishment (or ``leniency rate'') as an instrument for the trial judge's decision \citep{Bhuller2019}. Second, in Brazil, defendants will only fulfill their sentences after their judicial case is closed, implying that they will only be punished after the Appeals process or after they inform the Court System that they will not appeal. Consequently, using only trial judges' rulings to define which defendants were punished with an alternative sentence introduces a natural misclassification problem.\footnote{Within the empirical judge fixed effect literature, some authors construct their treatment variables based on trial judges' rulings --- e.g., \citet[California's dataset]{Kling2006}; \citet{Green2010,Bhuller2019}  --- while others construct their treatment variables based on the final ruling or on the actual sentence served by the defendant --- e.g., \citet[Florida's dataset]{Kling2006}; \citet{Arteaga2019}. Although the first group of authors is careful when interpreting their results as the impact of trial judge's decisions on future criminal and labor outcomes, we may affirm that there would be a misclassification problem if their focus had been on identifying the effect of final rulings.}

To better understand my methods' ability to identify the correctly measured MTE function, I also collect data on Appeals Court's decisions. By doing so, I can use the results based on the correctly measured punishment decision (each case's final ruling) to evaluate the bias in an analysis that ignores the misclassification problem. Moreover, I can compare these results with the results derived from the proposed set identification methods.

I find that the misclassification bias can be relatively large and complex. For example, I estimate that this type of bias can be as large as 10\% of the largest possible treatment effect. Furthermore, the misclassification bias can be either positive or negative depending on the instrument's value. Consequently, even an expert may have difficulties theorizing about the misclassification bias' direction and magnitude. For this reason, adopting methods that account for a possibly misclassified treatment variable is useful.

I also find that the proposed partial identification strategies work in practice. When bounding the MTE function, the estimated sets cover the estimated MTE function entirely in every court district.

Finally, using each case's final rulings as my treatment variable, I find that the effect of alternative sentences on recidivism is likely small. Additionally, the observable geographic heterogeneity across court districts appears to matter more than the unobservable heterogeneity across defendants' resistance to treatment.

Concerning its theoretical contribution, my work is inserted in the literature about identifying treatment effect parameters with measurement error. As illustrated by \cite{Hu2017}, this literature is vast and growing. Similarly to my work, four recent papers focused on identifying treatment effect parameters with heterogeneous effects, endogenous selection into treatment and misclassification of the treatment variable: \cite{Ura2018}, \cite{Calvi2019}, \cite{Tommasi2020} and \cite{Acerenza2021}.\footnote{Another important characteristic in this literature is whether the measurement error is differential or not. Appendix \ref{AppDifferential} provides a detailed discussion on this topic and illustrates that my framework allows for differential measurement error.}\textsuperscript{,}\footnote{Focusing on the difference-in-differences framework instead of the instrumental variable model, \cite{Denteh2022} and \cite{Negi2022} also contribute to the literature about misclassified treatment variables.}

These papers differ, for example, with respect to their target parameters. While the first three papers focus on the Local Average Treatment Effect (LATE) parameter, \cite{Acerenza2021} and I focus on the MTE function.

Moreover, unlike these papers, I do not assume the instrument is independent of the potential misclassified treatment variables or the misclassification decision. This flexibility is relevant in a variety of applied examples \citep{Bound2001,Haider2020}.

Allowing for co-dependence between the instrument and the misclassification decision is particularly important in my empirical application. Since sentences by extreme trial judges may be more frequently reversed than sentences by median trial judges, my instrument may be correlated with the misclassification decision. In fact, I find that stricter trial judges are more likely to have their sentences reversed in my empirical application (Subsection \ref{Sdescriptive}).

Even when \citet[Subsection 3.4]{Ura2018} and \citet[Appendix G]{Acerenza2021} extend their main bounds to allow for dependence between the instrument and the potential misclassified treatment variables, our strategies still complement each other. While I provide easy-to-derive bounds that can be used in a sensitivity analysis with respect to the degree of misclassification, they focus on worst-case bounds.\footnote{Appendix \ref{AppComparison} provides a detailed comparison between the bounds provided in Section \ref{Sbounds} and the bounds proposed by \citet[Appendix G]{Acerenza2021}. In particular, I show that, under some conditions, the bounds proposed by \citet[Appendix G]{Acerenza2021} contain zero while the bounds provided in Section \ref{Sbounds} exclude this value.}

Concerning its empirical contribution, my work is inserted in the literature about the effect of alternative sentences on future criminal behavior. Three recent papers in this field were written by \cite{Huttunen2020}, \cite{Giles2021} and \cite{Klaassen2021}. While the first group of authors uses data from Finland, the second uses data from Milwaukee (a city in the State of Wisconsin in the U.S.). Both sets of authors find that alternative sentences increase recidivism. Differently from them, \cite{Klaassen2021} finds that alternative sentences decrease recidivism in North Carolina (a state in the U.S.). Unlike these previous studies, my estimated treatment effect parameters are small and rarely statistically different from zero. Given the large amount of observable geographic heterogeneity in my estimated results, the difference between the recent literature and my findings may be due to different geographic contexts. A deeper understanding of the mechanisms behind these differences is beyond the scope of this work, even though they deserve further investigation.

This paper is organized as follows. Section \ref{Sframework} presents the structural model and the misclassification mechanism and discusses the identifying assumptions. In Section \ref{Sbounds}, I provide the identification results for the MTE bounds with a misclassified treatment variable under each set of assumptions. Moreover, Section \ref{Sestimation} briefly explains how to estimate the objects that are necessary to implement the identification strategy described in the previous section. Finally, Section \ref{Sempirical} describes the data and discusses the empirical results, while Section \ref{Sconclusion} concludes. This paper also contains an online supporting appendix.

\section{Econometric Framework}\label{Sframework}

In this section, I assume that I do not observe the final ruling in each case and develop an econometric model that simultaneously addresses endogenous selection-into-treatment and misclassification. To analyze the Marginal Treatment Effect (MTE) when the treatment variable is mismeasured, I start with the standard generalized selection model \citep{Heckman2006}, described in the potential outcome framework:
\begin{align}
Y & \label{EqY} = Y_{1} \cdot D + Y_{0} \cdot (1-D) \\
D & \label{EqD} = \mathbf{1}\left\lbrace U \leq P_{D}\left(Z\right) \right\rbrace
\end{align}
where $Z$ is an observable continuous instrumental variable (trial judge's leniency rate) with support given by a set $\mathcal{Z} \subset \mathbb{R}$, $P_{D} \colon \mathcal{Z} \rightarrow \mathbb{R}$ is an unknown function, $U$ is a latent heterogeneity variable (defendant's resistance to treatment or the amount of criminal evidence in her favor) and $D$ is the correctly classified treatment status (indicator that the defendant received some type of punishment --- non-prosecution agreement or conviction --- in her case's final ruling).\footnote{My framework can be adapted to binary instruments. This case, which bounds the LATE parameter, is detailed in Appendix \ref{AppLATE} and complements the work developed by \cite{Ura2018} and \cite{Calvi2019}.} Equation \eqref{EqD} models how the agent self-selects into treatment.\footnote{Note that, according to \cite{Vytlacil2002}, this threshold-crossing model is equivalent to the monotonicity assumption imposed in the LATE framework \citep{Imbens1994}.} Variable $Y$ is the realized outcome variable (recidivism indicator), while $Y_{0}$ and $Y_{1}$ are the potential outcomes when the agent is untreated (not punished) and treated (punished), respectively.

I augment this model with the possibly misclassified treatment status indicator $T$ (trial judge's sentence). Note that the binary nature of the treatment variable implies that the measurement error $\left(T - D\right)$ is non-classical, i.e., $Cov\left(T - D, D\right) < 0$. The misclassified treatment is relevant because the researcher observes only the vector $\left(Y, T, Z\right)$, while $Y_{1}$, $Y_{0}$, $D$ and $U$ are latent variables.\footnote{For brevity, I drop exogenous covariates from the model even though all results derived in the paper hold conditionally on covariates. Moreover, I assume that there is only one instrument even though all results hold with partial derivatives when there is a vector of continuous instrumental variables (Appendix \ref{AppExtensions}). This appendix also discusses a model with more than one misclassified treatment variable. Such an extension is useful to empirical applications whose treatment variable is defined based on a prediction algorithm \citep{ArellanoBover2020}.}

Following \cite{Heckman2006}, I impose four assumptions.
\begin{assumption}\label{ASindependence}
The latent variables $\left(Y_{0}, Y_{1}, U\right)$ are independent of the instrument $Z$, i.e., $Z \independent \left(Y_{0}, Y_{1}, U\right)$.
\end{assumption}

Assumption \ref{ASindependence} is an exogeneity assumption and is common in the literature about instrumental variables. In my empirical application, this assumption holds conditional on the court district because, in the State of São Paulo, Brazil, trial judges are randomly assigned to cases within each court district. Furthermore, this type of assumption is common in the judge fixed-effect literature \citep{Bhuller2019}, Education \citep{Cornelissen2018} and many other applied fields.

\begin{assumption}\label{ASrank}
	The derivatives of function $P_{D}$ and of the mismeasured propensity score are always different from zero, i.e., $\dfrac{dP_{D}\left(z\right)}{dz} \neq 0$ and $\dfrac{dP_{T}\left(z\right)}{dz} \neq 0$ for every $z \in \mathcal{Z}$, where $P_{T} \colon \mathcal{Z} \rightarrow \mathbb{R}$ is defined as $P_{T}\left(z\right) = \mathbb{P}\left[\left. T = 1 \right\vert Z = z\right]$ for any $z \in \mathcal{Z}$.
\end{assumption}

Assumption \ref{ASrank} is a rank condition, intuitively imposing that the instrument is locally relevant everywhere.\footnote{When combined with the continuous differentiability of $P_{D}$ and $P_{T}$ as implicitly imposed by my estimation method (Section \ref{Sestimation}), Assumption \ref{ASrank} implies that $P_{D}$ and $P_{T}$ are strictly monotone. Alternatively, I could impose that the set $\mathcal{Z}_{0} \coloneqq \left\lbrace z \in \mathcal{Z} \colon \dfrac{dP_{D}\left(z\right)}{dz} = 0 \text{ or } \dfrac{dP_{T}\left(z\right)}{dz} = 0 \right\rbrace$ has measure zero. In this case, $P_{D}$ and $P_{T}$ could be non-monotone.   Moreover, Propositions \ref{PropLIV}-\ref{ThmSharpMTE} and Corollaries \ref{CorSign}, \ref{CorOuterMTEsharp} and \ref{CorBoundTE} would still hold for any $z \in \mathcal{Z} \setminus \mathcal{Z}_{0}$ if I defined the domain of all relevant functions as $\mathcal{Z} \setminus \mathcal{Z}_{0}$ instead of $\mathcal{Z}$. I use Assumption \ref{ASrank} instead of its weaker version for notational ease.} Note also that Assumption \ref{ASrank} is stronger than the rank condition usually imposed in the literature about marginal treatment effects \citep[Assumption A-2]{Heckman2006}. In particular, since the correctly measured treatment variable is not observed, it is impossible to directly test that $\dfrac{dP_{D}\left(z\right)}{dz} \neq 0$ for every $z \in \mathcal{Z}$.  However, since the mismeasured propensity score is trivially identified, it is possible to test that $\dfrac{dP_{T}\left(z\right)}{dz} \neq 0$ for every $z \in \mathcal{Z}$. Observe also that Assumption \ref{ASrank} implies that $0 < \mathbb{P}\left[D = 1\right] < 1$, a support condition that is required for any evaluation estimator.

\begin{assumption}\label{AScontinuous}
	The distribution of the latent heterogeneity variable $U$ is absolutely continuous with respect to the Lebesgue measure.
\end{assumption}

Assumption \ref{AScontinuous} is a regularity condition that allows me to normalize the marginal distribution of $U$ to be the standard uniform. Consequently, I have that the true propensity score $\mathbb{P}\left[\left. D = 1 \right\vert Z = z\right]$ is equal to $P_{D}\left(z\right)$ for any $z \in \mathcal{Z}$. However, due to misclassification of the treatment variable, the correctly measured propensity score is not identified.

\begin{assumption}\label{ASfinite}
	The potential outcome variables have finite first moments, i.e., $\mathbb{E}\left[\left\vert Y_0 \right\vert \right] < \infty$ and $\mathbb{E}\left[\left\vert Y_1 \right\vert \right] < \infty$.
\end{assumption}

Assumption \ref{ASfinite} is a regularity condition that allows me to apply standard integration theorems and ensures that average treatment effects are well-defined.

Due to misclassification of the treatment variable, Assumptions \ref{ASindependence}-\ref{ASfinite} are not sufficient to identify the marginal treatment effect function (Proposition \ref{PropLIV}). To address this problem, I gradually impose two increasingly strong assumptions that allow me to derive increasingly strong identification results in Section \ref{Sbounds}. Consequently, I offer a menu of estimates whose credibility can be assessed by each reader based on the plausibility of each assumption \citep{Tamer2010}. Intuitively, these assumptions restrict the amount of measurement error by constraining the relationship between $D$, $T$ and $Z$.

To derive my first partial identification result (Corollary \ref{CorSign}), I require a weak sign restriction on the impact of $Z$ on the true treatment variable $D$ and on the misclassified treatment variable $T$.\footnote{Sign restrictions have been used previously in the misclassification literature to identify the treatment effect parameter in a linear model with homogeneous treatment effects \citep[Assumption 5]{Haider2020}.} Under Assumption \ref{ASsign}, it is possible to identify the sign of the marginal treatment effect.

\begin{assumption}\label{ASsign}
	The derivative of the correctly measured propensity score has the same sign as the derivative of the mismeasured propensity score, i.e.,
	$\text{sign}\left(\dfrac{dP_{D}\left(Z\right)}{dz}\right) = \text{sign}\left(\dfrac{dP_{T}\left(Z\right)}{dz}\right)$ for every $z \in \mathcal{Z}$.
\end{assumption}

In my empirical application, Assumption \ref{ASsign} imposes that tougher trial judges also increase the probability of receiving some type of punishment according to each case's final ruling. Intuitively, this assumption holds if tougher trial judges write more compelling rulings that are more likely to be affirmed by the Appeals Court. Alternatively, if only a small share of defendants appeal, this assumption may hold even if tougher trial judges' rulings are more likely to be reversed. Moreover, as discussed in Example \ref{Esign} (Appendix \ref{AppExamples}), this assumption holds if trial judges and Appeals judges have the same sentencing criteria, but face different information sets.\footnote{For another example about investment decisions, see Appendix \ref{AppSeq}. Additionally, in Appendix \ref{AppSuffSign}, I impose restrictions on the model's primitives that imply Assumptions \ref{ASsign} and \ref{ASbounded}. The sufficient conditions associated with Assumption \ref{ASsign} are connected with the framework proposed by \cite{Tommasi2020}.}

To derive a stronger partial identification result (Proposition \ref{CorOuterMTE}), I impose not only that the impact of $Z$ on $D$ and $T$ have the same sign, but also that those impacts are not arbitrarily different from each other. Under Assumption \ref{ASbounded}, it is possible to uniformly bound the marginal treatment effect function.

\begin{assumption}\label{ASbounded}
	The ratio between the derivatives of the correctly measured propensity score and of the mismeasured propensity score is bounded, i.e., there exists a known $c \in \left[1, +\infty\right)$ such that
	$\dfrac{\sfrac{dP_{D}\left(z\right)}{dz}}{\sfrac{dP_{T}\left(z\right)}{dz}} \in \left[\dfrac{1}{c}, c\right]$ for any value $z \in \mathcal{Z}$.
\end{assumption}

Note that using a smaller $c$ imposes a stronger restriction on the relationship between $D$, $T$ and $Z$.\footnote{I impose that the function $\dfrac{\sfrac{dP_{D}\left(\cdot\right)}{dz}}{\sfrac{dP_{T}\left(\cdot\right)}{dz}}$ is bounded by a constant for simplicity. Alternatively, I could impose that $\dfrac{\sfrac{dP_{D}\left(\cdot\right)}{dz}}{\sfrac{dP_{T}\left(\cdot\right)}{dz}}$ is bounded by $c\colon\mathcal{Z}\rightarrow\left[1,+\infty\right)$, where $c\left(\cdot\right)$ is a nontrivial function of the instrument. However, knowing an entire bounding function in any concrete empirical context may be difficult.}  In practice, the researcher may be uncertain about the value of $c$. After deriving Proposition \ref{CorOuterMTE}, I explain how to choose the largest value of $c$ that is compatible with the data and the other model assumptions. Alternatively, the researcher can follow a sensitivity analysis strategy \citep{Cinelli2019} and present results for different values of $c$.\footnote{If the researcher observes the correctly classified and the misclassified treatment variable for a subsample of her sample, she can use this subsample to estimate $c$ and use the estimated $c$ in Assumption \ref{ASbounded} to partially identify the MTE function in her entire sample.} 

In my empirical application, Assumption \ref{ASbounded} imposes that decreasing the trial judge's leniency increases the punishment probabilities according to the trial judge's ruling and according to the final ruling by similar amounts. As explained in Section \ref{Sresults}, the minimum valid $c$ in my empirical application is estimated to equal $1.13$.

To illustrate the theoretical plausibility of my framework in my empirical application, Example \ref{EboundedAppeal} (Appendix \ref{AppExamples}) describes a simple model of Appeals Courts reversing trial judges' rulings.\footnote{Appendix \ref{Ebounded} explains that Assumption \ref{ASbounded} may also be plausible when analyzing returns to education if having a college degree is randomly miscoded in a survey. Moreover, Appendix \ref{AppDifferential} illustrates that Assumptions \ref{ASindependence}-\ref{ASfinite} and \ref{ASbounded} are compatible with differential measurement error.}

In Appendix \ref{AppStandar}, I impose a stronger assumption that allows me to derive sharp uniform bounds around the MTE function and sharply bound any weighted integral of the MTE function (e.g., Average Treatment Effect (ATE), Average Treatment Effect on the Treated (ATT) and Average Treatment
Effect on the Untreated (ATU)). This extra condition imposes a restriction on the functional relationship between the correctly measured propensity score and the mismeasured propensity score in the sense that it connects the level and all the derivatives of those two objects.

\section{Partial Identification of the MTE with a Misclassified Treatment}\label{Sbounds}

My goal is to derive partial identification results for the Marginal Treatment Effect as a function of the value of the instrument (MTE function). Formally, I define this object as $\theta \colon \mathcal{Z} \rightarrow \mathbb{R}$ such that, for any $z \in \mathcal{Z}$,
\begin{equation}\label{EqMTE}
    \theta\left(z\right) = \mathbb{E}\left[\left. Y_{1} - Y_{0} \right\vert U = P_{D}\left(z\right)\right].
\end{equation}
Intuitively, this definition of the MTE function captures the effect of a treatment for the individual who is indifferent between taking the treatment or not, where the margin of indifference is defined by the value of the individual's instrument.\footnote{Given the equivalence result by \cite{Vytlacil2002}, it is possible to write the MTE function $\theta$ as depending on counterfactual treatment choices that satisfy the standard LATE monotonicity conditions. For any $z \in \mathcal{Z}$, define the random variable $D^{*}\left(z\right) = \mathbf{1}\left\lbrace U \leq P_{D}\left(z\right) \right\rbrace$. Note that $D^{*}\left(\cdot\right)$ satisfy the standard LATE monotonicity conditions and that the MTE function satisfies $\theta\left(z\right) = \mathbb{E}\left[\left. Y_{1} - Y_{0} \right\vert z = \arginf\limits_{z^{\prime} \colon D^{*}\left(z^{\prime} \right) = 1} P_{D}\left(z^{\prime}\right)\right]$.}

In my empirical application, the MTE function captures the effect of being punished with an alternative sentence on future criminal behavior for the defendant who is at the margin of being found guilty given her judge's leniency levels. Analyzing the MTE function at different margins of judge's leniency is important because punishment may have heterogeneous effects on the defendants and the heterogeneity may be correlated with judge's leniency. Consequently, understanding the heterogeneity of the impact of alternative sentences is key to understanding its benefits and costs.

Note that, while \cite{Heckman2006} define the MTE as a function of the latent heterogeneity $U$, I define the MTE as a function of the instrument.\footnote{In my work, I do not use the standard definition of the MTE function and its classic identification result $\left(\mathbb{E}\left[\left. Y_{1} - Y_{0} \right\vert U = p\right] = \dfrac{d \mathbb{E}\left[\left. Y \right\vert P_{D}\left(Z\right) = p\right]}{d p}\right)$ because I cannot point-identify $P_{D}$ due to misclassification of the treatment variable. Consequently, I am unable to associate an instrument value $z$ with a specific margin $u$ of the latent heterogeneity $U$, i.e., I do not know $u = P_{D}\left(z\right)$ even though I know that $\theta\left(z\right) = \mathbb{E}\left[\left. Y_{1} - Y_{0} \right\vert U = P_{D}\left(z\right)\right]$.} Consequently, different instrumental variables are associated with different MTE functions. Although the function $\theta\left(\cdot\right)$ is not policy-invariant according to the definition of \cite{Heckman2006}, I can still use it to compute interesting policy-relevant treatment effect parameters (PRTE) when the policy-maker can set the level of the instrument. For example, in my empirical application, I can still compute the treatment effect of making all judges as strict or as lenient as the strictest or most lenient judges.\footnote{Similar policy-relevant treatment effect parameters can be defined in any judge's lenience design study, one of the most common applications of the MTE.}

Moreover, in Appendix \ref{AppStandar}, I impose Assumptions \ref{ASsharp} and \ref{ASconstant} to derive a one-to-one map between my definition of the MTE function and its standard definition. Consequently, under these restrictive assumptions, I can derive bounds around common treatment effect parameters, such as the ATE, ATT, ATU and PRTE (Corollary \ref{CorBoundTE}).\footnote{In particular, Assumption \ref{ASconstant} is not valid in my empirical application.}

To partially identify the MTE function (Equation \eqref{EqMTE}), I analyze the consequences of a misclassified treatment variable on the Local Instrumental Variable (LIV) estimand and, then, derive increasingly strong identification results based on Assumptions \ref{ASsign} and \ref{ASbounded}. Analyzing the misclassification bias of the LIV estimand is important because this estimand is traditionally used to identify the MTE function in the previous literature.

If the researcher ignores that the treatment variable is misclassified, she can compute the LIV estimand using the misclassified treatment variable $T$ as if it was the actual treatment variable. In this case, the LIV estimand is defined as $f\colon\mathcal{Z}\rightarrow\mathbb{R}$ such that, for any $z \in \mathcal{Z}$, \begin{equation}
\label{EqLIV}
f\left(z\right) = \dfrac{\sfrac{d\mathbb{E}\left[\left. Y \right\vert Z = z\right]}{dz}}{\sfrac{d\mathbb{E}\left[\left. T \right\vert Z = z\right]}{dz}}
\end{equation}
following \cite{Chalak2017} and as $\tilde{f}\colon\mathcal{P}\rightarrow\mathbb{R}$ such that, for any $p$ in the support $\mathcal{P}$ of $P_{T}\left(Z\right)$,
\begin{equation}
\label{EqLIVHeckman}
\tilde{f}\left(p\right) = \dfrac{d\mathbb{E}\left[\left. Y \right\vert P_{T}\left(Z\right) = p \right]}{dp}
\end{equation} following \cite{Heckman2006}. The next proposition analyzes which object is identified by both definitions of the LIV estimand and clarifies two negative consequences of ignoring misclassification.

\begin{proposition}[LIV estimand]\label{PropLIV}
	Under Assumptions \ref{ASindependence}-\ref{ASfinite}, the LIV estimand $f$ satisfies
	\begin{equation}\label{EqLivChalak}
	f\left(z\right) = \dfrac{\sfrac{dP_{D}\left(z\right)}{dz}}{\sfrac{dP_{T}\left(z\right)}{dz}} \cdot \theta\left(z\right)
	\end{equation}
	for any $z \in \mathcal{Z}$.

	Moreover, if Assumptions \ref{ASindependence}-\ref{ASfinite} hold and the function $P_{T}$ is invertible, then the LIV estimand $\tilde{f}$ satisfies
	\begin{equation}\label{EqLivHUV}
	\dfrac{d\mathbb{E}\left[\left. Y \right\vert P_{T}\left(Z\right) = p \right]}{dp} = \dfrac{\sfrac{dP_{D}\left(P_{T}^{-1}\left(p\right)\right)}{dz}}{\sfrac{dP_{T}\left(P_{T}^{-1}\left(p\right)\right)}{dz}} \cdot \theta\left(P_{T}^{-1}\left(p\right)\right)
	\end{equation}
	for any $p$ in the support of $P_{T}\left(Z\right)$.
\end{proposition}

\begin{proof}
	The proof of this proposition follows \cite{Heckman2006} and is detailed in Appendix \ref{ProofLIV}.
\end{proof}

The first negative consequence shown by Proposition \ref{PropLIV} is that, when misclassification is ignored, two different definitions for the LIV estimand do not identify the MTE function. Similarly to the comparison between the LATE and the Wald Estimator when the treatment variable is misclassified \citep{Calvi2019,Tommasi2020}, there is a scaling factor connecting the LIV estimand and the MTE function.

Differently from the work done by \cite{Calvi2019} and \cite{Tommasi2020}, this scaling factor is a function, suggesting that the bias may be positive, negative or even zero depending on the point where the LIV estimand is evaluated. Interestingly, the LIV estimand's bias attenuates the true MTE function if $\dfrac{\sfrac{dP_{D}\left(z\right)}{dz}}{\sfrac{dP_{T}\left(z\right)}{dz}} < 1$ and enlarges the true MTE function if $\dfrac{\sfrac{dP_{D}\left(z\right)}{dz}}{\sfrac{dP_{T}\left(z\right)}{dz}} > 1$.  As illustrated by my empirical application (Section \ref{Sresults}), this phenomenon complicates an intuitive analysis of the misclassification bias where the researcher tries to guess its direction based on expert knowledge.

The second negative consequence shown by Proposition \ref{PropLIV} is that IV validity tests may fail if misclassification is ignored. For example, when the outcome variable has compact support, \citet[Section 3]{Fradsen2019} proposes to test the monotonicity condition (Equation \eqref{EqD}) by testing whether the function $\dfrac{d\mathbb{E}\left[\left. Y \right\vert P_{T}\left(Z\right) = p \right]}{dp}$ is bounded between the smallest and the largest possible treatment effects. Proposition \ref{PropLIV} shows that this test is not valid when the treatment variable is mismeasured because the scaling factor $\dfrac{\sfrac{dP_{D}\left(P_{T}^{-1}\left(p\right)\right)}{dz}}{\sfrac{dP_{T}\left(P_{T}^{-1}\left(p\right)\right)}{dz}}$ in Equation \eqref{EqLivHUV} is possibly unbounded without extra assumptions. A misclassified treatment variable also renders the monotonicity test proposed by \citet[Theorem 1]{Heckman2005} invalid as detailed in Appendix \ref{AppRejection}. Moreover, the mismeasured propensity score may not satisfy index sufficiency as explained in Appendix \ref{AppSufficiency}.

Furthermore, ignoring misclassification makes interpreting the usual IV estimand more difficult. Following \citet[Section III.B]{Heckman2006}, I can show that the naive IV estimand satisfies $$\dfrac{Cov\left(Z, Y\right)}{Cov\left(Z, T\right)} = \int_{0}^{1} \omega\left(u\right) \cdot \mathbb{E}\left[\left. Y_{1} - Y_{0} \right\vert U = u\right] \, du,$$ where $\omega\left(u\right) = \dfrac{\mathbb{E}\left[\left. Z - E\left[Z\right] \right\vert P_{D}\left(Z\right) \geq u\right] \cdot \mathbb{P}\left[P_{D}\left(Z\right) \geq u\right]}{Cov\left(Z,T\right)}$. Unless $Cov\left(Z,T\right) = Cov\left(Z,D\right)$, the weights $\omega\left(\cdot\right)$ do not integrate to one and the IV estimand do not identify a proper weighted average of the MTE even when the weights are positive. Additionally, since $Cov\left(Z,T\right) = Cov\left(Z,D\right)$ is equivalent to $Cov\left(Z,T - D\right) = 0$, this condition intuitively imposes a testable restriction in my empirical application: sentences by extreme trial judges are as likely to be reversed as sentences by median trial judges. Since I find that stricter trial judges are more likely to have their sentences reversed by the Appeals Court (Subsection \ref{Sdescriptive}), the naive IV estimand will have weighting problems in my empirical application.

Now, to derive increasingly strong identification results for $\theta\left(\cdot\right)$, I add Assumptions \ref{ASsign} and \ref{ASbounded}. The first identification result (Corollary \ref{CorSign}) shows that I can identify the sign of the MTE function under a weak assumption about the signs of the derivatives of the correctly measured propensity score function and of the mismeasured propensity score function.

\begin{corollary}[Identifying the sign of the MTE function]\label{CorSign}
	Under Assumptions \ref{ASindependence}-\ref{ASfinite} and \ref{ASsign}, the sign of $\theta\left(z\right)$ is identified for any $z \in \mathcal{Z}$.
\end{corollary}

\begin{proof}
	In Equation \eqref{EqLivChalak}, the scaling function that multiplies the MTE function $\theta\left(\cdot\right)$ is strictly positive under Assumptions \ref{ASrank} and \ref{ASsign}. Consequently, I have that $sign\left(\theta\left(z\right)\right) = sign\left(f\left(z\right)\right)$ for any $z \in \mathcal{Z}$.
\end{proof}

Knowing the sign of the MTE function $\theta\left(z\right)$ at a point $z \in \mathcal{Z}$ is important. If the instrument is policy-relevant, this result can be used to ensure that, at every choice margin given by the instrument, the expected benefit is positive. For example, in my empirical application, the policymaker can re-educate judges whose punishment rates are related to a positive effect on recidivism to change their punishment criteria to points $\theta\left(z\right)$ that are associated with a negative effect on average. Even when the instrument is not policy-relevant, knowing whether the MTE function $\theta\left(\cdot\right)$ is mostly positive or negative is useful to evaluate the pros and cons of a treatment.

The second identification result (Proposition \ref{CorOuterMTE}) shows that, under an assumption about the ratio between the derivatives of the correctly measured propensity score function and of the mismeasured propensity score function, I can uniformly bound the MTE function. Moreover, the distance between the true MTE function and any function in this set is bounded above by an identifiable constant under Assumptions \ref{ASindependence}-\ref{ASfinite} and \ref{ASbounded}.

\begin{proposition}[Uniform Outer Set for the MTE function]\label{CorOuterMTE}

Suppose Assumptions \ref{ASindependence}-\ref{ASfinite} and \ref{ASbounded} hold. I have that $\theta \in \Theta_{1}$, where
\begin{equation*}
	\Theta_{1} \coloneqq \left\lbrace \tilde{\theta}\colon \mathcal{Z} \rightarrow \mathbb{R} \left\vert \text{For any } z \in \mathcal{Z}, \tilde{\theta}\left(z\right) \in \left\lbrace \begin{array}{ll}
		\left[\dfrac{1}{c} \cdot f\left(z\right), c \cdot f\left(z\right) \right] & \text{if } f\left(z\right) \geq 0 \\
		\left[c \cdot f\left(z\right), \dfrac{1}{c} \cdot f\left(z\right) \right] & \text{if } f\left(z\right) < 0
	\end{array}  \right. . \right. \right\rbrace.
\end{equation*}

Moreover, for any $\tilde{\theta} \in \Theta_{1}$ and $d \in \left(0, +\infty\right]$, $\left\vert\left\vert \theta - \tilde{\theta} \right\vert\right\vert_{d} \leq \left(\dfrac{c^{2} - 1}{c} \right) \cdot \left\vert\left\vert f  \right\vert\right\vert_{d}$, where $\left\vert\left\vert \cdot \right\vert\right\vert_{d}$ is the $L_{d}$-norm if $d < \infty$ and $\left\vert\left\vert \cdot \right\vert\right\vert_{d}$ is the $\sup$-norm if $d = \infty$.

\end{proposition}

\begin{proof}
	The proof of the first part of Proposition \ref{CorOuterMTE} uses Equation \eqref{EqLivChalak} and the bounds imposed on the ratio $\dfrac{\sfrac{dP_{D}\left(z\right)}{dz}}{\sfrac{dP_{T}\left(z\right)}{dz}}$ by Assumption \ref{ASbounded} to uniformly bound the MTE function, while the proof of the second part of Proposition \ref{CorOuterMTE} uses the definition of the norm $\left\vert\left\vert \cdot \right\vert\right\vert_{d}$. The details are in Appendix \ref{ProofOuterMTE}.
\end{proof}

Proposition \ref{CorOuterMTE} is strictly stronger than Corollary \ref{CorSign} in the sense that not only it identifies the sign of $\theta\left(z\right)$ at any point $z \in \mathcal{Z}$, but it also provides the largest possible effect and the smallest possible effect for each value of the instrument. If the instrument is policy-relevant, this result can be used to ensure that, at every choice margin given by the instrument, the expected benefit is larger than the expected treatment cost. For example, in my empirical application, the policymaker can re-educate judges whose punishment rates are related to effects that are not large enough to compensate for punishment costs to change their punishment criteria to points associated with effects that pass, on average, a cost-benefit analysis. Even when the instrument is not policy-relevant, bounding the MTE function $\theta\left(\cdot\right)$ is useful to know whether most points pass, on average, a cost-benefit analysis.

Due to Proposition \ref{CorOuterMTE}, I can also adapt the test proposed by \cite{Fradsen2019} to test Assumptions \ref{ASindependence}-\ref{ASfinite} and \ref{ASbounded} when the support of $Y_{0}$ and $Y_{1}$ is bounded. To do so, I check whether $\sup_{\tilde{\theta} \in \Theta_{1}}\sup_{z \in \mathcal{Z}}\tilde{\theta}\left(z\right)$ is smaller than the largest possible effect and whether $\inf_{\tilde{\theta} \in \Theta_{1}}\inf_{z \in \mathcal{Z}}\tilde{\theta}\left(z\right)$ is larger than the smallest possible effect. If that is the case, I do not reject Assumptions \ref{ASindependence}-\ref{ASfinite} and \ref{ASbounded}. Note that this test can be used to define the largest value of $c \in \left[1, +\infty \right)$ that is plausible according to the data. To implement it, find $\overline{z}$ and $\underline{z}$ that respectively maximizes and minimizes the function $f\left(\cdot\right)$ and, then, find the largest $c$ such that $\left(\dfrac{1}{c} \cdot f\left(\overline{z}\right), c \cdot f\left(\overline{z}\right), \dfrac{1}{c} \cdot f\left(\underline{z}\right), c \cdot f\left(\underline{z}\right) \right) \in \left[\underline{\theta}, \overline{\theta}\right]^{4}$, where $\underline{\theta}$ is the smallest possible effect and $\overline{\theta}$ is the largest possible effect.

\section{Estimation}\label{Sestimation}

In this section, I briefly explain how to estimate the MTE function's sign (Corollary \ref{CorSign}) and its bounds (Proposition \ref{CorOuterMTE}). To estimate these objects, I need to estimate two objects: the mismeasured propensity score and the outcome equation's reduced-form model. Importantly, in my empirical application, these objects depend not only on the value $z$ of the instrument but also on the value $x$ of the covariates. These extra variables contain a full set of court district dummies and are included because, in São Paulo, trial judges are randomly allocated to criminal cases only after conditioning on the court district.

To estimate the mismeasured propensity score, I treat it as a purely reduced-form object. Consequently, I model the misclassified treatment variable's conditional expectation as a separable function between the instrument and the covariates, depending on a polynomial of the instrument and a full set of court district dummies. Under these parametric assumptions, this model can be estimated by OLS.

To estimate the outcome equation's reduced-form model, I treat it as an object derived from the economic model's primitives. Specifically, the correctly measured propensity score function and the conditional expectation of the treatment effect as a function of the latent resistance to treatment are separable between the instrument and the covariates, depending on a polynomial of the instrument and a full set of court district dummies. Consequently, the outcome equation's reduced-form model is a polynomial function that depends on the interaction between the instrument and the court district dummies. Under these parametric assumptions, this model can be estimated by OLS.

By combining the parameters from these two OLS regressions, I can estimate the mismeasured LIV estimand (Equation \eqref{EqLIV}) and use it to estimate the MTE function's sign and bounds.

Moreover, in my empirical application, I also need to estimate the correctly measured MTE function to use it as a benchmark against the analysis that ignores misclassification. To do so, I need to estimate the LIV estimand that uses the correctly measured propensity score in its denominator. This object's estimator is very similar to the mismeasured LIV estimand's estimator. The only difference is that I now use a parametric polynomial approximation for the conditional expectation of the correctly classified treatment variable.

To have a deeper understanding of the estimation methods and their performance in a Monte Carlo exercise, see Appendix \ref{AppEstimation}.

\section{Empirical Application}\label{Sempirical}

In my empirical application, I answer the question: ``Do alternative sentences impact recidivism?''. To answer this question, I collect data from all criminal cases brought to the Justice Court System in the State of São Paulo, Brazil, from 2010 to 2019. In Subsection \ref{Sdescriptive}, I briefly explain my dataset and provide key descriptive statistics. In Appendix \ref{AppDescriptive}, I provide detailed descriptive statistics. In Appendix \ref{AppData}, I provide a detailed explanation on how I constructed my dataset. Finally, in Subsection \ref{Sresults}, I describe the results of my empirical analysis.

\subsection{Data and Descriptive Statistics}\label{Sdescriptive}

I collect data from all criminal cases brought to the Justice Court System in the State of São Paulo, Brazil, between January 4\textsuperscript{th}, 2010, and December 3\textsuperscript{rd}, 2019. I restricted my sample to cases that started between 2010 and 2017 because the last two years are used only to define my outcome variable. Moreover, I focus on criminal cases whose maximum prison sentence is less than 4 years because, according to Brazilian Law, these cases must be punished with alternative sentences (fines and community service). Due to this sample restriction, the most common crime types in my sample are theft and domestic violence. After those two restrictions, my dataset contains 51,731 case-defendant pairs.

In my dataset, I observe the defendant's full name, the defendant's court district, the case's starting date, the assigned trial judge's full name, the trial judge's full sentence, the trial judge's sentence's date, whether the case went to the Appeals Court, the Appeals Court's ruling if there is one, and the Appeals Court's ruling's date if there is one. Based on those variables, I define my outcome variable ($Y = $``recidivism within 2 years of the final sentence''), my misclassified treatment variable ($T = $``trial judge's decision''), my correctly classified treatment variable ($D = $``final ruling''), my instrument ($Z = $ ``trial judge's leniency rate'') and my covariates (X = ``full set of court district dummies'').

My misclassified treatment variable $T$ is based only on trial judge's sentences and divides them into two groups. The first group (treated) receives a punishment, i.e., its defendants were fined or sentenced to community services because they were either convicted or signed a non-prosecution agreement. The second group (control) did not receive a punishment, i.e., its defendants were acquitted or its cases were dismissed.

My correctly classified treatment variable $D$ also divides the case-defendant pairs into the groups ``punished'' and ``not punished''. However, it considers the final ruling in each case. If the case did not go to the Appeals Court, this variable equals the misclassified treatment variable $T$. However, if the case went to the Appeals Court, this variable may differ from $T$ because it also considers the Appeals Court's decision. If the trial sentence was affirmed, then $D$ is equal to $T$. However, if the trial sentence was reversed, then $D$ is equal to $1 - T$.

Importantly, both treatment variables are based on a logistic LASSO that maps rulings into binary variables (Appendix \ref{AppData}). Although my prediction algorithms may make classification errors, I ignore this source of misclassification in my empirical analysis. Consequently, I focus exclusively on the misclassification problems generated by the appeals process. This source of misclassification is arguably more interesting because it is embedded in the economic nature of the problem instead of being mechanically created by a prediction algorithm.\footnote{Alternatively, I could focus on the misclassification error generated by the prediction algorithms. The method proposed in this paper partially identifies the $MTE$ function even if the final sentence is possibly misclassified. The drawback of this approach is the impossibility of estimating the misclassification bias because I would not observe the correctly measured final sentence if I took prediction errors into account. If I were to follow this approach, I could define one misclassified treatment variable for each prediction algorithm (e.g., random forest or logistic LASSO) and use the methods proposed in Appendix \ref{AppMoreT} for the case with more than one misclassified treatment variable.}

Table \ref{TabMisclassification} shows the joint distribution of the correctly classified treatment variable $D$ and the misclassified treatment variable $T$. Since most cases (67.3\%) do not go to the Appeals Court, most cases are correctly classified (95.6\%) as described in the main diagonal. The other two cells describe the cases that are misclassified when I ignore the Appeals Court's decisions. First, I find that 3.5\% of the defendants were punished by the Trial Judge and were able to reverse their sentences in the Appeals Court. Moreover, in 0.9\% of the cases, the defendant was not punished by the Trial Judge, but the prosecutor was able to appeal and reverse the decision.

My instrument $Z$ is the trial judge's leniency rate. This variable equals the leave-one-out rate of punishment for each trial judge, where the defendant's own decision is excluded from this average. To do so, I only use the 639 judges who analyzed more than 20 cases during my sample period.

\begin{figure}[h]
	\begin{center}
		\includegraphics[width = 0.47 \textwidth]{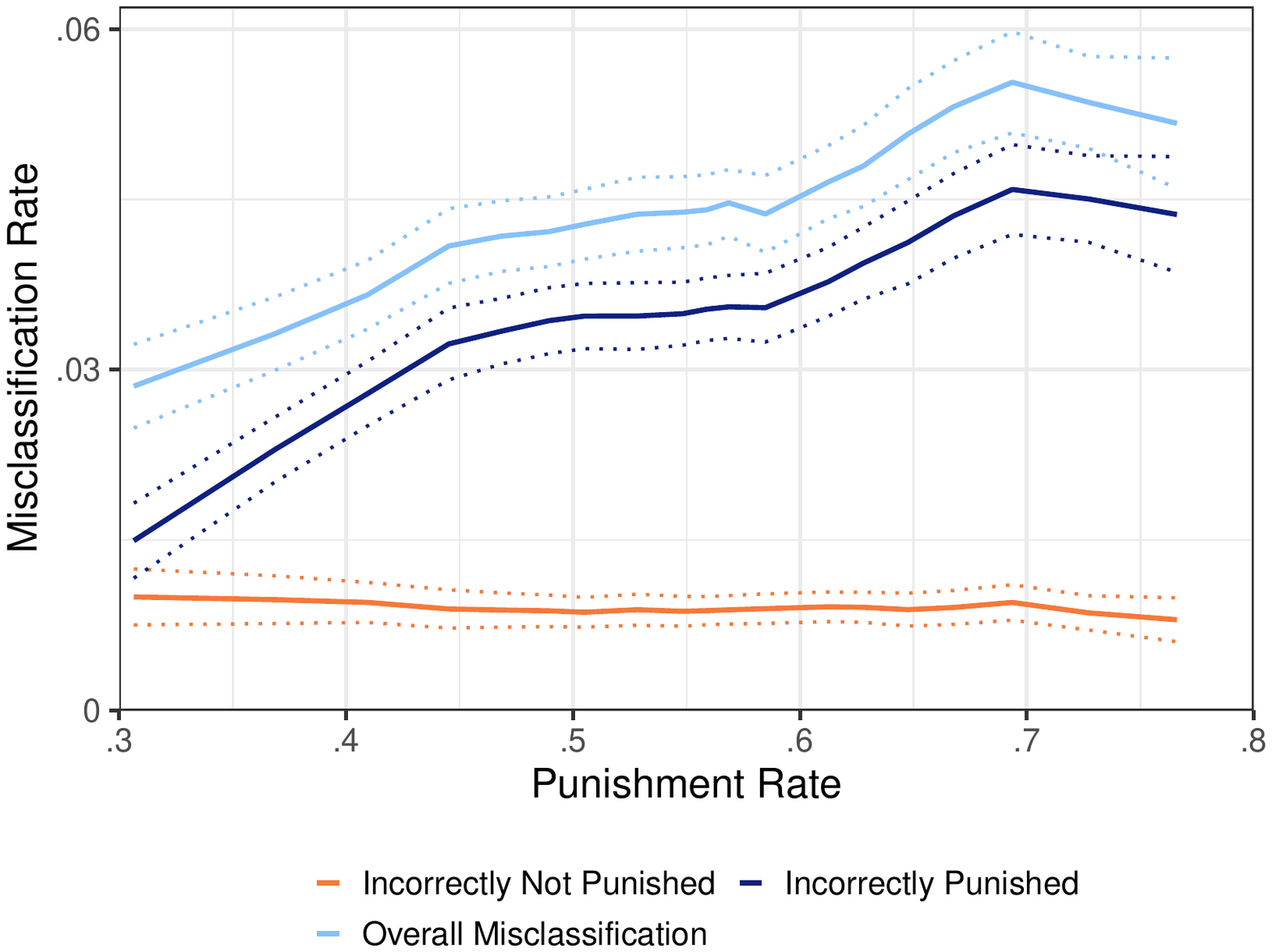}
		\caption{Misclassification Rate as a function of the Punishment Rate}
		\label{FigMisclassification}
	\end{center}
	\footnotesize{Notes: ``Incorrectly Not Punished'' refers to the cases with $T = 0$ and $D = 1$. ``Incorrectly Punished'' refers to the cases with $T = 1$ and $D = 0$. ``Overall Misclassification'' refers to the cases with $T \neq D$. All nonparametric functions in this figure were estimated using local linear regressions with an Epanechnikov kernel based on \cite{Calonico2019}. I used 19 evaluations points based on the 5\textsuperscript{th}, 10\textsuperscript{th}, $\ldots$, 95\textsuperscript{th} quantiles of the punishment rate $Z$. The bandwidth was optimally selected according to the MSE criterion. If the optimally selected bandwidth has less than 50 units, it is enlarged to contain at least 50 observations.}
\end{figure}

Having described the treatment and instrumental variables, I can now discuss their relationship. Figure \ref{FigMisclassification} shows three conditional probabilities where the conditioning variable is the instrument $Z$. The orange line is the share of defendants who were initially found not guilty by the trial judge and had their sentences reversed by the Appeals Court conditional on the punishment rate of the trial judge. The dark blue line is the share of defendants who were initially found guilty by the trial judge and had their sentences reversed by the Appeals Court conditional on the punishment rate of the trial judge. Finally, the light blue line is the share of defendants who had their sentences reversed by the Appeals Court conditional on the punishment rate of the trial judge. The dotted lines are robust bias-corrected 95\%-confidence intervals \citep{Calonico2019}.

Figure \ref{FigMisclassification} illustrates the importance of having more than one alternative set of assumptions when addressing misclassification issues. Although the orange line suggests that there is no clear dependency between being incorrectly not punished (i.e., $T = 0, D = 1$) and the trial judge's leniency rate, the dark blue and light blue lines suggest that being incorrectly punished (i.e., $T = 1, D = 0$) and having a misclassified sentence (i.e., $T \neq D$) depend positively on the trial judge's punishment rate. Consequently, the method proposed by \cite{Acerenza2021} is not appropriate to analyze this empirical problem because it imposes that having a misclassified sentence and the instrument are independent. Since my framework (Sections \ref{Sframework} and \ref{Sbounds}) relies on alternative restrictions on the relationship between $T$, $D$ and $Z$, my method may be appropriate to analyze this empirical question. Importantly, Assumptions \ref{ASsign} and \ref{ASbounded} seem valid in this application as discussed in Subsection \ref{Sresults}.

I, now, describe how I define my outcome variable ($Y = $``recidivism within 2 years of the final sentence''). A defendant $i$ in a case $j$ recidivated ($Y_{ij} = 1$) if and only if defendant $i$'s full name appears in a case $\bar{j}$ whose starting date is within 2 years after case $j$'s final sentence's date. Importantly, case $\bar{j}$ can be about any type of crime, including more severe crimes whose maximum sentence is greater than four years, while case $j$ has to be about a crime whose maximum sentence is at most 4 years. To match defendants' names across cases, I use the Jaro–Winkler similarity metric and define a match if the similarity between full names in two different cases is greater than or equal to 0.95.\footnote{\cite{Abramitzky2019} match full names in historical Censuses in the U.S. and Norway. They define a match between two individuals if the Jaro–Winkler similarity between their names is greater than or equal to 0.90 and if their dates of birth match exactly. Since I do not observe defendants' dates of birth, I adopt a stricter Jaro-Winkler similarity threshold to define a match in my dataset.}

Even though this fuzzy matching algorithm may misclassify the outcome variable of some case-defendant pairs, I assume that this type of error is negligible and do not account for it in my empirical analysis. This assumption is plausible because Brazilian names are frequently long, containing four or more words. For instance, in my dataset, 59.4\% of the case-defendant pairs have names with four or more words, and only 6.7\% of them have names with only two words.

My covariates contain a full set of court district dummies. Since my identification strategy leverages the random allocation of judges to criminal cases, I only use districts with two or more judges.

In Section \ref{Sresults}, I estimate my results using the entire sample, but I focus my discussion on one court district: Ribeirão Preto. This district has the second largest number of judges in my sample (9 judges) and illustrates most of the issues caused by a misclassified treatment variable when our target parameter is the MTE function. Moreover, Ribeirão Preto is the seventh largest city in the State of São Paulo with more than 700 thousand inhabitants.

At the end, I impose one final restriction in my dataset: common support between the treatment and control groups. To do so, I impose that the minimum and maximum values of the instrument $Z$ are the same across both treatment arms. My final sample has 43,468 case-defendant pairs when I use the correctly classified treatment variable $D$ and 43,461 case-defendant pairs when I use the misclassified treatment variable $T$.

\subsection{Empirical Results}\label{Sresults}

In this subsection, I describe the results of my empirical analysis in four parts. I start by describing the first-stage results in Part \ref{SubFirst}. Then, I report the results for the correctly estimated MTE function in Part \ref{SubCorrect}. Furthermore, I analyze the misclassification bias in my empirical application in Part \ref{SubBias}. Lastly, I discuss the estimated bounds around the MTE function in Part \ref{SubBounds}.

\subsubsection{First-Stage Results}\label{SubFirst}

I start by presenting the results of the first stage regression in my empirical analysis. In my model, the correctly classified treatment variable $D$ (``final ruling'') is a function of instrument $Z$ (``trial judge's punishment rate'') and court district fixed effects. Following Section \ref{Sestimation} and Appendix \ref{Sestcorrectly}, I use a polynomial series to approximate the correctly measured propensity score and report the estimated coefficients of linear and quadratic models in Columns (1) and (2) in Table \ref{TabFirstStage}. Even though the quadratic coefficient is not statistically significant, I use a quadratic propensity score in my analysis because this is the most parsimonious parametric model that allows for a non-constant scaling factor in Equation \eqref{EqLivChalak}. Moreover, in Appendix \ref{AppPSsemi}, I estimate the correctly measured propensity score semiparametrically and find that this function is well-behaved, implying that a quadratic model is a good approximation for the correctly measured propensity score.

\begin{figure}[p]
	\begin{center}
		\begin{subfigure}[t]{0.47\textwidth}
			\centering
			\includegraphics[width = \textwidth]{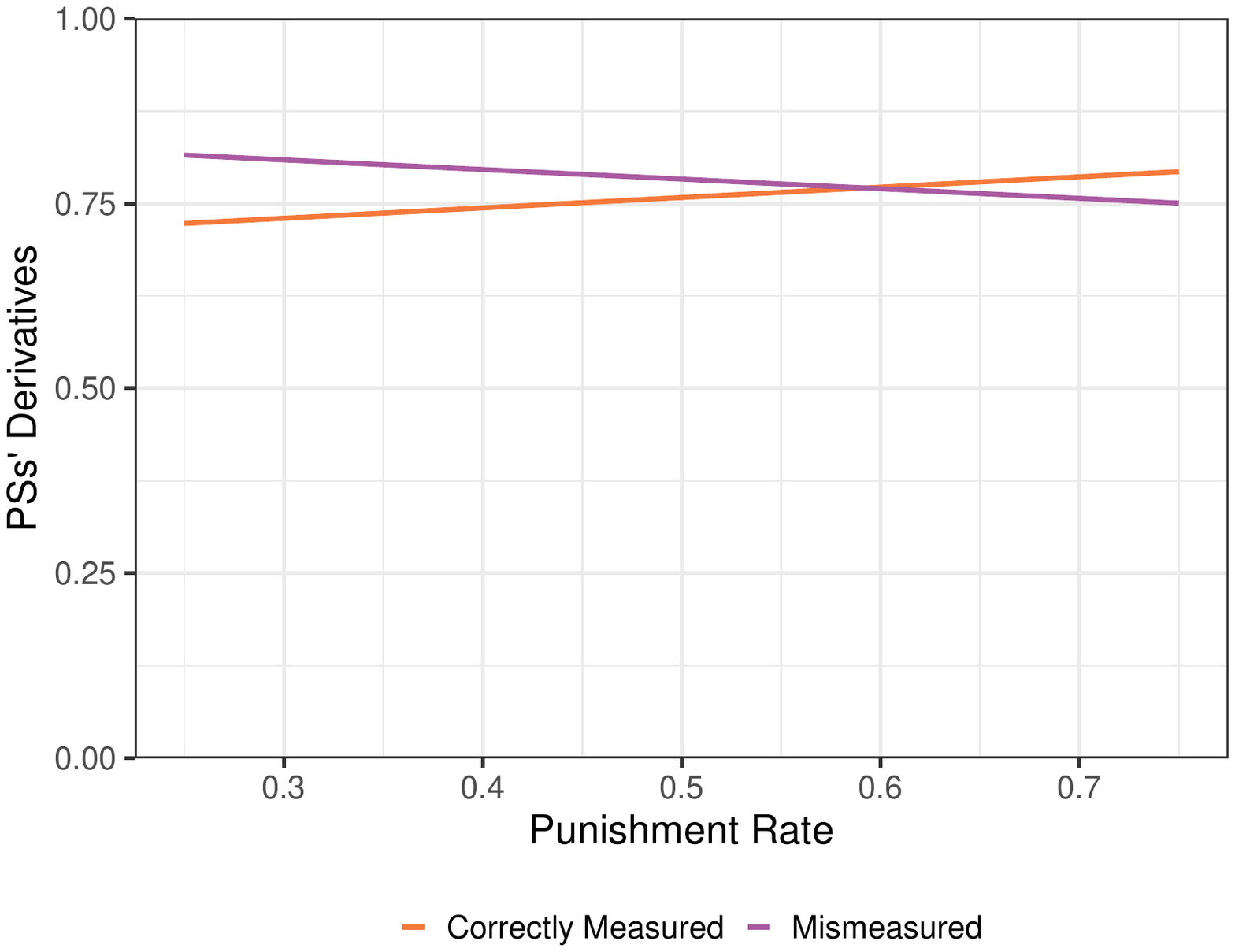}
			\caption{Propensity Score's Derivatives}
			\label{FigPSderivatives}
		\end{subfigure}
		\hfill
		\begin{subfigure}[t]{0.47\textwidth}
			\centering
			\includegraphics[width = \textwidth]{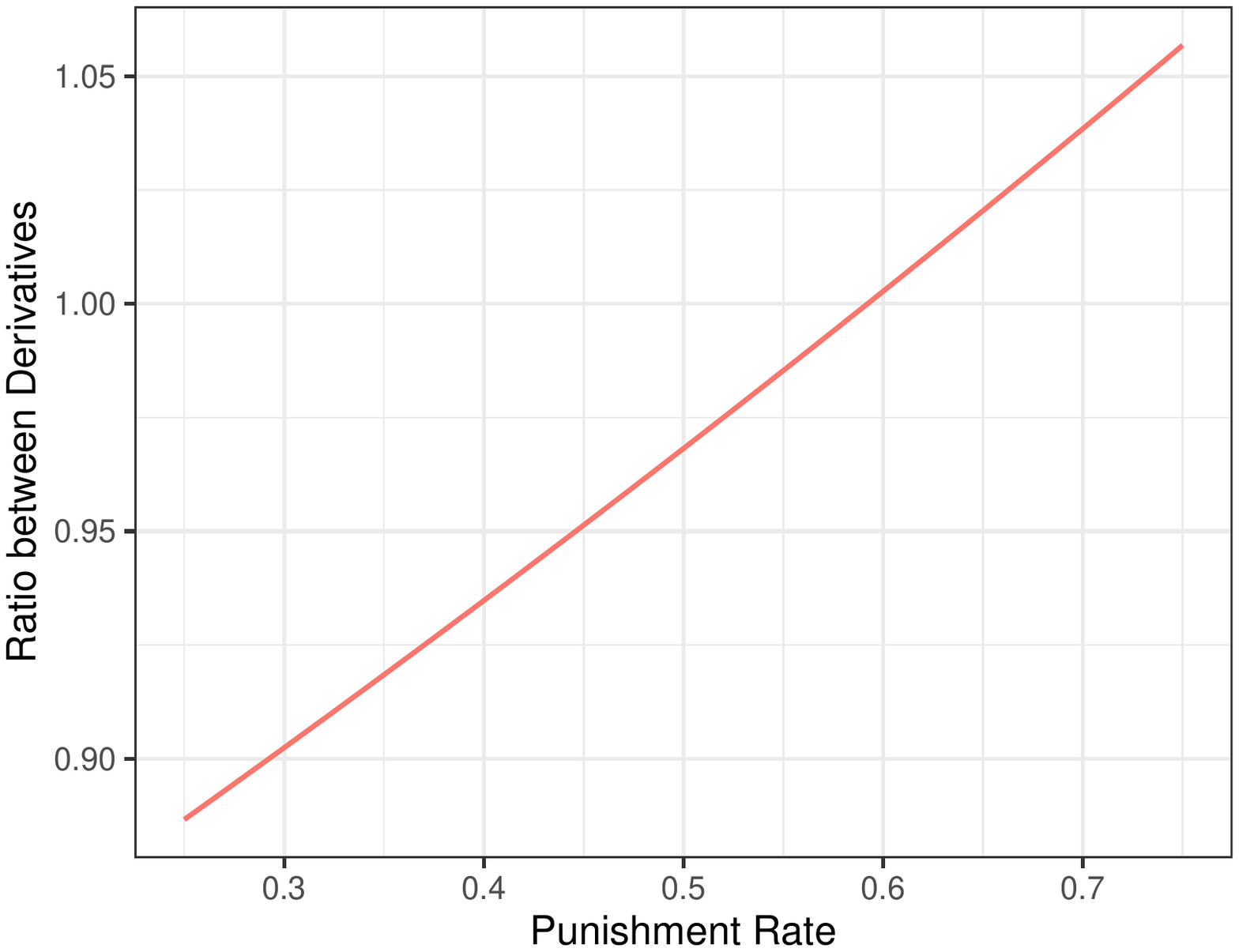}
			\caption{Ratio between Propensity Score's Derivatives $\left(\dfrac{\sfrac{dP_{D}\left(z\right)}{d z}}{\sfrac{dP_{T}\left(z\right)}{d z}}\right)$}
			\label{FigPSderivativesRatio}
		\end{subfigure}
		\caption[]{Characteristics of the Propensity Score Functions}
		\label{FigPSderiv}
	\end{center}
	\footnotesize{Notes: In Subfigure \ref{FigPSderivatives}, the orange line is the estimated derivative of the correctly measured propensity score $\left(\sfrac{dP_{D}\left(z,x\right)}{d z}\right)$ while the purple line is the estimated derivative of the mismeasured propensity score $\left(\sfrac{dP_{T}\left(z,x\right)}{d z}\right)$. Subfigure \ref{FigPSderivativesRatio} plots the ratio between these two derivatives $\left(\dfrac{\sfrac{dP_{D}\left(z,x\right)}{d z}}{\sfrac{dP_{T}\left(z,x\right)}{d z}}\right)$. All estimates are based on a quadratic version of the parametric model explained in Appendix \ref{AppEstimation}. In this model, the derivatives of the propensity score functions do not depend on the value of the covariates even though the levels of these functions depend on the value of the covariates.}
\end{figure}

Furthermore, estimating the correctly measured propensity score allows me to verify the validity of Assumption \ref{ASrank}. First, my instrument is relevant according to the F-statistic of the first-stage regression. Second, Subfigure \ref{FigPSderivatives} shows that the derivative of the correctly measured propensity score (orange line) is around 0.75. Both results imply that the first part of Assumption \ref{ASrank} is valid under the assumption that the correctly measured propensity score is quadratic, i.e., $\dfrac{dP_{D}\left(z,x\right)}{dz} \neq 0$ for every value $z$ of the instrument and every value $x$ of the covariates.\footnote{In Appendix \ref{AppPSsemi}, I estimate $P_{D}\left(\cdot,\cdot\right)$ semiparametrically. The steep inclines of the estimated functions also suggest that Assumption \ref{ASrank} holds.}

Additionally, when I assume that $D$ is not observable, my partial identification strategy (Section \ref{Sbounds}) relies on the mismeasured propensity score $\left(P_{T}\left(z,x\right) = \mathbb{E}\left[\left. T \right\vert Z = z, X = x \right]\right)$ to capture features of the MTE function $\theta\left(z,x\right)$. Following Section \ref{Sestimation} and Appendix \ref{Sestmisclassified}, I use parametric models to approximate the mismeasured propensity score and report the estimated coefficients in Columns (3) and (4) in Table \ref{TabFirstStage}.

Similarly to the correctly measured propensity score, I focus on the quadratic model for the mismeasured propensity score. First, note that my instrument is relevant according to the F-statistic of the first-stage regression. Second, Subfigure \ref{FigPSderivatives} shows that the derivative of the mismeasured propensity score (purple line) is around 0.75. Both results imply that the second part of Assumption \ref{ASrank} is valid under the assumption that the mismeasured propensity score is quadratic, i.e., $\dfrac{dP_{T}\left(z,x\right)}{dz} \neq 0$ for every value $z$ of the instrument and every value $x$ of the covariates.\footnote{In Appendix \ref{AppPSsemi}, I estimate $P_{T}\left(\cdot,\cdot\right)$ semiparametrically. The steep inclines of the estimated functions also suggest that Assumption \ref{ASrank} holds.}

More importantly, Subfigure \ref{FigPSderivativesRatio} shows that Assumptions \ref{ASsign} and \ref{ASbounded} are valid. Since the ratio between the derivatives of the propensity score functions $\left(\dfrac{\sfrac{dP_{D}\left(z\right)}{d z}}{\sfrac{dP_{T}\left(z\right)}{d z}}\right)$ is always positive, I know that these derivatives have the same sign. Note also that this ratio is bounded above by 1.13. For this reason and to be conservative, I impose that Assumption \ref{ASbounded} holds with $c = 1.2$ in my empirical analysis (Subfigure \ref{FigMTEbounds}).\footnote{When I semiparametrically estimate the correctly measured propensity score and the mismeasured propensity score (Appendix \ref{AppPSsemi}), I find that Assumption \ref{ASbounded} holds with $c = 1.19$. To be cautious about the correct value of $c$, my sensitivity analysis results impose a larger value ($c = 1.5$ in Figure \ref{FigMTESensitivity}). This difference illustrates the importance of conducting a sensitivity analysis \citep{Cinelli2019} where we gradually increase $c$ to understand the impact of allowing for a more intense misclassification problem.}

Subfigure \ref{FigPSderivativesRatio} also illustrates a key consequence of Proposition \ref{PropLIV}. The ratio between the propensity score functions can be greater or smaller than one depending on the instrument's value. Consequently, the LIV estimand's bias will attenuate or enlarge the true MTE function depending on the instrument's value.

\subsubsection{Results for the Correctly Estimated MTE Function}\label{SubCorrect}

Now, I report, in Figure \ref{FigRibeirao} the results associated with Ribeirão Preto's court district. In Subfigure \ref{FigMTEbounds}, the orange line is the correctly estimated MTE function $\theta\left(\cdot, \cdot\right)$ (Equation \eqref{EQcorrectlivest}), the dark blue line is the estimated misclassified LIV estimand (Equations \eqref{EqLIV} and \eqref{EQlivest}), and the light blue lines are the estimated upper and lower bounds of the set $\Theta_{1}$ (Proposition \ref{CorOuterMTE} and Equations \eqref{EQestupper} and \eqref{EQestlower}).\footnote{For a discussion about the estimated sign of the MTE function (Corollary \ref{CorSign}) for all court districts, see Appendix \ref{AppSign}.}

Focusing on the correctly estimated MTE function (orange line), I find that the marginal treatment effect depends on the value of the instrument. For example, my results suggest that, in Ribeirão Preto, alternative sentences (fines and community service) have almost no impact on agents who would be punished by most judges while they increase recidivism for agents who would be punished only by stricter judges (around a punishment rate equal to 0.6). However, this result should be interpreted cautiously because some of the point-estimates are larger than one, the largest possible effect given the support of the outcome variable. Importantly, when I compute bootstrapped 90\%-confidence bands (Subfigure \ref{FigMTECI}), I find that the confidence bands contain, at least partially, the zero function. This finding indicates that the effect of alternative sentences on recidivism is likely small.\footnote{A small effect of alternative sentences on recidivism is also supported by standard two-stage least squares (2SLS) regressions (Appendix \ref{App2SLS}) and by the standard analysis of the function $\mathbb{E}\left[\left. Y_{1} - Y_{0} \right\vert U = u, X = x \right]$ (Appendix \ref{AppStandardMTE}).}

\subsubsection{Analyzing the Misclassification Bias}\label{SubBias}

\begin{figure}[p]
	\begin{center}
		\begin{subfigure}[t]{0.47\textwidth}
			\centering
			\includegraphics[width = \textwidth]{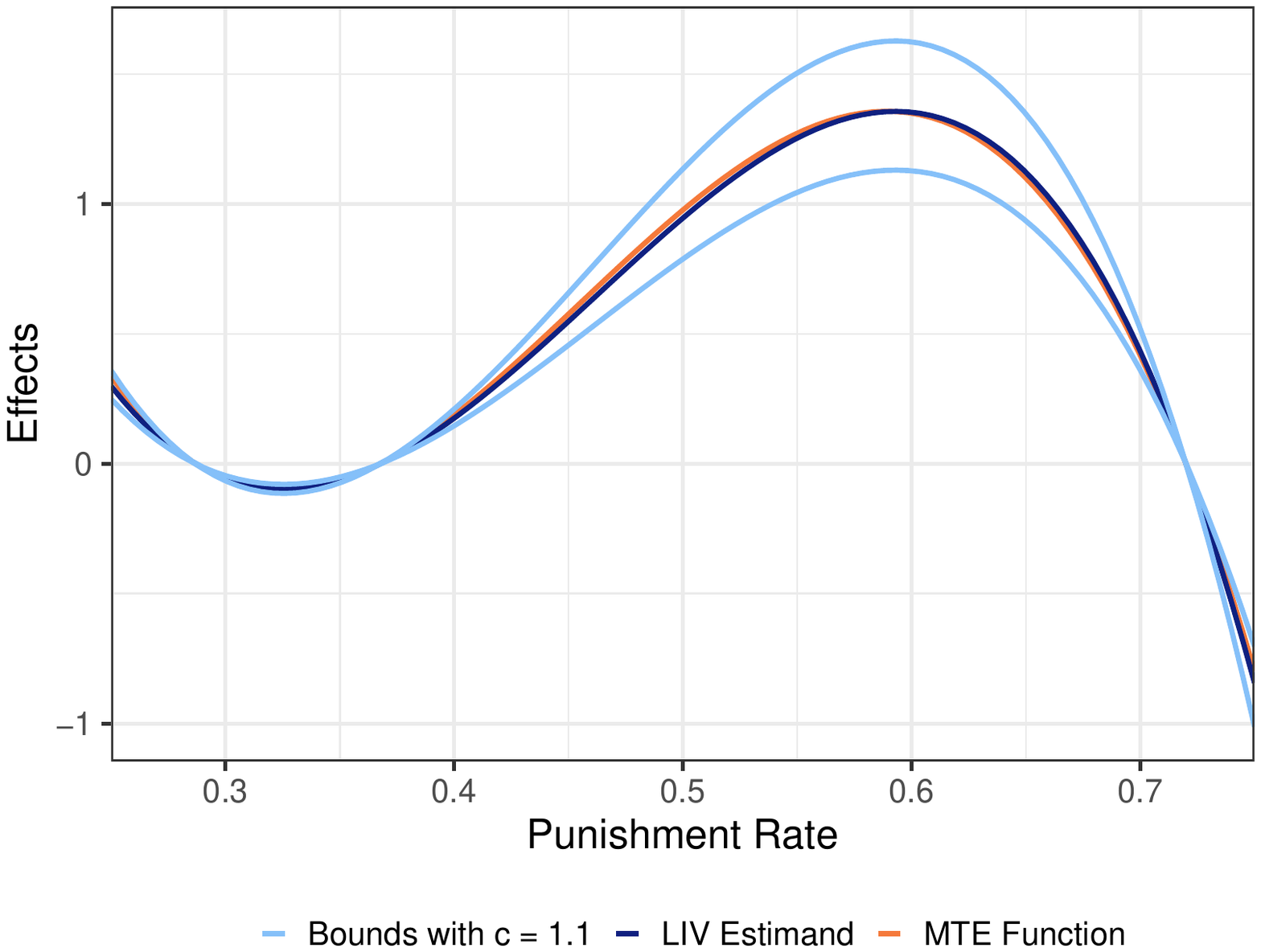}
			\caption{Bounds around the MTE function $\theta\left(\cdot, \cdot \right)$ --- Proposition \ref{CorOuterMTE}}
			\label{FigMTEbounds}
		\end{subfigure}
		\hfill
		\begin{subfigure}[t]{0.47\textwidth}
			\centering
			\includegraphics[width = \textwidth]{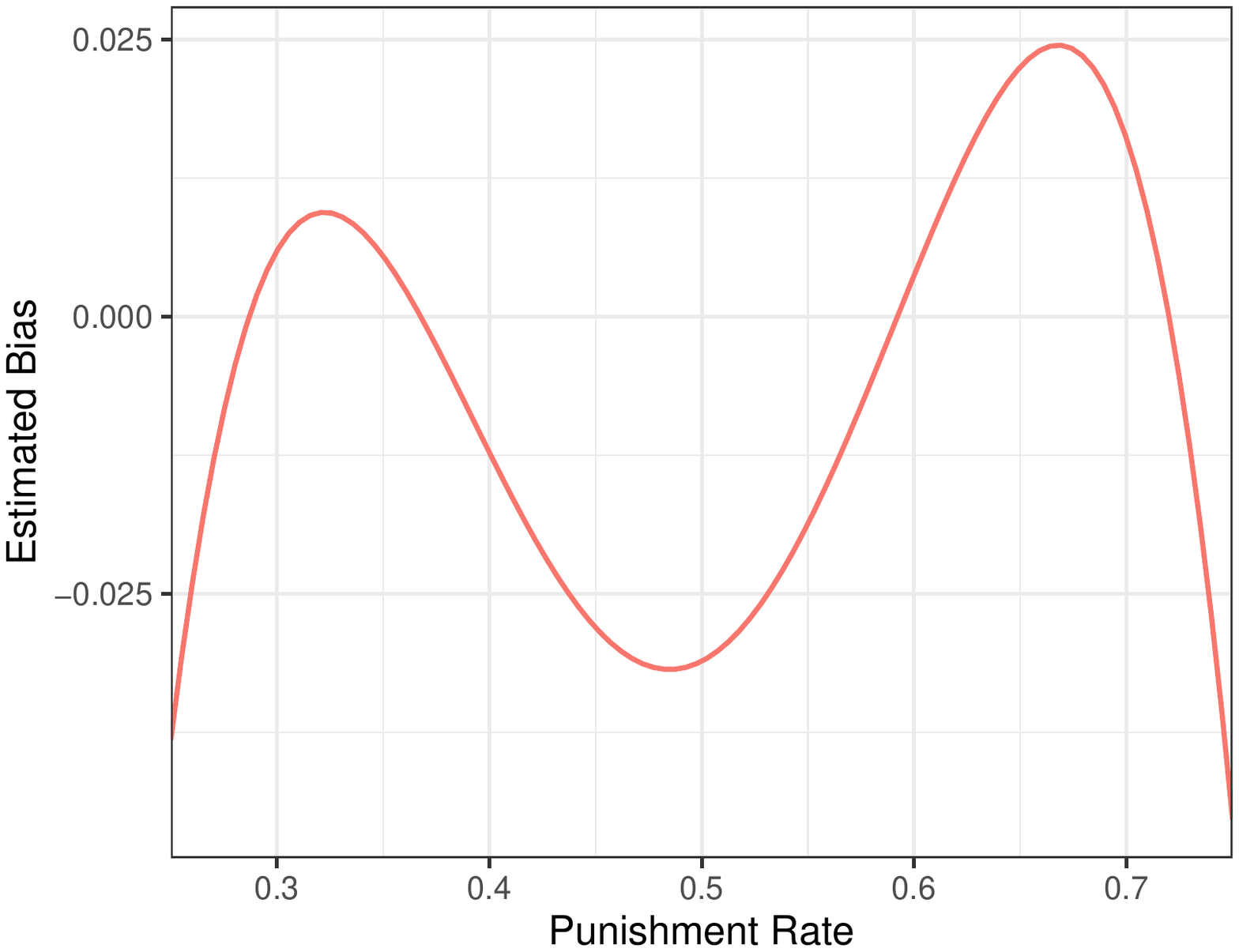}
			\caption{LIV estimand's bias}
			\label{FigMTEbias}
		\end{subfigure}
		\caption[]{Results for Ribeirão Preto}
		\label{FigRibeirao}
	\end{center}
	\footnotesize{Notes: In Subfigure \ref{FigMTEbounds}, the orange line is the estimated MTE function $\theta\left(\cdot, \cdot \right)$ based on the LIV estimator that uses the correctly classified treatment variable $D$ (Equation \eqref{EQcorrectlivest}). The dark blue line is the point-estimate of the LIV estimand that uses the misclassified treatment variable $T$ (Equations \eqref{EqLIV} and \eqref{EQlivest}). The light blue lines are the estimated upper and lower bounds of the set $\Theta_{1}$ (Proposition \ref{CorOuterMTE} and Equations \eqref{EQestupper} and \eqref{EQestlower}) based on a constant $c = 1.2$. Subfigure \ref{FigMTEbias} plots the estimated difference between the LIV estimand (dark blue line in the first subfigure) and the MTE function (orange line in the first subfigure).}
\end{figure}

Figure \ref{FigRibeirao} also illustrates the danger of ignoring misclassification of the treatment variable. By comparing the estimated MTE function (orange line) and the estimated misclassified LIV estimand (dark blue lines), I can estimate the bias that is generated by the misclassified treatment variable (Subfigure \ref{FigMTEbias}).

Note that this bias can be negative or positive depending on the sign of the true MTE function and on the size of the ratio between the derivatives of the propensity score functions $\left(\dfrac{\sfrac{dP_{D}\left(z\right)}{d z}}{\sfrac{dP_{T}\left(z\right)}{d z}}\right)$. First, when the MTE function is negative and $\dfrac{\sfrac{dP_{D}\left(z\right)}{d z}}{\sfrac{dP_{T}\left(z\right)}{d z}} < 1$ (around a punishment rate equal to 0.35), the bias is positive, implying that there is an attenuation bias. Second, when the MTE function is positive and $\dfrac{\sfrac{dP_{D}\left(z\right)}{d z}}{\sfrac{dP_{T}\left(z\right)}{d z}} < 1$ (around a punishment rate equal to 0.5), the bias is negative, implying that there is an attenuation bias. Third, when the MTE function is positive and $\dfrac{\sfrac{dP_{D}\left(z\right)}{d z}}{\sfrac{dP_{T}\left(z\right)}{d z}} > 1$ (around a punishment rate equal to 0.65), the bias is positive, implying that there is an exploding bias. Finally, when the MTE function is negative and $\dfrac{\sfrac{dP_{D}\left(z\right)}{d z}}{\sfrac{dP_{T}\left(z\right)}{d z}} > 1$ (around a punishment rate equal to 0.73), the bias is negative, implying that there is an exploding bias.

This result highlights the complexity of the misclassification bias when estimating entire functions. Note that its sign may change and it may move the estimates away from zero. Consequently, even an expert may have difficulties theorizing about its direction.

To also understand the magnitude of the misclassification bias, I report summary statistics for the LIV estimand's bias in Table \ref{TabBias}. Panel A reports the mean, the standard deviation, the minimum and the maximum of the raw differences between the estimated LIV estimand and the estimated MTE function for five values of the instrument $\left(z \in \left\lbrace .3, .4, .5, .6, .7 \right\rbrace\right)$ across 192 court districts. Note that the average bias varies for each value of the instrument and it achieves values almost as large as 3. More interestingly, the minimum bias across districts is always negative while the maximum bias is always positive.

Although Panel A shows that the estimated bias can be large, these concerning values may be due to the fact that polynomial models do not take into account that the treatment effect parameters must be between -1 and 1 in this empirical application. For this reason, Panel B reports the same summary statistics as Panel A, but, before taking the difference between the estimated LIV estimand and the estimated MTE function, it trims these estimates to be between the minimum possible treatment effect (-1) and the maximum possible treatment effect (1).

Observe that, despite a much smaller average bias, the minimum and maximum biases across districts can be as large as -.11 and .09, respectively. This result implies that the misclassification bias can be relatively large, reaching around 10\% of the maximum possible effect. Since it is a priori unknown whether any given empirical application resembles the small bias context of most court districts or the large bias context of some districts, this result illustrates the usefulness of adopting methods that account for misclassification bias when the treatment variable may be misclassified.

\subsubsection{Bounding the MTE Function with a Misclassified Treatment Variable}\label{SubBounds}

As argued in the last subsection, it is important to use methods that account for misclassification bias. When the target parameter is the MTE function, one of these methods is proposed in Sections \ref{Sframework} and \ref{Sbounds} and illustrated by the light blue lines in Subfigure \ref{FigMTEbounds}. Note that the identified set safely contains the estimated MTE function (orange line), exemplifying that the proposed method works in a specific real-world example.\footnote{Subfigure \ref{FigMTERibeiraoCI} shows bootstrapped 90\%-confidence bands around the identified set.} Moreover, the estimated sets cover the estimated MTE functions entirely in every court district in my sample.

Although successful in my empirical example, the proposed partial identification method may have worked only because my unique dataset allows me to estimate the correctly measured propensity score $P_{D}$ and find a constant $c$ that conservatively satisfies Assumption \ref{ASbounded}. Since most real-world applications do not have access to the correctly measured treatment variable $D$, this approach to choosing $c$ is frequently unfeasible.

For this reason, I propose two alternative ways to approximate the constant $c$ using data. The first one is described in Section \ref{Sframework} and illustrated in Figure \ref{FigMTESensitivity}. It consists simply of choosing different values of $c$ to understand the impact of allowing for a more intense misclassification problem. The second way to approximate the constant $c$ is described in Section \ref{Sbounds} and consists of choosing a $c$ that is associated with the most intense misclassification problem (largest plausible $c$) in applications whose set $\Theta_{1}$ is still contained within the set of possible treatment effects.\footnote{This method of choosing $c$ is a straightforward adaptation of the test proposed by \cite{Fradsen2019}.} Unfortunately, it is not possible to illustrate this method of choosing $c$ using Ribeirão Preto's results, because its LIV estimand is already greater than the largest possible treatment effect for some values of the instrument.

Finally, Subfigure \ref{FigPSderivativesRatio} suggests that Assumption \ref{ASconstant} is not valid. For this reason, I do not estimate the sharp bounds of set $\Theta_{2}$ in Proposition \ref{ThmSharpMTE} nor the bounds around standard treatment effect parameters (Corollary \ref{CorBoundTE}).

\section{Conclusion}\label{Sconclusion}

In this paper, I address a widespread empirical challenge: policy evaluation with a misclassified treatment variable. I propose a novel partial identification strategy to identify the MTE function with a misclassified treatment. This method explores restrictions on the relationship between the instrument, the misclassified treatment and the correctly measured treatment, allowing for dependence between the instrument and the misreporting decision.

As an illustration, I analyze whether alternative sentences (fines and community service) affect recidivism in Brazil. I find that the misclassification bias is empirically relevant, reaching 10\% of the largest possible treatment effect. I also find that the estimated bounds contain the correctly estimated MTE function entirely for every court district. Lastly, I find that the effect of alternative sentences on recidivism is likely small even though the point estimates present a large amount of observable geographic heterogeneity.

This result contrasts with recent findings in the empirical literature \citep{Huttunen2020,Giles2021,Klaassen2021}. While they find significant effects in different directions, my estimated parameters are rarely statistically different from zero. These differences may be due to different geographic contexts and deserve further investigation in future work.


\renewcommand{\refname}{References}

\bibliographystyle{chicago}
\bibliography{refs1}

\pagebreak

\newpage

\pagebreak

\section*{Figures}

 \clearpage

 \clearpage

 \clearpage


\pagebreak

\newpage

\pagebreak

\setcounter{table}{0}
\renewcommand\thetable{A.\arabic{table}}

\setcounter{figure}{0}
\renewcommand\thefigure{A.\arabic{figure}}

\setcounter{equation}{0}
\renewcommand\theequation{A.\arabic{equation}}

\appendix

\begin{center}
	\huge
	Supporting Information

	(Online Appendix)

\end{center}

\normalsize

All proofs are detailed in Appendix \ref{proofs}. Appendices \ref{AppRejection} and \ref{AppSufficiency} explain two problems generated by misclassification in the MTE framework. In Appendix \ref{AppAssumptions}, I discuss simple economic models that ensure that the identifying assumptions in Section \ref{Sframework} and Appendix \ref{AppStandar} hold. Furthermore, in Appendix \ref{AppExtensions}, I extend my model to encompass the cases with more than one instrumental variable or more than one misclassified treatment variable, while, in Appendix \ref{AppLATE}, I extend my analysis to a binary instrument and the LATE parameter, complementing the work done by \cite{Ura2018} and \cite{Calvi2019}. Moreover, in Appendix \ref{AppStandar}, I provide details on how to construct sharp uniform bounds around the MTE function and on how to connect these objects to bounds around common treatment effect parameters. In Appendix \ref{AppEstimation}, I provide detailed information on the estimation method discussed in Section \ref{Sestimation} and analyze the proposed estimators' performance in a Monte Carlo exercise. Furthermore, Appendices \ref{AppDataGeneral} and \ref{AppAddEmpirical} provide additional details about my empirical application. Moreover, in Appendix \ref{AppComparison}, I compare my partial identification results against the results derived by \citet[Appendix G]{Acerenza2021}. Finally, in Appendix \ref{AppDifferential}, I show that Assumptions \ref{ASindependence}-\ref{ASfinite} and \ref{ASbounded} are compatible with differential measurement error.

\clearpage

\section{Proofs of the main results} \label{proofs}

\setcounter{table}{0}
\renewcommand\thetable{A.\arabic{table}}

\setcounter{figure}{0}
\renewcommand\thefigure{A.\arabic{figure}}

\setcounter{equation}{0}
\renewcommand\theequation{A.\arabic{equation}}

\setcounter{theorem}{0}
\renewcommand\thetheorem{A.\arabic{theorem}}

\setcounter{proposition}{0}
\renewcommand\theproposition{A.\arabic{proposition}}

\setcounter{corollary}{0}
\renewcommand\thecorollary{A.\arabic{corollary}}

\setcounter{assumption}{0}
\renewcommand\theassumption{A.\arabic{assumption}}

\setcounter{remark}{0}
\renewcommand\theremark{A.\arabic{remark}}

\subsection{Proof of Proposition \ref{PropLIV}}\label{ProofLIV}
First, I prove Equation \eqref{EqLivChalak}.\footnote{In this entire appendix, LIE stands for Law of Iterated Expectations.} Fix $z \in \mathcal{Z}$ arbitrarily. Note that
\begin{align*}
	\mathbb{E}\left[\left. Y \right\vert Z = z\right] & = \mathbb{E}\left[\left. Y_{1} \cdot D + Y_{0} \cdot \left(1 - D\right) \right\vert Z = z\right] & \text{by Equation } \eqref{EqY} \\
	& = \mathbb{E}\left[\left. Y_{0} \right\vert Z = z\right] + \mathbb{E}\left[\left. \left(Y_{1} - Y_{0}\right) \cdot D \right\vert Z = z\right] \\
	& = \mathbb{E}\left[Y_{0} \right] + \mathbb{E}\left[\left. \left(Y_{1} - Y_{0}\right) \cdot D \right\vert Z = z\right] & \text{by Assumption } \ref{ASindependence} \\
	& = \mathbb{E}\left[Y_{0} \right] + \mathbb{E}\left[\left. \left(Y_{1} - Y_{0}\right) \cdot \mathbf{1}\left\lbrace U \leq P_{D}\left(Z\right) \right\rbrace \right\vert Z = z\right] & \text{by Equation } \eqref{EqD} \\
	& = \mathbb{E}\left[Y_{0} \right] + \mathbb{E}\left[\left. \left(Y_{1} - Y_{0}\right) \cdot \mathbf{1}\left\lbrace U \leq P_{D}\left(z\right) \right\rbrace \right\vert Z = z\right] \\
	& = \mathbb{E}\left[Y_{0} \right] + \mathbb{E}\left[\left(Y_{1} - Y_{0}\right) \cdot \mathbf{1}\left\lbrace U \leq P_{D}\left(z\right) \right\rbrace \right] & \text{by Assumption } \ref{ASindependence} \\
	& = \mathbb{E}\left[Y_{0} \right] + \mathbb{E}\left[\mathbf{1}\left\lbrace U \leq P_{D}\left(z\right) \right\rbrace \cdot \mathbb{E}\left[\left. Y_{1} - Y_{0}\right\vert U\right] \right] & \text{by the LIE} \\
	& = \mathbb{E}\left[Y_{0}\right] + \int_{0}^{P_{D}\left(z\right)} \mathbb{E}\left[\left. Y_{1} - Y_{0}\right\vert U = u\right] \, du,
\end{align*}
implying, by the Leibniz Integral Rule, that
\begin{equation*}
\dfrac{d\mathbb{E}\left[\left. Y \right\vert Z = z\right]}{dz} = \dfrac{dP_{D}\left(z\right)}{dz} \cdot \mathbb{E}\left[\left. Y_{1} - Y_{0} \right\vert U = P_{D}\left(z\right)\right].
\end{equation*}
Since $\dfrac{d\mathbb{E}\left[\left. T \right\vert Z = z\right]}{dz} = \dfrac{dP_{T}\left(z\right)}{dz}$ by definition, Equation \eqref{EqLivChalak} holds.

Now,  I prove Equation \eqref{EqLivHUV}. Fix $p$ in the support of $P_{T}\left(Z\right)$ arbitrarily. Note that
\begin{align*}
\mathbb{E}\left[\left. Y \right\vert P_{T}\left(Z\right) = p\right] & = \mathbb{E}\left[\left. Y \right\vert Z = P_{T}^{-1}\left(p\right)\right] \hspace{20pt} \text{ because } P_{T} \text{ is invertible} \\
& = \mathbb{E}\left[Y_{0}\right] + \int_{0}^{P_{D}\left(P_{T}^{-1}\left(p\right)\right)} \mathbb{E}\left[\left. Y_{1} - Y_{0}\right\vert U = u\right] \, du,
\end{align*}
implying, by the Leibniz Integral Rule, the Chain Rule and the Inverse Function Theorem, that Equation \eqref{EqLivHUV} holds.

\begin{remark}
    Invertibility of function $P_{T}$ is fundamental to derive Equation \eqref{EqLivHUV}. Without this assumption, the set $A\left(p\right) \coloneqq \left\lbrace z \in \mathcal{Z} \colon P_{T}\left(z\right) = p \right\rbrace$ may not be a singleton. In this case, taking the derivative of $\mathbb{E}\left[\left. Y \right\vert P_{T}\left(Z\right) = p\right]$ with respect to $p$ moves more than one margin of choice. In particular, all agent types such that $U = P_{D}\left(z\right)$ for some $z \in A\left(p\right)$ may change their choices. Consequently, allowing $P_{T}$ to be non-invertible makes the interpretation of the LIV estimand $\left(\dfrac{d\mathbb{E}\left[\left. Y \right\vert P_{T}\left(Z\right) = p \right]}{dp}\right)$ even harder than the one found in Equation \eqref{EqLivHUV}.
\end{remark}

\subsection{Proof of Proposition \ref{CorOuterMTE}}\label{ProofOuterMTE}
I, first, show that $\theta \in \Theta_{1}$. Fix $z \in \mathcal{Z}$ arbitrarily.

Observe that:
\begin{align*}
\theta\left(z\right) & = \dfrac{\sfrac{dP_{T}\left(z\right)}{dz}}{\sfrac{dP_{D}\left(z\right)}{dz}} \cdot f\left(z\right) & \text{according to Equation } \eqref{EqLivChalak} \\
& \in \left\lbrace \begin{array}{ll}
	\left[\dfrac{1}{c} \cdot f\left(z\right), c \cdot f\left(z\right) \right] & \text{if } f\left(z\right) \geq 0 \\
	\left[c \cdot f\left(z\right), \dfrac{1}{c} \cdot f\left(z\right) \right] & \text{if } f\left(z\right) < 0
\end{array}  \right. & \text{by Assumptions } \ref{ASrank} \text{ and } \ref{ASbounded}.
\end{align*}
As a consequence, I have that $\theta \in \Theta_{1}$.

For the second part of Proposition \ref{CorOuterMTE}, observe that, for any $\tilde{\theta} \in \Theta_{1}$,
\begin{align*}
\left\vert\left\vert \theta - \tilde{\theta} \right\vert\right\vert_{\infty} & = \sup_{z \in \mathcal{Z}} \left\lbrace \left\vert \theta\left(z\right) - \tilde{\theta}\left(z\right) \right\vert \right\rbrace & \text{by definition of the norm } \left\vert\left\vert \cdot \right\vert\right\vert_{\infty} \\
& \leq \sup_{z \in \mathcal{Z}} \left\lbrace \left\vert c \cdot f\left(z\right) - \dfrac{1}{c} \cdot f\left(z\right) \right\vert \right\rbrace & \text{by definition of } \Theta_{1} \\
& = \sup_{z \in \mathcal{Z}} \left\lbrace \left\vert c - \dfrac{1}{c} \right\vert \cdot \left\vert f\left(z\right) \right\vert \right\rbrace & \\
& = \left\vert c - \dfrac{1}{c} \right\vert \cdot \sup_{z \in \mathcal{Z}} \left\lbrace \left\vert f\left(z\right) \right\vert \right\rbrace & \\
& = \left\vert \dfrac{c^{2} - 1}{c} \right\vert \cdot \left\vert\left\vert f \right\vert\right\vert_{\infty} & \text{by definition of the norm } \left\vert\left\vert \cdot \right\vert\right\vert_{\infty}
\end{align*}
and, for any $d \in \mathbb{R}_{++}$,
\begin{align*}
\left\vert\left\vert \theta - \tilde{\theta} \right\vert\right\vert_{d} & = \left( \int \left\vert \theta\left(z\right) - \tilde{\theta}\left(z\right) \right\vert^{d} \, dz \right)^{\sfrac{1}{d}} & \text{by definition of the norm } \left\vert\left\vert \cdot \right\vert\right\vert_{d} \\
& \leq \left( \int \left\vert c \cdot f\left(z\right) - \dfrac{1}{c} \cdot f\left(z\right) \right\vert^{d} \, dz \right)^{\sfrac{1}{d}} & \text{by definition of } \Theta_{1} \\
& = \left( \int \left\vert c - \dfrac{1}{c} \right\vert^{d} \cdot \left\vert f\left(z\right) \right\vert^{d} \, dz \right)^{\sfrac{1}{d}} \\
& = \left\vert c - \dfrac{1}{c} \right\vert \cdot \left( \int \left\vert f\left(z\right) \right\vert^{d} \, dz \right)^{\sfrac{1}{d}} & \\
& = \left\vert \dfrac{c^{2} - 1}{c} \right\vert \cdot \left\vert\left\vert f \right\vert\right\vert_{d} & \text{by definition of the norm } \left\vert\left\vert \cdot \right\vert\right\vert_{d}.
\end{align*}

\newpage

\section{Incorrect Rejection of the Generalized Roy Model}\label{AppRejection}

\setcounter{table}{0}
\renewcommand\thetable{B.\arabic{table}}

\setcounter{figure}{0}
\renewcommand\thefigure{B.\arabic{figure}}

\setcounter{equation}{0}
\renewcommand\theequation{B.\arabic{equation}}

\setcounter{theorem}{0}
\renewcommand\thetheorem{B.\arabic{theorem}}

\setcounter{proposition}{0}
\renewcommand\theproposition{B.\arabic{proposition}}

\setcounter{corollary}{0}
\renewcommand\thecorollary{B.\arabic{corollary}}

\setcounter{assumption}{0}
\renewcommand\theassumption{B.\arabic{assumption}}

Without measurement error, \citet[Theorem 1]{Heckman2005} derive testable implications of Equations \eqref{EqY} and \eqref{EqD} and Assumptions \ref{ASindependence}-\ref{ASfinite}. For example, they show that $\mathbb{E}\left[\left. D \cdot g\left(Y\right) \right\vert P_{D}\left(Z\right) = p\right]$ is an increasing function of $p$, where $g \colon \mathbb{R} \rightarrow \mathbb{R}_{+}$. However, when using the mismeasured treatment variable and the mismeasured propensity score, this testable restriction of the model may not hold when naively using the mismeasured treatment variable and propensity score instead of the correctly measured ones. To show this result, I find a joint distribution for $\left(T, D, Y_{0}, Y_{1}, U, Z\right)$ such that Equations \eqref{EqY} and \eqref{EqD} and Assumptions \ref{ASindependence}-\ref{ASfinite} hold and $\mathbb{E}\left[\left. T \cdot g\left(Y\right) \right\vert P_{T}\left(Z\right) = p\right]$ is not an increasing function of $p$.

Let $Z \sim Uniform\left[0, 1\right]$, $U \sim Uniform\left[0, 1\right]$, $Y_{0} \sim Bernouille\left(\sfrac{1}{5}\right)$ and $Y_{1} \sim Bernouille\left(\sfrac{4}{5}\right)$ be mutually independent random variables. Define $g \colon \mathbb{R} \rightarrow \mathbb{R}_{+}$ such that $g(y) = y$ for any $y \in \mathbb{R}$, $D \coloneqq \mathbf{1}\left\lbrace U \leq Z \right\rbrace$ and $T \coloneqq \mathbf{1}\left\lbrace U \leq 1 - 2 \cdot Z \right\rbrace$. Note that $P_{D}\left(z\right) = z$ and $P_{T}\left(z\right) = 1 - 2 \cdot z$ for any $z \in \left[0, \sfrac{1}{2}\right]$. Observe that
\begin{align*}
& \mathbb{E}\left[\left. T \cdot Y \right\vert P_{T}\left(Z\right) = \sfrac{1}{2}\right] - \mathbb{E}\left[\left. T \cdot Y \right\vert P_{T}\left(Z\right) = \sfrac{1}{3}\right] \\
& \hspace{20pt} = \mathbb{E}\left[\left. T \cdot Y \right\vert Z = \sfrac{1}{4}\right] - \mathbb{E}\left[\left. T \cdot Y \right\vert Z = \sfrac{1}{3}\right] \\
& \hspace{20pt} = \mathbb{E}\left[\left. T \cdot \left(D \cdot Y_{1} + \left(1 - D\right) \cdot Y_{0} \right) \right\vert Z = \sfrac{1}{4}\right] - \mathbb{E}\left[\left. T \cdot \left(D \cdot Y_{1} + \left(1 - D\right) \cdot Y_{0} \right) \right\vert Z = \sfrac{1}{3}\right] \\
& \hspace{20pt} = \mathbb{E}\left[\left. T \cdot Y_{0} \right\vert Z = \sfrac{1}{4}\right] + \mathbb{E}\left[\left. T \cdot D \cdot \left(  Y_{1} -  Y_{0} \right) \right\vert Z = \sfrac{1}{4}\right] \\
& \hspace{40pt} - \mathbb{E}\left[\left. T \cdot Y_{0} \right\vert Z = \sfrac{1}{3}\right] - \mathbb{E}\left[\left. T \cdot D \cdot \left(  Y_{1} -  Y_{0} \right) \right\vert Z = \sfrac{1}{3}\right] \\
& \hspace{20pt} = \mathbb{E}\left[\left. \mathbf{1}\left\lbrace U \leq 1 - 2 \cdot Z \right\rbrace \cdot Y_{0} \right\vert Z = \sfrac{1}{4}\right] + \mathbb{E}\left[\left. \mathbf{1}\left\lbrace U \leq 1 - 2 \cdot Z \right\rbrace \cdot \mathbf{1}\left\lbrace U \leq Z \right\rbrace \cdot \left(  Y_{1} -  Y_{0} \right) \right\vert Z = \sfrac{1}{4}\right] \\
& \hspace{40pt} - \mathbb{E}\left[\left. \mathbf{1}\left\lbrace U \leq 1 - 2 \cdot Z \right\rbrace \cdot Y_{0} \right\vert Z = \sfrac{1}{3}\right] - \mathbb{E}\left[\left. \mathbf{1}\left\lbrace U \leq 1 - 2 \cdot Z \right\rbrace \cdot \mathbf{1}\left\lbrace U \leq Z \right\rbrace \cdot \left(  Y_{1} -  Y_{0} \right) \right\vert Z = \sfrac{1}{3}\right] \\
& \hspace{20pt} = \mathbb{E}\left[\mathbf{1}\left\lbrace U \leq 1 - 2 \cdot \sfrac{1}{4} \right\rbrace \cdot Y_{0} \right] + \mathbb{E}\left[ \mathbf{1}\left\lbrace U \leq 1 - 2 \cdot \sfrac{1}{4} \right\rbrace \cdot \mathbf{1}\left\lbrace U \leq \sfrac{1}{4} \right\rbrace \cdot \left(  Y_{1} -  Y_{0} \right) \right] \\
& \hspace{40pt} - \mathbb{E}\left[\mathbf{1}\left\lbrace U \leq 1 - 2 \cdot \sfrac{1}{3} \right\rbrace \cdot Y_{0} \right] - \mathbb{E}\left[\mathbf{1}\left\lbrace U \leq 1 - 2 \cdot \sfrac{1}{3} \right\rbrace \cdot \mathbf{1}\left\lbrace U \leq \sfrac{1}{3} \right\rbrace \cdot \left(  Y_{1} -  Y_{0} \right) \right] \\
& \hspace{20pt} = \dfrac{\mathbb{E}\left[Y_{0}\right]}{2} + \dfrac{\mathbb{E}\left[Y_{1} - Y_{0}\right]}{4} - \dfrac{\mathbb{E}\left[Y_{0}\right]}{3} - \dfrac{\mathbb{E}\left[Y_{1} - Y_{0}\right]}{3} \\
& \hspace{20pt} = \dfrac{\mathbb{E}\left[Y_{0}\right]}{4} - \dfrac{\mathbb{E}\left[Y_{1}\right]}{12} = \dfrac{1}{20} - \dfrac{1}{15} = -\dfrac{1}{60},
\end{align*}
implying  that $\mathbb{E}\left[\left. T \cdot Y \right\vert P_{T}\left(Z\right) = p\right]$ is not an increasing function of $p$.

\newpage

\section{Failure of Index Sufficiency}\label{AppSufficiency}
\setcounter{table}{0}
\renewcommand\thetable{C.\arabic{table}}

\setcounter{figure}{0}
\renewcommand\thefigure{C.\arabic{figure}}

\setcounter{equation}{0}
\renewcommand\theequation{C.\arabic{equation}}

\setcounter{theorem}{0}
\renewcommand\thetheorem{C.\arabic{theorem}}

\setcounter{proposition}{0}
\renewcommand\theproposition{C.\arabic{proposition}}

\setcounter{corollary}{0}
\renewcommand\thecorollary{C.\arabic{corollary}}

\setcounter{assumption}{0}
\renewcommand\theassumption{C.\arabic{assumption}}

Without measurement error, \cite{Heckman2006} show that the propensity score satisfies index sufficiency, i.e., $\mathbb{E}\left[\left. Y \right\vert Z = z\right] = \mathbb{E}\left[\left. Y \right\vert P_{D}\left(Z\right) = p\right]$ for any $z \in \mathcal{Z}$ and $p \in \left[0, 1\right]$ such that $p = P_{D}\left(z\right)$. However, the mismeasured propensity score may not satisfy this property. To prove this result, I find a joint distribution for $\left(T, D, Y_{0}, Y_{1}, U, Z\right)$ such that Equations \eqref{EqY} and \eqref{EqD} and Assumptions \ref{ASindependence}-\ref{ASfinite} hold and $\mathbb{E}\left[\left. Y \right\vert Z = z\right] \neq \mathbb{E}\left[\left. Y \right\vert P_{T}\left(Z\right) = p\right]$ for some $z \in \mathcal{Z}$ and $p \in \left[0, 1\right]$ such that $p = P_{T}\left(z\right)$.

Let $Z \sim Uniform\left[0, 1\right]$, $U \sim Uniform\left[0, 1\right]$, $Y_{0} \sim N\left(0, 1\right)$ and $Y_{1} \sim N\left(1, 1\right)$ be mutually independent random variables. Define $D \coloneqq \mathbf{1}\left\lbrace U \leq Z \right\rbrace$ and $T \coloneqq \mathbf{1}\left\lbrace U \leq 4 \cdot Z^{2} - 4 \cdot Z + 1 \right\rbrace$. Note that $P_{D}\left(z\right) = z$ and $P_{T}\left(z\right) = 4 \cdot z^{2} - 4 \cdot z + 1$ for any $z \in \left[0, 1\right]$. Observe that $P_{T}\left(0\right) = 1$, $P_{T}\left(1\right) = 1$,
\begin{equation*}
\mathbb{E}\left[\left. Y \right\vert Z = 1\right] = \mathbb{E}\left[\left. Y \right\vert P_{D}\left(Z\right) = 1\right] = \mathbb{E}\left[ Y_{1} \right] = 1
\end{equation*}
and
\begin{align*}
\mathbb{E}\left[\left. Y \right\vert P_{T}\left(Z\right) = 1\right] & = \mathbb{E}\left[\left. Y \right\vert Z \in \left\lbrace 0, 1\right\rbrace\right] \\
& = \mathbb{E}\left[\left. Y \right\vert Z = 0\right] \cdot \mathbb{P}\left[\left. Z = 0 \right\vert Z \in \left\lbrace 0, 1\right\rbrace \right] + \mathbb{E}\left[\left. Y \right\vert Z = 1\right] \cdot \mathbb{P}\left[\left. Z = 1 \right\vert Z \in \left\lbrace 0, 1\right\rbrace \right] \\
& = \dfrac{\mathbb{E}\left[ Y_{0} \right]}{2} + \dfrac{\mathbb{E}\left[ Y_{1} \right]}{2} \\
& = \dfrac{1}{2},
\end{align*}
implying that the mismeasured propensity score does not satisfy index sufficiency.

\newpage

\section{Simple Sufficient Conditions for Assumptions \ref{ASsign}, \ref{ASbounded} and \ref{ASsharp}}\label{AppAssumptions}
\setcounter{table}{0}
\renewcommand\thetable{D.\arabic{table}}

\setcounter{figure}{0}
\renewcommand\thefigure{D.\arabic{figure}}

\setcounter{equation}{0}
\renewcommand\theequation{D.\arabic{equation}}

\setcounter{theorem}{0}
\renewcommand\thetheorem{D.\arabic{theorem}}

\setcounter{proposition}{0}
\renewcommand\theproposition{D.\arabic{proposition}}

\setcounter{corollary}{0}
\renewcommand\thecorollary{D.\arabic{corollary}}

\setcounter{assumption}{0}
\renewcommand\theassumption{D.\arabic{assumption}}

\setcounter{example}{0}
\renewcommand\theexample{D.\arabic{example}}

In this appendix, I state simple sufficient conditions that, when combined with Equations \eqref{EqY}-\eqref{EqD} and Assumptions \ref{ASindependence}-\ref{ASfinite}, ensure that Assumptions \ref{ASsign}, \ref{ASbounded} and \ref{ASsharp} hold. Moreover, these conditions are stated with specific empirical contexts to illustrate the broad applicability of my theoretical framework to many applied problems. At the end, I impose restrictions on the model's primitives that imply Assumptions \ref{ASsign}, \ref{ASbounded} and \ref{ASsharp}.

\subsection{Main Examples}\label{AppExamples}

\begin{example}[Collecting New Evidence in a Criminal Case]\label{Esign}
	Let $U_{T}$ be the amount of evidence in favor of the defendant when the trial judge makes her decision and $U$ be the amount of evidence in favor of the defendant in the Appeals Court. I assume that either the district attorney or the defendant's lawyers can collect new evidence between the trial judge's decision and the Appeals judge's analysis, and I define $V \coloneqq U - U_{T}$ as the new evidence that was collected. Moreover, I assume that the trial judge and the Appeals judge are equally strict, i.e., their sentences only differ because of the new evidence.\footnote{Assuming that two judges are equally strict is plausible. For example, consider the case when the defendant is arrested before the trial and is sent to prison while awaiting trial. In Brazil, if a lawyer petitions the court for a writ of \emph{habeas corpus} multiple times and brings new evidence each time, the same trial judge will analyze each petition.} Mathematically, I write that $$D \coloneqq \mathbf{1}\left\lbrace U \leq P_{D}\left(Z\right) \right\rbrace$$ and $$T \coloneqq \mathbf{1}\left\lbrace U_{T} \leq P_{D}\left(Z\right) \right\rbrace = \mathbf{1}\left\lbrace U - V \leq P_{D}\left(Z\right) \right\rbrace.$$

	Assuming that $Z \independent \left(U_{T}, U\right)$ and denoting $F_{U_{T}}$ as the distribution function of $U_{T}$ and $f_{U_{T}}$ as the density function of $U_{T}$, I can show that, for any $z \in \mathcal{Z}$, $$P_{T}\left(z\right) \coloneqq \mathbb{P}\left[\left. T = 1 \right\vert Z = z\right] = F_{U_{T}}\left(P_{D}\left(z\right)\right),$$ implying, by the chain rule, that $$\dfrac{d P_{T}\left(z\right)}{d z} = f_{U_{T}}\left(P_{D}\left(z\right)\right) \cdot \dfrac{d P_{D}\left(z\right)}{d z}.$$ Consequently, Assumption \ref{ASsign} holds because a density function is always positive.
\end{example}

\begin{example}[Reversing Trial Judges' Rulings]\label{EboundedAppeal}
	Let $Z$ be the punishment rate of the trial judge with $\mathcal{Z} = \left[0,1\right]$. Let $D$ be the final ruling in each case and $T$ be the trial judge's ruling. The trial judge's ruling is reversed if $V > P_{NR}\left(z\right)$, where $P_{NR}:\left[0,1\right]\rightarrow\left[0,1\right]$ captures the quality of the trial judge's ruling and is continuously differentiable with $P_{NR}\left(z\right) \geq \dfrac{1}{2} + \epsilon$ and $\dfrac{dP_{NR}\left(z\right)}{dz} \geq \epsilon$ for some $\epsilon \in \left(0, \sfrac{1}{3}\right)$ and for any $z \in \left[0,1\right]$, and $V \sim Uniform\left[0, 1\right]$ captures the minimum trial judge's ruling's quality that Appeals Court finds acceptable with $V \independent \left(Z, D\right)$. Mathematically, we have that $$T = \mathbf{1}\left\lbrace V \leq P_{NR}\left(z\right) \right\rbrace \cdot D + \mathbf{1}\left\lbrace V > P_{NR}\left(z\right) \right\rbrace \cdot \left(1 - D\right).$$ Moreover, define $P_{D}:\left[0,1\right]\rightarrow\left[0,1\right]$ such that $P_{D}\left(z\right) = \mathbb{P}\left[\left. D = 1 \right\vert Z = z\right]$ for any $z \in \left[0,1\right]$ and assume that $P_{D}$ is continuously differentiable with $P_{D}\left(z\right) \geq \dfrac{1}{2} + \epsilon$ and $\dfrac{dP_{D}\left(z\right)}{dz} \geq \epsilon$ for any $z \in \left[0,1\right]$.

	Note that, for any $z \in \left[0,1\right]$
	\begin{align*}
	P_{T}\left(z\right) & \coloneqq \mathbb{P}\left[\left. T = 1 \right\vert Z = z\right] \\
	& = \mathbb{P}\left[\left. T = 1 \right\vert Z = z, D = 1\right] \cdot \mathbb{P}\left[\left. D = 1 \right\vert Z = z\right] + \mathbb{P}\left[\left. T = 1 \right\vert Z = z, D = 0\right] \cdot \mathbb{P}\left[\left. D = 0 \right\vert Z = z\right] \\
	& = \mathbb{P}\left[\left. V \leq P_{NR}\left(z\right) \right\vert Z = z, D = 1\right] \cdot P_{D}\left(z\right) + \mathbb{P}\left[\left. V > P_{NR}\left(z\right) \right\vert Z = z, D = 0\right] \cdot \left(1 - P_{D}\left(z\right)\right) \\
	& = P_{NR}\left(z\right) \cdot P_{D}\left(z\right) + \left(1 - P_{NR}\left(z\right)\right) \cdot \left(1 - P_{D}\left(z\right)\right),
	\end{align*}
	implying that $$\dfrac{dP_{T}\left(z\right)}{dz} = \left(- 1 + 2 \cdot P_{NR}\left(z\right)\right) \cdot \dfrac{dP_{D}\left(z\right)}{dz} + \left(- 1 + 2 \cdot P_{D}\left(z\right)\right) \cdot \dfrac{dP_{NR}\left(z\right)}{dz}.$$

	Observe that, given the functional form assumptions made in this example, $\dfrac{dP_{T}\left(z\right)}{dz} \geq 4 \cdot \epsilon^{2} > 0$. Consequently, we can define $g:\left[0,1\right]\rightarrow\mathbb{R}_{++}$ such that, for any $z \in \left[0,1\right]$, $$g\left(z\right) = \dfrac{\sfrac{dP_{D}\left(z\right)}{dz}}{\sfrac{dP_{T}\left(z\right)}{dz}} = \dfrac{1}{- 1 + 2 \cdot P_{NR}\left(z\right)} - \left(\dfrac{- 1 + 2 \cdot P_{D}\left(z\right)}{- 1 + 2 \cdot P_{NR}\left(z\right)}\right) \cdot \dfrac{\sfrac{dP_{NR}\left(z\right)}{dz}}{\sfrac{dP_{T}\left(z\right)}{dz}}.$$ Observe that, given the functional form assumptions made in this example, $g$ is positive and continuous. Since $Z = \left[0, 1\right]$ is compact, $g$ attains its maximum and minimum, implying that there exists a constant $c \in \left[1, + \infty\right)$ such that Assumption \ref{ASbounded} holds.

	Moreover, note that the smallest constant $c \in \left[1, + \infty\right)$ that satisfies Assumption \ref{ASbounded} depends on the maximum and the minimum of function $g$. Consequently, this constant depends on the misclassification rate through the function $P_{NR}$. This dependence implies that varying the misclassification rate may require varying the width of the bounds in Proposition \ref{CorOuterMTE}.
\end{example}

\begin{example}[Misreporting due to Stigma]\label{Esharp}
	\cite{Hernandez2007} documents that welfare participation is measured with error. I am interested in evaluating the marginal treatment effect of welfare participation on children's health outcomes. Since people are likely to lie about welfare participation because of social stigma, I assume that the measurement error is unidirectional, i.e., the person will truthfully report no participation ($T = 0$) if they do not receive welfare benefits ($D = 0$), but the person may report participation ($T = 1$) or no participation ($T = 0$) if they receive welfare benefits ($D = 1$). Formally, I assume that $$T = \mathbf{1}\left\lbrace V \leq 0.95\right\rbrace \cdot D,$$ where $V \sim Uniform\left[0, 1\right]$ captures how much the individual cares about others' opinions and $V \independent \left(Z, D\right)$. In this case, I have that, for any $z \in \mathcal{Z}$, $P_{T}\left(z\right) = 0.95 \cdot P_{D}\left(z\right)$, implying that, if $P_{T}$ is invertible, then Assumption \ref{ASconstant} holds for $c = \dfrac{20}{19}$. Moreover, if $P_{T}$ is not invertible, then Assumptions \ref{ASsharp} and \ref{ASconstant} cannot hold in this context.
\end{example}

\subsection{Assumption \ref{ASsign}: Repeated Decision Making}\label{AppSeq}
Similarly to Example \ref{Esign}, I analyze another problem when agents with similar behavior have to take the same decision at two points in time and new information is acquired between these two time periods.

Suppose that an entrepreneur wants to borrow from a bank to implement a project and talks to her account manager every time she needs a loan. In this context, $U_{T}$ is information about the project's profitability when the entrepreneur talks to her account manager for the first time, $U$ is information about the project's profitability when she talks to her account manager for the second time, $Z$ is the account's manager willingness to lend, $T$ is whether the entrepreneur got a loan the first time and $D$ is whether the entrepreneur got a loan at some point in time. Assumption \ref{ASsign} holds in this context under the same assumptions explained in Example \ref{Esign}. Moreover, Assumption \ref{ASbounded} may hold depending on the distribution of $U_{T}$.

\subsection{Assumption \ref{ASbounded}: Random Miscoding}\label{Ebounded}

\cite{Black2003} documents that schooling is measured with error in the American Census and in the Current Population Survey (CPS). I want to identify the marginal treatment effect of college attendance ($D$) on wages. To account for measurement error in the treatment variable, I assume that all errors are due to individual inattention when filling out the survey questionnaire, i.e., the individual checked the wrong box by mistake. Formally, I assume that the individual's answer to the question ``Have you ever attended college?'' ($T$) is given by  $$T = \mathbf{1}\left\lbrace V \leq 0.95 \right\rbrace \cdot D + \mathbf{1}\left\lbrace V > 0.95 \right\rbrace \cdot \left(1 - D\right),$$ where $V \sim Uniform\left[0, 1\right]$ captures the individual attention and $V \independent \left(Z, D\right)$. Note that, in this case, 5\% of the individuals check the wrong box due to inattention. Observe also that, for any $z \in \mathcal{Z}$,
\begin{align*}
P_{T}\left(z\right) & \coloneqq \mathbb{P}\left[\left. T = 1 \right\vert Z = z\right] \\
& = \mathbb{P}\left[\left. T = 1 \right\vert Z = z, D = 1\right] \cdot \mathbb{P}\left[\left. D = 1 \right\vert Z = z\right] + \mathbb{P}\left[\left. T = 1 \right\vert Z = z, D = 0\right] \cdot \mathbb{P}\left[\left. D = 0 \right\vert Z = z\right] \\
& = 0.95 \cdot P_{D}\left(z\right) + 0.05 \cdot \left(1 - P_{D}\left(z\right)\right) \\
& = 0.05 + 0.9 \cdot  P_{D}\left(z\right),
\end{align*}
implying that $\dfrac{\sfrac{d P_{D}\left(z\right)}{d z}}{\sfrac{d P_{T}\left(z\right)}{d z}} = \dfrac{10}{9}$. Consequently, Assumption \ref{ASbounded} holds for $c = \dfrac{10}{9}$, implying that Assumption \ref{ASsign} holds too.

\subsection{Assumption \ref{ASconstant}: Dynamic Setting}\label{AppDynamic}
In Example \ref{Esharp}, I show that Assumption \ref{ASconstant} holds if $T = \mathbf{1}\left\lbrace V \leq 0.95\right\rbrace \cdot D$, $V \sim Uniform\left[0, 1\right]$, $V \independent \left(Z, D\right)$ and $P_{T}$ is invertible. In this subsection, I provide two dynamic examples that fit into the same framework.

\subsubsection{Eviction Order and Execution}
\citet[Appendix B]{Humphries2019} discuss the effect of an eviction order when the sheriff may or may not execute the order after the judge's decision. In this case, there are three counterfactual outcomes: $Y_{0}$ is the outcome of interest when there is no eviction order; $Y_{1}$ is the outcome of interest when there is an eviction order, but the sheriff does not execute the order; and $Y_{2}$ is the outcome of interest when there is an eviction order and the sheriff executes the order. Let $D$ denote the eviction order and $T$ denote the execution of the eviction order. Consequently, the observed outcome variable is given by $Y = Y_{0} \cdot \left(1 - D\right) + Y_{1} \cdot D \cdot \left(1 - T\right) + Y_{2} \cdot D \cdot T$.

\begin{figure}[!htb]
	\begin{center}
		\subfloat[Eviction Order and Execution\label{FigEviction}]{\includegraphics[width = .45 \columnwidth]{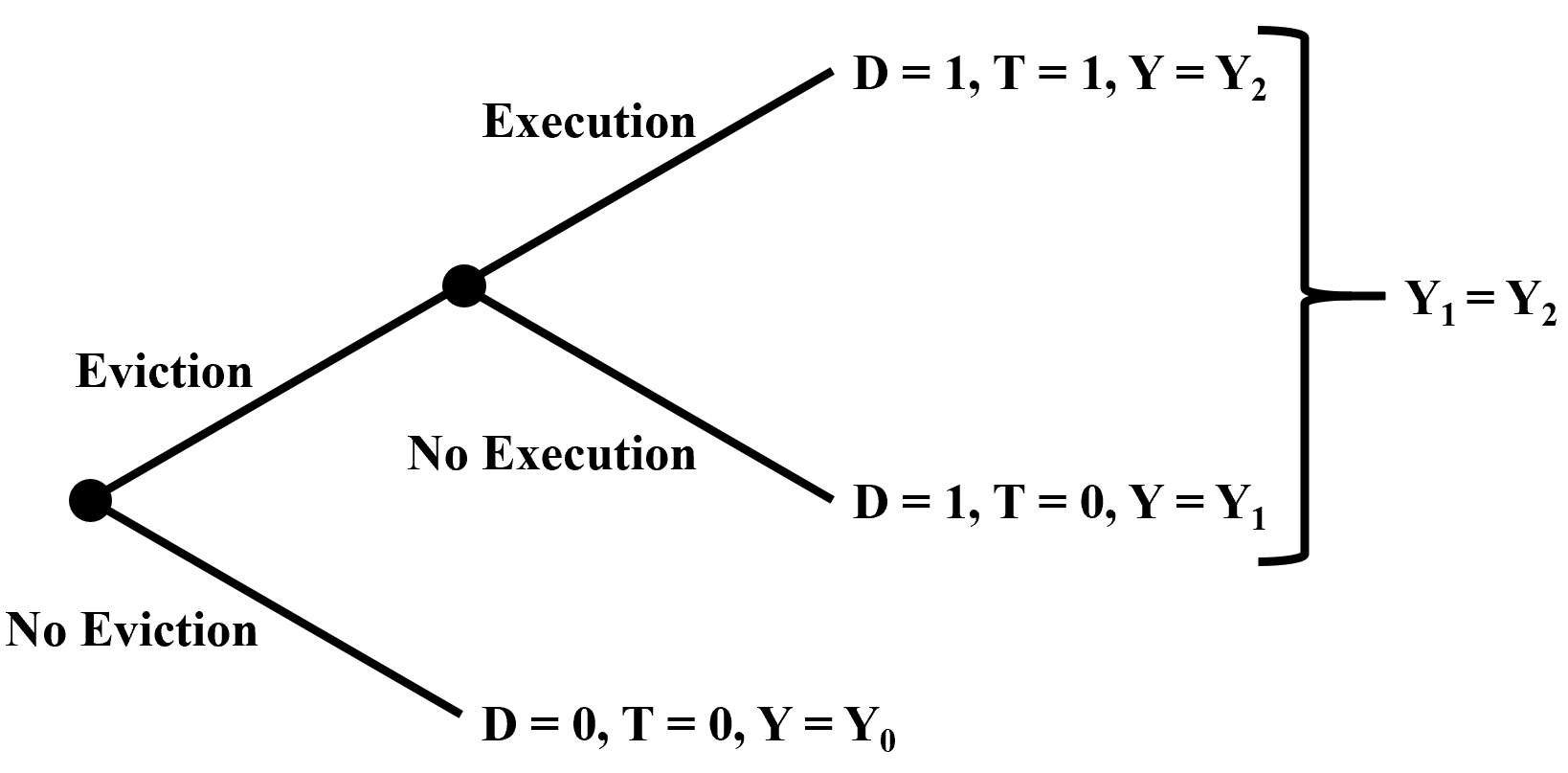}} \quad
		\subfloat[Returns to High School Education\label{FigHighSchool}]{\includegraphics[width = .45 \columnwidth]{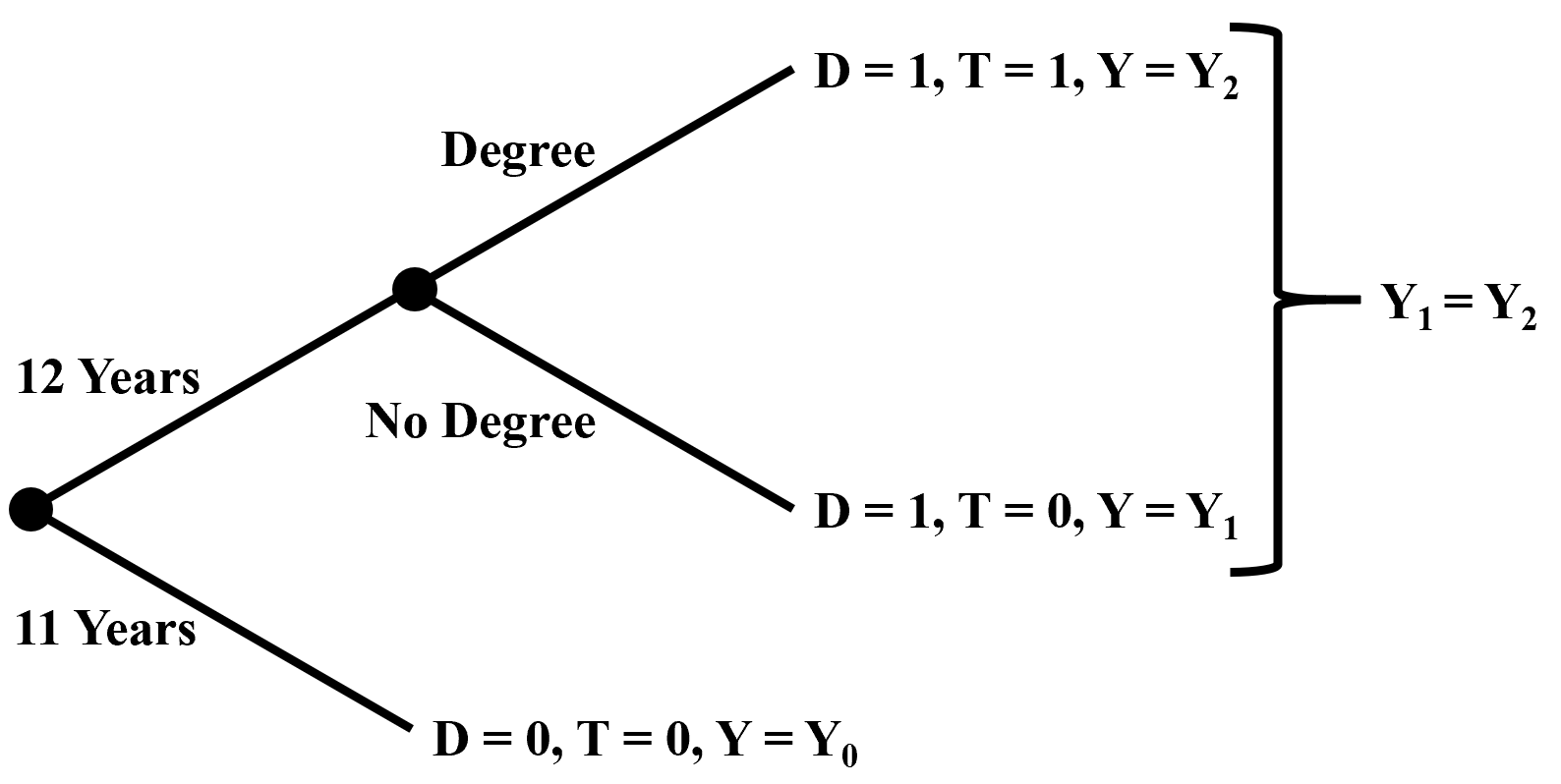}}
	\end{center}
	\caption{Examples: Dynamic Setting}\label{FigDynamic}
\end{figure}

For simplicity, I assume that the sheriff's execution does not matter to the individual, i.e., $Y_{1} = Y_{2}$. Consequently, the treatment variable of interest is $D$, as defined by \cite{Humphries2019}. However, I decide to erroneously use $T$ as my treatment variable. As Figure \ref{FigEviction} makes clear, the measurement error in this problem is similar to the one described in Example \ref{Esharp}, i.e., $D = 0$ is always correctly measured, but $D = 1$ may be mismeasured. Consequently, Assumption \ref{ASconstant} holds in this dynamic setting under the same assumptions described in Example \ref{Esharp}.

\subsubsection{Returns to High School Education}
Assume that I want to compare the wages of three types of individuals: $Y_{0}$ is the counterfactual wage of an agent with 11 years of education; $Y_{1}$ is the counterfactual wage of an agent with 12 years of education and no high school diploma; and $Y_{2}$ is the counterfactual wage of an agent with 12 years of education and a high school diploma. Let $D$ denote whether the person has 12 ($D = 1$) or 11 years of education and $T$ denote whether the person has a high school diploma ($T = 1$) or not. Consequently, the observed outcome variable is given by $Y = Y_{0} \cdot \left(1 - D\right) + Y_{1} \cdot D \cdot \left(1 - T\right) + Y_{2} \cdot D \cdot T$.

Following the empirical evidence gathered by \cite{Clark2014}, I assume that having a high school diploma has no effect, $Y_{1} = Y_{2}$. Consequently, the treatment variable of interest is $D$. However, I decide to erroneously use $T$ as my treatment variable. As Figure \ref{FigHighSchool} makes clear, the measurement error in this problem is similar to the one described in Example \ref{Esharp}, i.e., $D = 0$ is always correctly measured, but $D = 1$ may be mismeasured. Consequently, Assumption \ref{ASconstant} holds in this dynamic setting under the same assumptions described in Example \ref{Esharp}.

\subsection{Assumptions \ref{ASsign}, \ref{ASbounded} and \ref{ASsharp}: Restrictions on the Model's Primitives}\label{AppSuffSign}
In this subsection, I adapt to the MTE framework two assumptions that were used by \cite{Tommasi2020} to analyze the LATE parameter. Proposition \ref{PropSign} shows that these two extra assumptions are sufficient conditions for Assumption \ref{ASsign}. I also propose one extra assumption that is a sufficient condition for Assumptions \ref{ASbounded} and \ref{ASsharp}.

\begin{assumption}\label{ASAppError}
	The instrumental variable is independent of the measurement error conditional on the latent heterogeneity, i.e., $\left. Z \independent T \right\vert U, D$.
\end{assumption}

\begin{assumption}\label{ASAppProxy}
	The misclassified treatment variable is an informative proxy about the true treatment conditional on the latent heterogeneity, i.e., for any $u \in \left[0, 1\right]$, $\mathbb{P}\left[\left. T = 1 \right\vert U = u, D = 1\right] > \mathbb{P}\left[\left. T = 1 \right\vert U = u, D = 0\right]$.
\end{assumption}

\begin{assumption}\label{ASAppStrongProxy}
	There is a known lower bound on the amount of information contained in the misclassified treatment variable, i.e., there exists a known constant $c \in \left[1, +\infty\right)$ such that $\mathbb{P}\left[\left. T = 1 \right\vert U = u, D = 1\right] - \mathbb{P}\left[\left. T = 1 \right\vert U = u, D = 0\right] \geq \dfrac{1}{c}$ for any $u \in \left[0, 1\right]$.
\end{assumption}

\begin{proposition}\label{PropSign}
	If Assumptions \ref{ASindependence}-\ref{ASfinite}, \ref{ASAppError} and \ref{ASAppProxy} hold, then Assumption \ref{ASsign} holds. Moreover, if Assumption \ref{ASAppStrongProxy} holds too, then Assumptions \ref{ASbounded} and \ref{ASsharp} hold.
\end{proposition}

\begin{proof}
	Fix $z \in \mathcal{Z}$ arbitrarily. Note that
	\begin{align*}
	P_{T}\left(z\right) & = \mathbb{P}\left[\left. T = 1 \right\vert Z = z\right] \text{ by definition} \\
	& =  \int\mathbb{P}\left[\left. T = 1 \right\vert Z = z, U = u\right] d F_{\left. U \right\vert Z} \left(\left. u \right\vert z\right) \text{ by the LIE} \\
	& =  \int_{0}^{1} \mathbb{P}\left[\left. T = 1 \right\vert Z = z, U = u\right] d u \text{ by Assumptions \ref{ASindependence} and \ref{AScontinuous}} \\
	& = \int_{0}^{P_{D}\left(z\right)} \mathbb{P}\left[\left. T = 1 \right\vert Z = z, U = u, D = 1\right] d u \\
	& \hspace{20pt} + \int_{P_{D}\left(z\right)}^{1} \mathbb{P}\left[\left. T = 1 \right\vert Z = z, U = u, D = 0\right] d u \text{ by the LIE} \\
	& = \int_{0}^{P_{D}\left(z\right)} \mathbb{P}\left[\left. T = 1 \right\vert U = u, D = 1\right] d u \\
	& \hspace{20pt} + \int_{P_{D}\left(z\right)}^{1} \mathbb{P}\left[\left. T = 1 \right\vert U = u, D = 0\right] d u \text{ by Assumption \ref{ASAppError}},
	\end{align*}
	implying, by the Leibniz Integral Rule, that
	\begin{equation}\label{EqDifferentialEquation}
	\dfrac{dP_{T}\left(z\right)}{dz} = \dfrac{dP_{D}\left(z\right)}{dz} \cdot \left(\mathbb{P}\left[\left. T = 1 \right\vert U = P_{D}\left(z\right), D = 1\right] - \mathbb{P}\left[\left. T = 1 \right\vert U = P_{D}\left(z\right), D = 0\right]\right).
	\end{equation}
	Since $\left(\mathbb{P}\left[\left. T = 1 \right\vert U = P_{D}\left(z\right), D = 1\right] - \mathbb{P}\left[\left. T = 1 \right\vert U = P_{D}\left(z\right), D = 0\right]\right) > 0$ by Assumption \ref{ASAppProxy}, I have that Assumption \ref{ASsign} holds.

	Moreover, if Assumption \ref{ASAppStrongProxy} holds, then $$\dfrac{\sfrac{dP_{D}\left(z\right)}{dz}}{\sfrac{dP_{T}\left(z\right)}{dz}} \in \left[1,c\right],$$ implying that Assumption \ref{ASbounded} holds.

	Additionally, the Fundamental Theorem of Calculus and Equation \eqref{EqDifferentialEquation} imply that there exists $b \in \mathbb{R}$ such that
	\begin{align*}
	P_{D}\left(z\right) & = \int_{-\infty}^{z} \dfrac{d P_{D}\left(\tilde{z}\right)}{dz} \, d\tilde{z} + b \\
	& = \int_{-\infty}^{z} \left(\mathbb{P}\left[\left. T = 1 \right\vert U = P_{D}\left(\tilde{z}\right), D = 1\right] - \mathbb{P}\left[\left. T = 1 \right\vert U = P_{D}\left(\tilde{z}\right), D = 0\right]\right)^{-1} \cdot \dfrac{d P_{T}\left(\tilde{z}\right)}{dz} \, d\tilde{z} + b.
	\end{align*}
	Consequently, Assumption \ref{ASsharp} holds too.
\end{proof}

\newpage

\section{Extensions}\label{AppExtensions}
\setcounter{table}{0}
\renewcommand\thetable{E.\arabic{table}}

\setcounter{figure}{0}
\renewcommand\thefigure{E.\arabic{figure}}

\setcounter{equation}{0}
\renewcommand\theequation{E.\arabic{equation}}

\setcounter{theorem}{0}
\renewcommand\thetheorem{E.\arabic{theorem}}

\setcounter{proposition}{0}
\renewcommand\theproposition{E.\arabic{proposition}}

\setcounter{corollary}{0}
\renewcommand\thecorollary{E.\arabic{corollary}}

\setcounter{assumption}{0}
\renewcommand\theassumption{E.\arabic{assumption}}

In this appendix, I explore two extensions to the model explained in the main text. In the first subsection, I discuss the case when I observe more than one instrumental variable. In the second subsection, I explain the case when I observe more than one misclassified treatment variable. For brevity, I focus on extending Proposition \ref{CorOuterMTE} in both cases.

\subsection{Extension: More than One Instrumental Variable}\label{AppMoreIV}

To analyze the case with more than one instrumental variable, I assume that I observe $M \in \mathbb{N}$ instruments: $Z_{1},\ldots,Z_{M}$.\footnote{With multiple instruments, the single-index threshold crossing model in Equation \eqref{EqD} imposes strong uniformity restrictions as discussed by \cite{Mogstad2021}. In a context with a correctly classified treatment variable and multiple instruments, \cite{Mogstad2021a} partially identifies the marginal treatment response function using a weaker treatment choice model. Extending their results to a context with a misclassified treatment variable is an interesting line of future research.} Consequently, $Z$ is now a random vector and $\mathcal{Z} \subset \mathbb{R}^{M}$ denotes its support. I also modify Assumptions \ref{ASrank} and \ref{ASbounded} to account for more than one instrumental variable.

\begin{assumption}\label{ASEXTrank}
	All partial derivatives of functions $P_{D}$ and $P_{T}$ are always different from zero, i.e., $\dfrac{\partial P_{D}\left(z\right)}{\partial z_{m}} \neq 0$ and $\dfrac{\partial P_{T}\left(z\right)}{\partial z_{m}} \neq 0$ for every $z \in \mathcal{Z}$ and every $m \in \left\lbrace 1, \ldots, M \right\rbrace$.
\end{assumption}

\begin{assumption}\label{ASEXTbounded}
	The ratios between the partial derivatives of $P_{D}$ and $P_{T}$ are bounded, i.e., for each $m \in \left\lbrace 1, \ldots, M \right\rbrace$, there exists a known $c_{m} \in \left[1, +\infty\right)$ such that
	$\dfrac{\sfrac{\partial P_{D}\left(z\right)}{\partial z_{m}}}{\sfrac{\partial P_{T}\left(z\right)}{\partial z_{m}}} \in \left[\dfrac{1}{c_{m}}, c_{m}\right]$ for any value $z \in \mathcal{Z}$.
\end{assumption}

Moreover, I define one LIV estimand $f_{m}$ for each instrumental variable $Z_{m}$. For any $m \in \left\lbrace 1,\ldots,M\right\rbrace$, $f_{m}\colon\mathcal{Z}\rightarrow\mathbb{R}$ is a function such that, for any $z \in \mathcal{Z}$, $$f_{m}\left(z\right) = \dfrac{\sfrac{\partial \mathbb{E}\left[\left. Y \right\vert Z = z\right]}{\partial z_{m}}}{\sfrac{\partial \mathbb{E}\left[\left. T \right\vert Z = z\right]}{\partial z_{m}}}.$$

I, now, extend Proposition \ref{CorOuterMTE} to the case with more than one instrumental variable.

\begin{corollary}\label{CorAppOuterMTE}

	Suppose Assumptions \ref{ASindependence}, \ref{ASEXTrank}, \ref{AScontinuous}, \ref{ASfinite} and \ref{ASEXTbounded} hold. For each $m \in \left\lbrace 1, \ldots, M \right\rbrace$, I have that $\theta \in \Theta_{1}^{m}$, where
	\begin{equation*}
		\Theta_{1}^{m} \coloneqq \left\lbrace \tilde{\theta}\colon \mathcal{Z} \rightarrow \mathbb{R} \left\vert \text{For any } z \in \mathcal{Z}, \tilde{\theta}\left(z\right) \in \left\lbrace \begin{array}{ll}
			\left[\dfrac{1}{c_{m}} \cdot f_{m}\left(z\right), c_{m} \cdot f_{m}\left(z\right) \right] & \text{if } f_{m}\left(z\right) \geq 0 \\
			\left[c_{m} \cdot f_{m}\left(z\right), \dfrac{1}{c_{m}} \cdot f_{m}\left(z\right) \right] & \text{if } f_{m}\left(z\right) < 0
		\end{array}  \right. . \right. \right\rbrace.
	\end{equation*}

	Moreover, for any $m \in \left\lbrace 1, \ldots, M \right\rbrace$, $\tilde{\theta} \in \Theta_{1}^{m}$ and $d \in \left(0, +\infty\right]$, $\left\vert\left\vert \theta - \tilde{\theta} \right\vert\right\vert_{d} \leq \left(\dfrac{c_{m}^{2} - 1}{c_{m}} \right) \cdot \left\vert\left\vert f_{m}  \right\vert\right\vert_{d}$, where $\left\vert\left\vert \cdot \right\vert\right\vert_{d}$ is the $L_{d}$-norm if $d < \infty$ and $\left\vert\left\vert \cdot \right\vert\right\vert_{d}$ is the $\sup$-norm if $d = \infty$.

	Consequently, I have that $$\theta \in \cap_{m = 1}^{M}\Theta_{1}^{m}$$ and $$\left\vert\left\vert \theta - \tilde{\theta} \right\vert\right\vert_{d} \leq \min_{m \in \left\lbrace 1, \ldots, M \right\rbrace} \left\lbrace \left(\dfrac{c_{m}^{2} - 1}{c_{m}} \right) \cdot \left\vert\left\vert f_{m}  \right\vert\right\vert_{d} \right\rbrace$$ for any $\tilde{\theta} \in \cap_{m = 1}^{M}\Theta_{1}^{m}$ and $d \in \left(0, +\infty\right]$.

\end{corollary}

As a consequence of Corollary \ref{CorAppOuterMTE}, observe that, if the researcher has access to multiple instrumental variables, combining them in her analysis is more efficient. By doing so, the researcher will find a smaller outer set around the MTE function, i.e, $\cap_{m = 1}^{M}\Theta_{1}^{m} \subset \Theta_{1}^{m^{\prime}}$ for any $m^{\prime} \in \left\lbrace 1, \ldots, M \right\rbrace$.

Finally, note that Corollary \ref{CorAppOuterMTE} can be used to test the identifying assumptions in the extended model. If $\cap_{m = 1}^{M}\Theta_{1}^{m} = \emptyset$, then at least one of Assumptions \ref{ASindependence}, \ref{ASEXTrank}, \ref{AScontinuous}, \ref{ASfinite} and \ref{ASEXTbounded} is invalid.

\subsection{Extension: More than One Misclassified Treatment Variable}\label{AppMoreT}

Analyzing the case with multiple misclassified treatment variables is potentially interesting to many empirical applications. For instance, many studies use some prediction method to infer the treatment status of each observation \citep{ArellanoBover2020}. In those cases, the authors may define one treatment variable for each plausible prediction method, and each treatment variable is potentially misclassified due to prediction errors. Even the empirical application in Section \ref{Sempirical} could benefit from this extension if I were interested in analyzing the impact of being punished according to the trial judge's ruling and misclassification was due to the prediction error of different algorithms being used to convert textual trial judge's rulings to binary treatment variables.

To analyze the case with more than one misclassified treatment variable, I assume that I observe $Q \in \mathbb{N}$ misclassified treatment variables: $T_{1},\ldots, T_{Q}$. I also modify Assumptions \ref{ASrank} and \ref{ASbounded} to account for more than one misclassified treatment variable.

\begin{assumption}\label{ASEXTrankT}
	The derivatives of function $P_{D}$ and of the mismeasured propensity scores are always different from zero, i.e., $\dfrac{dP_{D}\left(z\right)}{dz} \neq 0$ and $\dfrac{dP_{T}^{q}\left(z\right)}{dz} \neq 0$ for every $z \in \mathcal{Z}$ and $q \in \left\lbrace 1, \ldots, Q \right\rbrace$, where $P_{T}^{q} \colon \mathcal{Z} \rightarrow \mathbb{R}$ is defined as $P_{T}^{q}\left(z\right) = \mathbb{P}\left[\left. T_{q} = 1 \right\vert Z = z\right]$ for any $z \in \mathcal{Z}$.
\end{assumption}

\begin{assumption}\label{ASEXTboundedT}
	The ratios between the derivatives of $P_{D}$ and $P_{T}^{q}$ are bounded, i.e., for each $q \in \left\lbrace 1, \ldots, Q \right\rbrace$, there exists a known $c_{q} \in \left[1, +\infty\right)$ such that
	$\dfrac{\sfrac{d P_{D}\left(z\right)}{d z}}{\sfrac{d P_{T}^{q}\left(z\right)}{d z}} \in \left[\dfrac{1}{c_{q}}, c_{q}\right]$ for any value $z \in \mathcal{Z}$.
\end{assumption}

Moreover, I define one LIV estimand $f_{q}$ for each misclassified treatment variable $T_{q}$. For any $q \in \left\lbrace 1,\ldots,Q\right\rbrace$, $f_{q}\colon\mathcal{Z}\rightarrow\mathbb{R}$ is a function such that, for any $z \in \mathcal{Z}$, $$f_{q}\left(z\right) = \dfrac{\sfrac{d \mathbb{E}\left[\left. Y \right\vert Z = z\right]}{d z}}{\sfrac{d \mathbb{E}\left[\left. T_{q} \right\vert Z = z\right]}{d z}}.$$

I, now, extend Proposition \ref{CorOuterMTE} to the case with more than one misclassified treatment variable.

\begin{corollary}\label{CorAppOuterMTEt}

	Suppose Assumptions \ref{ASindependence}, \ref{ASEXTrankT}, \ref{AScontinuous}, \ref{ASfinite} and \ref{ASEXTboundedT} hold. For each $q \in \left\lbrace 1, \ldots, Q \right\rbrace$, I have that $\theta \in \Theta_{1}^{q}$, where
	\begin{equation*}
		\Theta_{1}^{q} \coloneqq \left\lbrace \tilde{\theta}\colon \mathcal{Z} \rightarrow \mathbb{R} \left\vert \text{For any } z \in \mathcal{Z}, \tilde{\theta}\left(z\right) \in \left\lbrace \begin{array}{ll}
			\left[\dfrac{1}{c_{q}} \cdot f_{q}\left(z\right), c_{q} \cdot f_{q}\left(z\right) \right] & \text{if } f_{q}\left(z\right) \geq 0 \\
			\left[c_{q} \cdot f_{q}\left(z\right), \dfrac{1}{c_{q}} \cdot f_{q}\left(z\right) \right] & \text{if } f_{q}\left(z\right) < 0
		\end{array}  \right. . \right. \right\rbrace.
	\end{equation*}

	Moreover, for any $q \in \left\lbrace 1, \ldots, Q \right\rbrace$, $\tilde{\theta} \in \Theta_{1}^{q}$ and $d \in \left(0, +\infty\right]$, $\left\vert\left\vert \theta - \tilde{\theta} \right\vert\right\vert_{d} \leq \left(\dfrac{c_{q}^{2} - 1}{c_{q}} \right) \cdot \left\vert\left\vert f_{q}  \right\vert\right\vert_{d}$, where $\left\vert\left\vert \cdot \right\vert\right\vert_{d}$ is the $L_{d}$-norm if $d < \infty$ and $\left\vert\left\vert \cdot \right\vert\right\vert_{d}$ is the $\sup$-norm if $d = \infty$.

	Consequently, I have that $$\theta \in \cap_{q = 1}^{Q}\Theta_{1}^{q}$$ and $$\left\vert\left\vert \theta - \tilde{\theta} \right\vert\right\vert_{d} \leq \min_{q \in \left\lbrace 1, \ldots, Q \right\rbrace} \left\lbrace \left(\dfrac{c_{q}^{2} - 1}{c_{q}} \right) \cdot \left\vert\left\vert f_{q}  \right\vert\right\vert_{d} \right\rbrace$$ for any $\tilde{\theta} \in \cap_{q = 1}^{Q}\Theta_{1}^{q}$ and $d \in \left(0, +\infty\right]$.

\end{corollary}

As a consequence of Corollary \ref{CorAppOuterMTEt}, observe that, if the researcher has access to multiple misclassified treatment variables, it is more efficient to combine all of them in her analysis. By doing so, the researcher will find a smaller outer set around the MTE function, i.e, $\cap_{q = 1}^{Q}\Theta_{1}^{q} \subset \Theta_{1}^{q^{\prime}}$ for any $q^{\prime} \in \left\lbrace 1, \ldots, Q \right\rbrace$.

Finally, note that Corollary \ref{CorAppOuterMTEt} can be used to test the identifying assumptions in the extended model. If $\cap_{q = 1}^{Q}\Theta_{1}^{q} = \emptyset$, then at least one of Assumptions \ref{ASindependence}, \ref{ASEXTrankT}, \ref{AScontinuous}, \ref{ASfinite} and \ref{ASEXTboundedT} is invalid.

\newpage

\section{Partial Identification of the LATE Parameter}\label{AppLATE}
\setcounter{table}{0}
\renewcommand\thetable{F.\arabic{table}}

\setcounter{figure}{0}
\renewcommand\thefigure{F.\arabic{figure}}

\setcounter{equation}{0}
\renewcommand\theequation{F.\arabic{equation}}

\setcounter{theorem}{0}
\renewcommand\thetheorem{F.\arabic{theorem}}

\setcounter{proposition}{0}
\renewcommand\theproposition{F.\arabic{proposition}}

\setcounter{corollary}{0}
\renewcommand\thecorollary{F.\arabic{corollary}}

\setcounter{assumption}{0}
\renewcommand\theassumption{F.\arabic{assumption}}

In this Appendix, I adapt my framework in Section \ref{Sframework} and my results in Section \ref{Sbounds} to the binary instrument case, complementing the work developed by \citet[Subsection 3.4]{Ura2018} with a sensitivity analysis tool. To analyze the Local Average Treatment Effect (LATE) when the treatment variable is mismeasured, I start with the standard generalized selection model \citep{Heckman2006}.
\begin{align}
Y & \label{EqYApp} = Y_{1} \cdot D + Y_{0} \cdot (1-D) \\
D & \label{EqDApp} = \mathbf{1}\left\lbrace U \leq P_{D}\left(Z\right) \right\rbrace
\end{align}
where $Z$ is an observable binary instrumental variable with support given by $\mathcal{Z} \subset \left\lbrace 0, 1\right\rbrace$, $P_{D} \colon \mathcal{Z} \rightarrow \mathbb{R}$ is a unknown function, $U$ is a latent heterogeneity variable and $D$ is the correctly measured treatment. The variable $Y$ is the realized outcome variable, while $Y_{0}$ and $Y_{1}$ are the potential outcomes when the agent is untreated and treated, respectively.

I augment this model with the possibly mismeasured treatment status indicator, $T$. The mismeasured treatment is relevant because the researcher observes only the vector $\left(Y, T, Z\right)$, while $Y_{1}$, $Y_{0}$, $D$ and $U$ are latent variables.

I use the following assumptions in order to derive my partial identification results.
\begin{assumption}\label{ASAppindependence}
	$Z \independent \left(Y_{0}, Y_{1}, U\right)$
\end{assumption}

\begin{assumption}\label{ASApprank}
	The minimal required rank assumption is
	\begin{equation*}
	\begin{array}{l}
	\Delta T \coloneqq \mathbb{P}\left[\left. T = 1 \right\vert Z = 1\right] - \mathbb{P}\left[\left. T = 1 \right\vert Z = 0\right] \neq 0 \text{ and} \\
	\Delta D \coloneqq \mathbb{P}\left[\left. D = 1 \right\vert Z = 1\right] - \mathbb{P}\left[\left. D = 1 \right\vert Z = 0\right] \neq 0.
	\end{array}
	\end{equation*}
	However, for ease of notation, I impose
	\begin{equation*}
	\begin{array}{l}
	\Delta T \coloneqq \mathbb{P}\left[\left. T = 1 \right\vert Z = 1\right] - \mathbb{P}\left[\left. T = 1 \right\vert Z = 0\right] \neq 0 \text{ and} \\
	\Delta D \coloneqq \mathbb{P}\left[\left. D = 1 \right\vert Z = 1\right] - \mathbb{P}\left[\left. D = 1 \right\vert Z = 0\right] > 0.
	\end{array}
	\end{equation*}
\end{assumption}

\begin{assumption}\label{ASAppcontinuous}
	The distribution of the latent heterogeneity variable $U$ is absolutely continuous with respect to the Lebesgue measure.
\end{assumption}

\begin{assumption}\label{ASAppfinite}
	$\mathbb{E}\left[\left\vert Y_0 \right\vert \right] < \infty$ and $\mathbb{E}\left[\left\vert Y_1 \right\vert \right] < \infty$
\end{assumption}

\begin{assumption}\label{ASAppsign}
	$sign\left(\Delta T\right) > 0$
\end{assumption}

\begin{assumption}\label{ASAppbounded}
	$\dfrac{\Delta D}{\Delta T} \in \left[\dfrac{1}{c}, c\right]$ for a known $c \in \left[1, +\infty \right)$
\end{assumption}

\begin{assumption}\label{ASAppconstant}
	For a known $c \in \left[0, \dfrac{1}{\overline{p}} \right]$, there exists an unknown $\alpha \in \left[\dfrac{1}{c}, c\right]$ such that $P_{D}\left(z\right) = \alpha \cdot P_{T}\left(z\right)$, where $\overline{p} \coloneqq \max \left\lbrace P_{T}\left(0\right), P_{T}\left(1\right) \right\rbrace = P_{T}\left(1\right)$.
\end{assumption}
My goal is to derive partial identification results for the Local Average Treatment Effect (LATE) given by $\beta \coloneqq \mathbb{E}\left[\left. Y_{1} - Y_{0} \right\vert U \in \left[P_{D}\left(0\right), P_{D}\left(1\right) \right]\right].$ To achieve this goal, I, first, analyze the consequences of a mismeasured treatment variable on the Wald estimand and, then, derive increasingly strong identification results based on Assumptions \ref{ASAppsign}-\ref{ASAppconstant}.

If the researcher ignores that the treatment variable is mismeasured, she can compute the Wald estimand using the mismeasured treatment variable $T$ as if it was the actual treatment variable. The next proposition analyzes which object is identified by this naive approach.

\begin{proposition}[Wald Estimand]\label{PropWald}
	Under Assumptions \ref{ASAppindependence}-\ref{ASAppfinite}, the Wald estimand satisfies
	\begin{equation}\label{EqWald}
	\dfrac{\Delta Y}{\Delta T} = \dfrac{\Delta D}{\Delta T} \cdot \beta,
	\end{equation}
	where $\Delta Y \coloneqq \mathbb{E}\left[\left. Y \right\vert Z = 1\right] - \mathbb{E}\left[\left. Y \right\vert Z = 0\right]$.
\end{proposition}

Proposition \ref{PropWald} shows that, when measurement error is ignored, the Wald estimand does not identify the LATE parameter. This result is discussed in depth by \cite{Ura2018} and \cite{Calvi2019}.

Now, adding Assumptions \ref{ASAppsign}-\ref{ASAppconstant}, I derive increasingly strong identification results for the LATE parameter.

\begin{corollary}[Identifying the sign of the LATE parameter]\label{CorAppSign}
	If Assumptions \ref{ASAppindependence}-\ref{ASAppfinite} and \ref{ASAppsign} hold, then $sign\left(\beta\right) = sign\left(\dfrac{\Delta Y}{\Delta T}\right)$.
\end{corollary}

\begin{proof}
	Under Assumption \ref{ASAppsign}, I have that $\dfrac{\Delta D}{\Delta T} > 0$ in Equation \eqref{EqWald}.
\end{proof}

\begin{corollary}[Bounds for the LATE parameter]\label{CorAppBound}

	If Assumptions \ref{ASAppindependence}-\ref{ASAppfinite} and \ref{ASAppbounded} hold, then
	\begin{equation*}
	\beta \in B \coloneqq \left\lbrace \begin{array}{cl}
	\left[\dfrac{1}{c} \cdot \dfrac{\Delta Y}{\Delta T}, c \cdot \dfrac{\Delta Y}{\Delta T} \right] & \text{if } \dfrac{\Delta Y}{\Delta T} > 0 \\
	& \\
	\left\lbrace 0 \right\rbrace & \text{if } \dfrac{\Delta Y}{\Delta T} = 0 \\
	& \\
	\left[c \cdot \dfrac{\Delta Y}{\Delta T}, \dfrac{1}{c} \cdot \dfrac{\Delta Y}{\Delta T} \right] & \text{if } \dfrac{\Delta Y}{\Delta T} < 0
	\end{array} \right. .
	\end{equation*}
\end{corollary}

\begin{proof}
	The proof uses Equation \eqref{EqWald} and the bounds imposed on the ratio $\dfrac{\Delta D}{\Delta T}$ by Assumption \ref{ASAppbounded} to bound the LATE parameter.
\end{proof}

\begin{proposition}[Sharp Identified Set for the LATE Parameter]\label{ThmSharpLATE} Suppose Assumptions \ref{ASAppindependence}-\ref{ASAppfinite} and \ref{ASAppconstant} hold. In addition, suppose that the outcome variable $Y$ is binary and that $B \subseteq \left[-1, 1\right]$. Then, $B$ is the sharp identified set for $\beta$ in the sense that, for any $\tilde{\beta} \in B$, there exist candidate random variables $\left(\tilde{U}, \tilde{Y}_{0}, \tilde{Y}_{1}, \tilde{D}\right)$ and a function $P_{\tilde{D}}\colon \mathcal{Z} \rightarrow \left[0, 1\right]$ such that
	\begin{enumerate}
		\item $\tilde{D}$ is monotonic with respect to $Z$ and its index is given by $P_{\tilde{D}}$, i.e.,
		\begin{equation}\label{EqAppDtilde}
		\tilde{D} = \mathbf{1}\left\lbrace \tilde{U} \leq P_{\tilde{D}}\left(Z\right) \right\rbrace;
		\end{equation}

		\item $\left(Z, \tilde{U}, \tilde{Y}_{0}, \tilde{Y}_{1}\right)$ and $P_{\tilde{D}}$ achieve the candidate target parameter, i.e.,
		\begin{equation}\label{EqBetaTilde}
		\tilde{\beta} = \mathbb{E}\left[\left. \tilde{Y}_{1} - \tilde{Y}_{0} \right\vert \tilde{U} \in \left[P_{\tilde{D}}\left(0\right), P_{\tilde{D}}\left(1\right)\right] \right];
		\end{equation}

		\item $\left(Z, \tilde{U}, \tilde{Y}_{0}, \tilde{Y}_{1}\right)$ and $P_{\tilde{D}}$ satisfy the data restriction given by
		\begin{equation}\label{EqAppDataRestriction}
		\Delta Y = \Delta \tilde{Y},
		\end{equation} where $\Delta \tilde{Y} \coloneqq \mathbb{P}\left[\left. \tilde{Y} = 1 \right\vert Z = 1\right] - \mathbb{P}\left[\left. \tilde{Y} = 1 \right\vert Z = 0\right]$ and $\tilde{Y} = \tilde{Y}_{1} \cdot \tilde{D} + \tilde{Y}_{0} \left(1 - \tilde{D}\right)$;

		\item $\tilde{Y}_{0}$ and $\tilde{Y}_{1}$ satisfy the support condition given by
		\begin{equation}\label{EqAppSupport}
		\tilde{\mathcal{Y}}_{0} \subseteq \mathcal{Y}_{0} = \left\lbrace 0, 1 \right\rbrace \text{ and } \tilde{\mathcal{Y}}_{1} \subseteq \mathcal{Y}_{1} = \left\lbrace 0, 1 \right\rbrace,
		\end{equation}
		where $\tilde{\mathcal{Y}}_{0}$, $\mathcal{Y}_{0}$ $\tilde{\mathcal{Y}}_{1}$ and $\mathcal{Y}_{1}$ are the support of $\tilde{Y}_{0}$, $Y_{0}$, $\tilde{Y}_{1}$ and $Y_{1}$, respectively;

		\item $\left(Z, \tilde{U}, \tilde{Y}_{0}, \tilde{Y}_{1}, \tilde{D}\right)$ satisfy Assumptions \ref{ASAppindependence}-\ref{ASAppfinite} and Assumption \ref{ASAppconstant}.
	\end{enumerate}
\end{proposition}

\begin{proof}
	The proof is by construction. For each $\tilde{\beta} \in B$, I define the density of the candidate random variables $\left(\tilde{U}, \tilde{Y}_{0}, \tilde{Y}_{1}, \tilde{D}\right)$ to ensure that they satisfy the five restrictions of Proposition \ref{ThmSharpLATE}.

	Fix $\tilde{\beta} \in B$ arbitrarily. By definition, there exists $a \in \left[\dfrac{1}{c}, c\right]$ such that $\tilde{\beta} = \dfrac{1}{a} \cdot \dfrac{\Delta Y}{\Delta T}$, where $c \in \left[1, \dfrac{1}{\overline{p}} \right]$. I break the construction of the candidate random variables $\left(\tilde{U}, \tilde{Y}_{0}, \tilde{Y}_{1}, \tilde{D}\right)$ in six steps.

	\begin{enumerate}[label={Step \theenumi. }, ref = {Step \theenumi}]
		\item Define $P_{\tilde{D}} \colon \mathcal{Z} \rightarrow \left[0, 1\right]$ such that $P_{\tilde{D}}\left(z\right) = a \cdot P_{T}\left(z\right)$ for any $z \in \mathcal{Z}$, ensuring that Assumptions \ref{ASApprank} and \ref{ASAppconstant} hold.

		\item\label{StepAppUtilde} Define $\tilde{U} \sim Uniform\left[0, 1\right]$ and $\tilde{D} \coloneqq \mathbf{1}\left\lbrace \tilde{U} \leq P_{\tilde{D}}\left(Z\right) \right\rbrace$, ensuring that Assumption \ref{ASAppcontinuous} and Equation \eqref{EqAppDtilde} hold.

		\item Since $f_{Z,\tilde{U},\tilde{Y}_{0},\tilde{Y}_{1}} = \mathbb{P}\left[Z = \cdot\right] \cdot f_{\left. \tilde{U} \right\vert Z} \cdot \mathbb{P}\left[\left. \tilde{Y}_{0} = \cdot, \tilde{Y}_{1} = \cdot \right\vert \tilde{U}, Z\right]$, I define the joint density function of $\left(Z, \tilde{U}, \tilde{Y}_{0}, \tilde{Y}_{1}\right)$ through its components $\mathbb{P}\left[Z = \cdot\right]$, $f_{\left. \tilde{U} \right\vert Z}$ and $\mathbb{P}\left[\left. \tilde{Y}_{0} = \cdot, \tilde{Y}_{1} = \cdot \right\vert \tilde{U}, Z\right]$. Fix $\left(z, u, y_{0}, y_{1}\right) \in \left\lbrace 0, 1\right\rbrace \times \mathbb{R} \times \left\lbrace 0, 1\right\rbrace^{2}$ arbitrarily.
		\begin{enumerate}
			\item Note that $\mathbb{P}\left[Z = \cdot\right]$ is identified. Consequently, $\mathbb{P}\left[Z = \cdot\right]$ is defined according to the data.

			\item Define $f_{\left. \tilde{U} \right\vert Z}\left(\left. u \right\vert z\right) = f_{\tilde{U}}\left(u\right)$ and $\mathbb{P}\left[\left. \tilde{Y}_{0} = y_{0}, \tilde{Y}_{1} = y_{1} \right\vert \tilde{U} = u, Z = z\right] = \mathbb{P}\left[\left. \tilde{Y}_{0} = y_{0}, \tilde{Y}_{1} = y_{1} \right\vert \tilde{U} = u\right]$, ensuring that Assumption \ref{ASAppindependence} holds. Consequently, I only have to define $f_{\tilde{U}}\left(u\right)$ and $\mathbb{P}\left[\left. \tilde{Y}_{0} = y_{0}, \tilde{Y}_{1} = y_{1} \right\vert \tilde{U} = u\right]$.

			\item Note that $f_{\tilde{U}}$ is defined in \ref{StepAppUtilde}.

			\item I impose $\mathbb{P}\left[\left. \tilde{Y}_{0} = y_{0}, \tilde{Y}_{1} = y_{1} \right\vert \tilde{U} = u\right] = \mathbb{P}\left[\left. \tilde{Y}_{0} = y_{0} \right\vert \tilde{U} = u\right] \cdot \mathbb{P}\left[\left. \tilde{Y}_{1} = y_{1} \right\vert \tilde{U} = u\right]$ for simplicity. Define $\mathbb{P}\left[\left. \tilde{Y}_{0} = 0 \right\vert \tilde{U} = u\right] = 1 - \mathbb{P}\left[\left. \tilde{Y}_{0} = 1 \right\vert \tilde{U} = u\right]$ and $\mathbb{P}\left[\left. \tilde{Y}_{1} = 0 \right\vert \tilde{U} = u\right] = 1 - \mathbb{P}\left[\left. \tilde{Y}_{1} = 1 \right\vert \tilde{U} = u\right]$. Consequently, I only have to define $\mathbb{P}\left[\left. \tilde{Y}_{0} = 1 \right\vert \tilde{U} = u\right]$ and $\mathbb{P}\left[\left. \tilde{Y}_{1} = 1 \right\vert \tilde{U} = u\right]$.

			\item Define the set $\mathcal{U} \coloneqq \left[P_{\tilde{D}}\left(0\right), P_{\tilde{D}}\left(1\right) \right] \subseteq \left[0, 1\right]$.

			\item If $u \notin \mathcal{U}$, $\mathbb{P}\left[\left. \tilde{Y}_{0} = 1 \right\vert \tilde{U} = u\right] = 0$ and $\mathbb{P}\left[\left. \tilde{Y}_{1} = 1 \right\vert \tilde{U} = u\right] = 0$.

			\item From now on, suppose that $u \in \mathcal{U}$. Define $$\mathbb{P}\left[\left. \tilde{Y}_{0} = 1 \right\vert \tilde{U} = u\right] = -\tilde{\beta} \cdot \mathbf{1}\left\lbrace \tilde{\beta} < 0 \right\rbrace$$ and $$\mathbb{P}\left[\left. \tilde{Y}_{1} = 1 \right\vert \tilde{U} = u\right] = \tilde{\beta} \cdot \mathbf{1}\left\lbrace \tilde{\beta} > 0 \right\rbrace.$$
		\end{enumerate}

		\item Note that $\tilde{Y}_{0}$ and $\tilde{Y}_{1}$ satisfy Assumption \ref{ASAppfinite} and Equation \eqref{EqAppSupport}.

		\item Observe that $\mathbb{E}\left[\left. \tilde{Y}_{1} - \tilde{Y}_{0} \right\vert \tilde{U} \in \left[P_{\tilde{D}}\left(0\right), P_{\tilde{D}}\left(1\right) \right] \right] = \tilde{\beta}$, ensuring that Equation \eqref{EqBetaTilde} holds.

		\item Define $\Delta \tilde{D} \coloneqq \mathbb{P}\left[\left. \tilde{D} = 1 \right\vert Z = 1\right] - \mathbb{P}\left[\left. \tilde{D} = 1 \right\vert Z = 0\right]$. Note that
		\begin{align*}
		\Delta \tilde{Y} & = \Delta \tilde{D} \cdot \tilde{\beta} \\
		& = a \cdot \Delta T \cdot \tilde{\beta} \\
		& = a \cdot \Delta T \cdot \dfrac{1}{a} \cdot \dfrac{\Delta Y}{\Delta T} \\
		& = \Delta Y,
		\end{align*}
		implying that Equation \eqref{EqAppDataRestriction} holds.
	\end{enumerate}
\end{proof}

\newpage

\section{Bounds around Standard Treatment Effect Parameters}\label{AppStandar}
\setcounter{table}{0}
\renewcommand\thetable{G.\arabic{table}}

\setcounter{figure}{0}
\renewcommand\thefigure{G.\arabic{figure}}

\setcounter{equation}{0}
\renewcommand\theequation{G.\arabic{equation}}

\setcounter{theorem}{0}
\renewcommand\thetheorem{G.\arabic{theorem}}

\setcounter{proposition}{0}
\renewcommand\theproposition{G.\arabic{proposition}}

\setcounter{corollary}{0}
\renewcommand\thecorollary{G.\arabic{corollary}}

\setcounter{assumption}{0}
\renewcommand\theassumption{G.\arabic{assumption}}

\setcounter{remark}{0}
\renewcommand\theremark{G.\arabic{remark}}

In this appendix, my final goal is to bound standard treatment effect parameters (ATE, ATT, ATU and PRTE). Firstly, I start by deriving sharp uniform bounds around the MTE function (Appendix \ref{AppSharp}).\footnote{My definition of sharpness imposes that I recover a derivative of the data distribution instead of the level of the data distribution. For a more detailed discussion, see Remark \ref{RemarkSharp}.} To do so, I impose that the outcome variable is binary and that there is a functional relationship between the mismeasured probability of treatment and the correctly measured probability of treatment. Secondly, I derive bounds for any treatment effect parameter that can be written as a weighted integral of the marginal treatment effect (Appendix \ref{AppConstant}). Lastly, I present the formal proofs of this Appendix's results in Appendix \ref{ProofSharpMTE}.

\subsection{Sharp Uniform Bounds around the MTE Function }\label{AppSharp}

In this appendix, I derive my strictest partial identification results (Corollary \ref{CorOuterMTEsharp}, Proposition \ref{ThmSharpMTE} and Corollary \ref{CorBoundTE}).\footnote{When proving these results in Appendix \ref{ProofSharpMTE}, I use an important consequence of Assumption \ref{ASrank}: function $P_{D}$ is invertible according to Rolle's Theorem.} To do so, I impose a restriction on the functional relationship between the correctly measured propensity score and the mismeasured propensity score in the sense that it connects the level and all the derivatives of those two objects.

\begin{assumption}\label{ASsharp}
	For a known $c \in \left[1, +\infty\right)$, a known set $\mathcal{A} \subseteq \mathcal{G} \times \mathbb{R}$ and an unknown pair $\left(\alpha, b\right) \in \mathcal{A}$, $P_{D}\left(z\right) = \bigint_{-\infty}^{z} \alpha\left(\tilde{z}\right) \cdot \dfrac{d P_{T}\left(\tilde{z}\right)}{d z} \, d\tilde{z} + b$, where $\mathcal{G} \coloneqq \left\lbrace g \colon \mathcal{Z} \rightarrow \left[\dfrac{1}{c}, c\right]  \right\rbrace$.
\end{assumption}

If set $\mathcal{A}$ is left unrestricted (i.e., $\mathcal{A} = \mathcal{G}\times\mathbb{R}$), then Assumption \ref{ASsharp} is equivalent to Assumption \ref{ASbounded}. However, Assumption \ref{ASsharp} will be stronger than Assumption \ref{ASbounded} if I restrict the set $\mathcal{A}$ to be a strict subset of $\mathcal{G}\times\mathbb{R}$. For instance, an easy-to-interpret version of Assumption \ref{ASsharp} imposes that the mismeasured propensity score is a rescaled version of the true propensity score by defining $\mathcal{A} \coloneqq \left\lbrace g \in \mathcal{G} \text{ and } g \text{ is constant.} \right\rbrace \times \left\lbrace 0 \right\rbrace$.\footnote{Example \ref{Esharp} (Appendix \ref{AppExamples}) explains that Assumption \ref{ASsharp} may be plausible when analyzing welfare benefits if misclassification is unidirectional. For two other examples in a dynamic setting, see Appendix \ref{AppDynamic}. Moreover, in Appendix \ref{AppSuffSign}, I impose restrictions on the model's primitives that imply Assumption \ref{ASsharp}.}\textsuperscript{,}\footnote{While Assumption \ref{ASsharp} in its general form can be seen as a high-level condition that is weaker than the restrictions imposed by \cite{Acerenza2021}, its more intuitive version is more restrictive than the method proposed by these authors in their third section.} This more interpretable version of Assumption \ref{ASsharp} is discussed in Appendix \ref{AppConstant}.

Under Assumption \ref{ASsharp}, I derive Corollary \ref{CorOuterMTEsharp}, which achieves a stricter uniform bound for the MTE function than the bound described in Proposition \ref{CorOuterMTE}.

\begin{corollary}[Uniform Outer Set for the MTE Function with Assumption \ref{ASsharp}] \label{CorOuterMTEsharp}
	Suppose Assumptions \ref{ASindependence}-\ref{ASfinite} and \ref{ASsharp} hold. I have that $\theta \in \Theta_{2}$, where
        \begin{equation*}
	\Theta_{2} \coloneqq \left\lbrace \tilde{\theta}\colon \mathcal{Z} \rightarrow \mathbb{R} \left\vert \begin{array}{l}
        \tilde{\theta}\left(z\right) = \dfrac{1}{a\left(z\right)} \cdot f\left(z\right) \text{ for any } z \in \mathcal{Z} \text{ where } \left(a, \tilde{b}\right) \in \mathcal{A} \text{ satisfy} \\
        \text{the following property: the function } P_{\tilde{D}}\colon\mathcal{Z}\rightarrow\mathbb{R} \text{ defined as} \\
        \hspace{20pt} P_{\tilde{D}}\left(z\right) = \bigint_{-\infty}^{z} a\left(\tilde{z}\right) \cdot \dfrac{d P_{T}\left(\tilde{z}\right)}{d z} \, d\tilde{z} + \tilde{b} \text{ for any } z \in \mathcal{Z} \\
        \text{is bounded between 0 and 1.}
	\end{array} \right. \right\rbrace.
	\end{equation*}
\end{corollary}

\begin{proof}
	Under Assumption \ref{ASsharp}, there exists $(\alpha, b) \in \mathcal{A}$ such that $P_{D}\left(z\right) = \bigint_{-\infty}^{z} \alpha\left(\tilde{z}\right) \cdot \dfrac{d P_{T}\left(\tilde{z}\right)}{d z} \, d\tilde{z} + b$. Consequently, Equation \eqref{EqLivChalak} implies that $\theta\left(z\right) = \dfrac{1}{\alpha\left(z\right)} \cdot f\left(z\right)$ for any $z \in \mathcal{Z}$. Furthermore, the normalization connected to Assumption \ref{AScontinuous} implies that the function $P_{D}$ is bounded between 0 and 1. I can, then, conclude that $\theta \in \Theta_{2}$.
\end{proof}

Corollary \ref{CorOuterMTEsharp} is stronger than Proposition \ref{CorOuterMTE} in the sense that it imposes more discipline in the acceptable MTE functions given the data restrictions. For example, Proposition \ref{CorOuterMTE} still accepts MTE functions that are equal to $f\left(z\right)$ times a non-constant scaling factor, while those functions are ruled out in Corollary \ref{CorOuterMTEsharp} when $\mathcal{A} \coloneqq \left\lbrace g \in \mathcal{G} \text{ and } g \text{ is constant.} \right\rbrace \times \left\lbrace 0 \right\rbrace$. Importantly, if $f\left(z\right)$ has different signs for different values of $z$, the upper contour and the lower contour of the bounds in Proposition \ref{CorOuterMTE} are not included in the set $\Theta_{2}$ when $\mathcal{A} \coloneqq \left\lbrace g \in \mathcal{G} \text{ and } g \text{ is constant.} \right\rbrace \times \left\lbrace 0 \right\rbrace$. Consequently, treatments that pass a cost-benefit analysis under Proposition \ref{CorOuterMTE} may not pass the same analysis under Corollary \ref{CorOuterMTEsharp}.

Although outer sets --- as the ones described in Proposition \ref{CorOuterMTE} and Corollary \ref{CorOuterMTEsharp} --- provide reliable information about the true value of the function $\theta\left(\cdot\right)$, it is theoretically interesting to investigate when an outer set exhaust all the information about $\theta\left(\cdot\right)$ based on the data and the model restrictions. Following this line of inquiry, the fourth identification result (Proposition \ref{ThmSharpMTE}) shows that the uniform bound in Corollary \ref{CorOuterMTEsharp} is sharp when the outcome variable is binary.

\begin{proposition}[Sharp Identified Set for the MTE Function]\label{ThmSharpMTE} Suppose Assumptions \ref{ASindependence}-\ref{ASfinite} and \ref{ASsharp} hold. In addition, suppose that the outcome variable $Y$ is binary, $\sup_{\tilde{\theta} \in \Theta_{2}}\sup_{z \in \mathcal{Z}}\tilde{\theta}\left(z\right) \leq 1$ and $\inf_{\tilde{\theta} \in \Theta_{2}}\inf_{z \in \mathcal{Z}}\tilde{\theta}\left(z\right) \geq -1$. Then, $\Theta_{2}$ is the sharp identified set for $\theta$ in the sense that, for any $\tilde{\theta} \in \Theta_{2}$, there exist candidate random variables $\left(\tilde{U}, \tilde{Y}_{0}, \tilde{Y}_{1}, \tilde{D}\right)$ and a function $P_{\tilde{D}}\colon \mathcal{Z} \rightarrow \left[0, 1\right]$ such that
	\begin{enumerate}
		\item $\tilde{D}$ is monotonic with respect to $Z$ and its index is given by $P_{\tilde{D}}$, i.e.,
		\begin{equation}\label{EqDtilde}
		\tilde{D} = \mathbf{1}\left\lbrace \tilde{U} \leq P_{\tilde{D}}\left(Z\right) \right\rbrace;
		\end{equation}

		\item $\left(Z, \tilde{U}, \tilde{Y}_{0}, \tilde{Y}_{1}\right)$ and $P_{\tilde{D}}$ achieve the candidate target parameter, i.e.,
		\begin{equation}\label{EqThetaTilde}
		\tilde{\theta}\left(z\right) = \mathbb{E}\left[\left. \tilde{Y}_{1} - \tilde{Y}_{0} \right\vert \tilde{U} = P_{\tilde{D}}\left(z\right) \right]
		\end{equation} for any $z \in \mathcal{Z}$;

		\item $\left(Z, \tilde{U}, \tilde{Y}_{0}, \tilde{Y}_{1}\right)$ and $P_{\tilde{D}}$ satisfy the data restriction given by
		\begin{equation}\label{EqDataRestriction}
		\dfrac{d\mathbb{E}\left[\left. Y \right\vert Z = z\right]}{dz} = \dfrac{d\mathbb{E}\left[\left. \tilde{Y} \right\vert Z = z\right]}{dz}
		\end{equation} for any $z \in \mathcal{Z}$, where $\tilde{Y} = \tilde{Y}_{1} \cdot \tilde{D} + \tilde{Y}_{0} \cdot \left(1 - \tilde{D}\right)$;

		\item $\tilde{Y}_{0}$ and $\tilde{Y}_{1}$ satisfy the support condition given by
		\begin{equation}\label{EqSupport}
		\tilde{\mathcal{Y}}_{0} \subseteq \mathcal{Y}_{0} = \left\lbrace 0, 1 \right\rbrace \text{ and } \tilde{\mathcal{Y}}_{1} \subseteq \mathcal{Y}_{1} = \left\lbrace 0, 1 \right\rbrace,
		\end{equation}
		where $\tilde{\mathcal{Y}}_{0}$, $\mathcal{Y}_{0}$ $\tilde{\mathcal{Y}}_{1}$ and $\mathcal{Y}_{1}$ are the support of $\tilde{Y}_{0}$, $Y_{0}$, $\tilde{Y}_{1}$ and $Y_{1}$, respectively;

		\item $\left(Z, \tilde{U}, \tilde{Y}_{0}, \tilde{Y}_{1}, \tilde{D}\right)$ satisfy Assumptions \ref{ASindependence}-\ref{ASfinite} and Assumption \ref{ASsharp}.
	\end{enumerate}
\end{proposition}

\begin{proof}
	To prove that the set $\Theta_{2}$ is the sharp identified set for $\theta$, I construct, for each $\tilde{\theta} \in \Theta_{2}$, the candidate random variables that satisfy the five restrictions of Proposition \ref{ThmSharpMTE}. The details of this proof are in Appendix \ref{ProofSharpMTE}.
\end{proof}

\begin{remark}\label{RemarkSharp}
	The definition of sharpness in Proposition \ref{ThmSharpMTE} is weaker than the definition of sharpness proposed by \cite{Canay2017} because it does not ensure that the data $\left(Z, \tilde{Y}\right)$ generated by the candidate random variables $\left(\tilde{U}, \tilde{Y}_{0}, \tilde{Y}_{1}, \tilde{D}\right)$ has the same distribution of the data $\left(Z, Y \right)$ that is generated by the true latent variables $\left(U, Y_{0}, Y_{1}, D\right)$. However, my definition still uses all the restrictions (Equation \eqref{EqDataRestriction}) directly imposed by the data on the MTE function.
\end{remark}

\begin{remark}
	While assuming that the outcome variable $Y$ is binary may be limiting, many empirical questions have a natural binary outcome variable and my partial identification strategy can be used to derive sharp bounds around the MTE function in those contexts. For example, analyzing the impact of alternative sentences on recidivism (Section \ref{Sempirical}) is a research question whose outcome variable is naturally binary. Beyond my empirical application, studying the effect of college attendance on employment or the effect of welfare program participation on child malnourishment are research questions whose outcome variables are naturally binary too. Moreover, if $Y$ is a continuous variable, I can analyze the dummy variable $\mathbf{1}\left\lbrace Y \leq y \right\rbrace$ for any $y \in \mathbb{R}$ and derive point-wise sharp bounds for the distributional marginal treatment effect function.
\end{remark}

Corollary \ref{CorOuterMTEsharp} and Proposition \ref{ThmSharpMTE} are useful because they provide bounds for any treatment effect parameter that can be written as a weighted integral of the MTE. Corollary \ref{CorBoundTE} in Appendix \ref{AppConstant} summarizes this result, illustrating that bounding the MTE function offers a flexible tool to answer many research questions that rely on other treatment effect parameters.

Unfortunately, the bounds in Corollary \ref{CorBoundTE} may be hard to compute. For this reason, I additionally impose Assumption \ref{ASconstant}. This strong condition restricts the mismeasured propensity score to be a re-scaled version of the correctly measured propensity score. The advantage of imposing such a restrictive assumption is that it simplifies the bounds in Corollary \ref{CorBoundTE} substantially, providing easy-to-compute bounds around standard treatment effect parameters.

Furthermore, when interest lies exclusively on a uni-dimensional treatment effect parameter (e.g., the ATE or some PRTE), Corollary \ref{CorBoundTE} can be used to implement the breakdown analysis proposed by \cite{Kline2013}. To do so, it is necessary that some desired conclusion (e.g., ATE is positive) holds when $c = 1$. Then, I increase the value of $c \in \left[1, +\infty \right)$ in Assumption \ref{ASsharp} until the desired conclusion ceases to hold for some values in the identified set of the parameter of interest. This value of $c$ that starts breaking down the desired conclusion is known as the breakdown point.

\subsection{Bounds around Weighted Integrals of the MTE}\label{AppConstant}

In this appendix, Corollary \ref{CorBoundTE} explains how to derive bounds for any treatment effect parameter that can be written as a weighted integral of the marginal treatment effect. Moreover, I provide explicit formulas for the weighting functions $\omega$ associated with the Average Treatment Effect (ATE), the Average Treatment Effect on the Treated (ATT), the Average Treatment Effect on the Untreated (ATU) and any Policy Relevant Treatment Effect (PRTE, \citealp{Heckman2006}).\footnote{For other treatment effect parameters that can be defined as weighted integrals of the MTE function, see \cite{Heckman2006}.} I also derive numerically easy-to-compute bounds around these treatment effect parameters by imposing that the mismeasured propensity score function is a re-scaled version of the true propensity score.

\begin{corollary}[Bounds on Treatment Effects Parameters]\label{CorBoundTE}
	Suppose Assumptions \ref{ASindependence}-\ref{ASfinite} and \ref{ASsharp} hold. Define a treatment effect parameter $TE_{\omega}$ as $$TE_{\omega} \coloneqq \int_{0}^{1} \theta\left(P_{D}^{-1}\left(u\right)\right) \cdot \omega\left(u, P_{D}\right) \, du,$$ where $\omega \colon \left[0,1\right] \times \mathbb{R}^{\mathbb{R}} \rightarrow \mathbb{R}$ is a known weighting function.

	Then, the treatment effect parameter $TE_{\omega}$ is bounded by
	\begin{align*}
	& \inf_{\left(a, \tilde{b}, P_{\tilde{D}}, \tilde{\theta}\right) \in \tilde{\mathcal{A}}} \bigintss_{0}^{1} \tilde{\theta}\left(P_{\tilde{D}}^{-1}\left(u\right)\right) \cdot \omega\left(u, P_{\tilde{D}}\right) \, du \leq TE_{\omega} \\
	& \hspace{60pt} \leq \sup_{\left(a, \tilde{b}, P_{\tilde{D}}, \tilde{\theta}\right) \in \tilde{\mathcal{A}}} \bigintss_{0}^{1} \tilde{\theta}\left(P_{\tilde{D}}^{-1}\left(u\right)\right) \cdot \omega\left(u, P_{\tilde{D}}\right) \, du;
	\end{align*}

	where $\tilde{\mathcal{A}} \coloneqq \left\lbrace \begin{array}{l}
	\left(a, \tilde{b}, P_{\tilde{D}}, \tilde{\theta}\right) \in \mathcal{A} \times \left\lbrace g \colon \mathcal{Z} \rightarrow \left[0, 1\right] \right\rbrace \times \left\lbrace g \colon \mathcal{Z} \rightarrow \mathbb{R} \right\rbrace \text{ such that } \\
	P_{\tilde{D}}\left(z\right) = \bigint_{-\infty}^{z} a\left(\tilde{z}\right) \cdot \dfrac{d P_{T}\left(\tilde{z}\right)}{d z} \, d\tilde{z} + \tilde{b} \text{ and } \tilde{\theta}\left(z\right) = \dfrac{1}{a\left(z\right)} \cdot f\left(z\right) \text{ for any } z \in \mathcal{Z}
	\end{array}  \right\rbrace$.

	If, additionally, the outcome variable $Y$ is binary, $\sup_{\tilde{\theta} \in \Theta_{2}}\sup_{z \in \mathcal{Z}}\tilde{\theta}\left(z\right) \leq 1$ and $\inf_{\tilde{\theta} \in \Theta_{2}}\inf_{z \in \mathcal{Z}}\tilde{\theta}\left(z\right) \geq -1$, then the above bounds are sharp in the sense defined in Proposition \ref{ThmSharpMTE}.
\end{corollary}

\begin{proof}
	Note that, under Assumption \ref{ASsharp}, $\theta\left(P_{D}^{-1}\left(u\right)\right) = \mathbb{E}\left[\left. Y_{1} - Y_{0} \right\vert U = u\right]$, implying that Corollary \ref{CorBoundTE} is a direct consequence of Corollary \ref{CorOuterMTEsharp} and Proposition \ref{ThmSharpMTE}.
\end{proof}

Now, I provide explicit formulas for the weighting functions $\omega$ associated with the ATE, the ATT, the ATU and any PRTE. According to \citet[Tables IA and IB]{Heckman2005}, we can write the $ATE \coloneqq \mathbb{E}\left[Y_{1} - Y_{0}\right]$, the $ATT \coloneqq \mathbb{E}\left[\left. Y_{1} - Y_{0} \right\vert D = 1\right]$, the $ATU \coloneqq \mathbb{E}\left[\left. Y_{1} - Y_{0} \right\vert D = 0\right]$ and any PRTE using the following weighting functions:
	\begin{align*}
	\omega_{ATE}\left(u, P_{D} \right) & = 1 \\
	\omega_{ATT}\left(u, P_{D} \right) & = \dfrac{1 - F_{P_{D}}\left(u\right)}{\mathbb{E}\left[P_{D}\left(Z\right)\right]} \\
	\omega_{ATU}\left(u, P_{D} \right) & = \dfrac{F_{P_{D}}\left(u\right)}{1 - \mathbb{E}\left[P_{D}\left(Z\right)\right]} \\
	\omega_{PRTE}\left(u, P_{D} \right) & = \dfrac{F_{P^{*}}\left(u\right) - F_{P_{D}}\left(u\right)}{\mathbb{E}\left[P_{D}\left(Z\right)\right] - \mathbb{E}\left[P^{*}\right]},
	\end{align*}
	where $u \in \left[0,1\right]$ and $P^{*}$ is the probability of receiving the treatment under an alternative policy regime.

To derive easy-to-compute bounds around these treatment effect parameters, I strengthen Assumption \ref{ASsharp} in the following way:
\begin{assumption}\label{ASconstant}
	The function $P_{T}$ is invertible and, for a known $c \in \left[1, \dfrac{1}{\overline{p}} \right]$, there exists an unknown $\alpha \in \left[\dfrac{1}{c}, c\right]$ such that $P_{D}\left(z\right) = \alpha \cdot P_{T}\left(z\right)$, where $\overline{p} \coloneqq \sup_{z \in \mathcal{Z}} P_{T}\left(z\right)$.
\end{assumption}

Note that this assumption imposes that the mismeasured propensity score function is a re-scaled version of the true propensity score. Mathematically, it implies that $\mathcal{A} = \left\lbrace g \in \mathcal{G} \text{ and } g \text{ is constant.} \right\rbrace \times \left\lbrace 0 \right\rbrace$. Consequently, under Assumption \ref{ASconstant}, the bounds in Corollary \ref{CorOuterMTEsharp} become
\begin{equation*}
\Theta_{2} \coloneqq \left\lbrace \tilde{\theta}\colon \mathcal{Z} \rightarrow \mathbb{R} \left\vert \begin{array}{l}
\text{For some } a \in \left[\dfrac{1}{c}, c\right], \tilde{\theta}\left(z\right) = \dfrac{1}{a} \cdot f\left(z\right) \text{ for any } z \in \mathcal{Z}.
\end{array} \right. \right\rbrace.
\end{equation*}
Note that this result is a more restrictive version of the results derived by \citet[Section 4]{Acerenza2021}. Moreover, the bounds in Corollary \ref{CorBoundTE} simplify to
\begin{equation*}
\begin{array}{rcccl}
\inf\limits_{a \in \left[\dfrac{1}{c}, c\right]} \bigint_{0}^{1} \dfrac{1}{a} \cdot  f\left(P_{T}^{-1}\left(\sfrac{u}{a}\right)\right) \, du & \leq & ATE & \leq & \sup\limits_{a \in \left[\dfrac{1}{c}, c\right]} \bigint_{0}^{1} \dfrac{1}{a} \cdot  f\left(P_{T}^{-1}\left(\sfrac{u}{a}\right)\right) \, du;
\end{array}
\end{equation*}

\begin{align*}
& \inf_{a \in \left[\dfrac{1}{c}, c\right]} \bigintss_{0}^{1} \dfrac{1}{a} \cdot  f\left(P_{T}^{-1}\left(\sfrac{u}{a}\right)\right) \cdot \left(\dfrac{1 - F_{P_{T}}\left(\sfrac{u}{a}\right)}{a\cdot\mathbb{E}\left[P_{T}\left(Z\right)\right]}\right) \, du \leq ATT \\
& \hspace{60pt} \leq \sup_{a \in \left[\dfrac{1}{c}, c\right]} \bigintss_{0}^{1} \dfrac{1}{a} \cdot  f\left(P_{T}^{-1}\left(\sfrac{u}{a}\right)\right) \cdot \left(\dfrac{1 - F_{P_{T}}\left(\sfrac{u}{a}\right)}{a \cdot \mathbb{E}\left[P_{T}\left(Z\right)\right]}\right) \, du;
\end{align*}

\begin{align*}
& \inf_{a \in \left[\dfrac{1}{c}, c\right]} \bigintss_{0}^{1} \dfrac{1}{a} \cdot  f\left(P_{T}^{-1}\left(\sfrac{u}{a}\right)\right) \cdot \left(\dfrac{F_{P_{T}}\left(\sfrac{u}{a}\right)}{1 - a \cdot \mathbb{E}\left[P_{T}\left(Z\right)\right]}\right) \, du \leq ATU \\
& \hspace{60pt} \leq \sup_{a \in \left[\dfrac{1}{c}, c\right]} \bigintss_{0}^{1} \dfrac{1}{a} \cdot  f\left(P_{T}^{-1}\left(\sfrac{u}{a}\right)\right) \cdot \left(\dfrac{F_{P_{T}}\left(\sfrac{u}{a}\right)}{1 - a \cdot  \mathbb{E}\left[P_{T}\left(Z\right)\right]}\right) \, du;
\end{align*}

\begin{align*}
& \inf_{a \in \left[\dfrac{1}{c}, c\right]} \bigintss_{0}^{1} \dfrac{1}{a} \cdot  f\left(P_{T}^{-1}\left(\sfrac{u}{a}\right)\right) \cdot \left(\dfrac{F_{P^{*}}\left(u\right) - F_{P_{T}}\left(\sfrac{u}{a}\right)}{a \cdot \mathbb{E}\left[P_{T}\left(Z\right)\right] - \mathbb{E}\left[P^{*}\right]}\right) \, du \leq PRTE \\
& \hspace{60pt} \leq \sup_{a \in \left[\dfrac{1}{c}, c\right]} \bigintss_{0}^{1} \dfrac{1}{a} \cdot  f\left(P_{T}^{-1}\left(\sfrac{u}{a}\right)\right) \cdot \left(\dfrac{F_{P^{*}}\left(u\right) - F_{P_{T}}\left(\sfrac{u}{a}\right)}{a \cdot \mathbb{E}\left[P_{T}\left(Z\right)\right] - \mathbb{E}\left[P^{*}\right]}\right) \, du.
\end{align*}

\begin{remark}
    Imposing that function $P_{T}$ is invertible in Assumption \ref{ASconstant} may be considered restrictive in some applications. Importantly, this condition is always testable because $P_{T}\left(z\right) \coloneqq \mathbb{P}\left[\left. T = 1 \right\vert Z = z\right]$ is an observable reduced-form object.
\end{remark}

\subsection{Proof of Proposition \ref{ThmSharpMTE}}\label{ProofSharpMTE}

The proof is by construction. For each $\tilde{\theta} \in \Theta_{2}$, I define the density of the candidate random variables $\left(\tilde{U}, \tilde{Y}_{0}, \tilde{Y}_{1}, \tilde{D}\right)$ to ensure that they satisfy the five restrictions of Proposition \ref{ThmSharpMTE}.

Fix $\tilde{\theta} \in \Theta_{2}$ arbitrarily. By definition, there exist $\left(a, \tilde{b}\right) \in \mathcal{A}$ such that $\tilde{\theta}\left(z\right) = \dfrac{1}{a\left(z\right)} \cdot f\left(z\right)$ for any $z \in \mathcal{Z}$ and that the function $P_{\tilde{D}}\colon\mathcal{Z}\rightarrow\mathbb{R}$ defined as $P_{\tilde{D}}\left(z\right) = \bigint_{-\infty}^{z} a\left(\tilde{z}\right) \cdot \dfrac{d P_{T}\left(\tilde{z}\right)}{d z} \, d\tilde{z} + \tilde{b}$ for any $z \in \mathcal{Z}$ is bounded between 0 and 1. I break the construction of the candidate random variables $\left(\tilde{U}, \tilde{Y}_{0}, \tilde{Y}_{1}, \tilde{D}\right)$ in six steps.

\begin{enumerate}[label={Step \theenumi. }, ref = {Step \theenumi}]
	\item\label{StepPS}  Note that, by definition, function $P_{\tilde{D}}$ satisfies Assumption \ref{ASsharp}. Observe also that $\dfrac{dP_{\tilde{D}}\left(z\right)}{dz} = a\left(z\right) \cdot \dfrac{dP_{T}\left(z\right)}{dz}$ for any $z \in \mathcal{Z}$, implying that Assumption \ref{ASrank} holds for $P_{\tilde{D}}$. Consequently, function $P_{\tilde{D}}$ is invertible according to Rolle's Theorem.

	\item\label{StepUtilde} Define $\tilde{U} \sim Uniform\left[0, 1\right]$ and $\tilde{D} \coloneqq \mathbf{1}\left\lbrace \tilde{U} \leq P_{\tilde{D}}\left(Z\right) \right\rbrace$, ensuring that Assumption \ref{AScontinuous} and Equation \eqref{EqDtilde} hold.

	\item Since $f_{Z,\tilde{U},\tilde{Y}_{0},\tilde{Y}_{1}} = f_{Z} \cdot f_{\left. \tilde{U} \right\vert Z} \cdot \mathbb{P}\left[\left. \tilde{Y}_{0} = \cdot, \tilde{Y}_{1} = \cdot \right\vert \tilde{U}, Z\right]$, I define the joint density function of $\left(Z, \tilde{U}, \tilde{Y}_{0}, \tilde{Y}_{1}\right)$ through its components $f_{Z}$, $f_{\left. \tilde{U} \right\vert Z}$ and $\mathbb{P}\left[\left. \tilde{Y}_{0} = \cdot, \tilde{Y}_{1} = \cdot \right\vert \tilde{U}, Z\right]$. Fix $\left(z, u, y_{0}, y_{1}\right) \in \mathbb{R}^{2} \times \left\lbrace 0, 1\right\rbrace^{2}$ arbitrarily.
	\begin{enumerate}
		\item Note that $f_{Z}$ is identified. Consequently, $f_{Z}$ is defined according to the data.

		\item Define $f_{\left. \tilde{U} \right\vert Z}\left(\left. u \right\vert z\right) = f_{\tilde{U}}\left(u\right)$ and $\mathbb{P}\left[\left. \tilde{Y}_{0} = y_{0}, \tilde{Y}_{1} = y_{1} \right\vert \tilde{U} = u, Z = z\right] = \mathbb{P}\left[\left. \tilde{Y}_{0} = y_{0}, \tilde{Y}_{1} = y_{1} \right\vert \tilde{U} = u\right]$, ensuring that Assumption \ref{ASindependence} holds. Consequently, I only have to define $f_{\tilde{U}}\left(u\right)$ and $\mathbb{P}\left[\left. \tilde{Y}_{0} = y_{0}, \tilde{Y}_{1} = y_{1} \right\vert \tilde{U} = u\right]$.

		\item Note that $f_{\tilde{U}}$ is defined in \ref{StepUtilde}.

		\item I impose $\mathbb{P}\left[\left. \tilde{Y}_{0} = y_{0}, \tilde{Y}_{1} = y_{1} \right\vert \tilde{U} = u\right] = \mathbb{P}\left[\left. \tilde{Y}_{0} = y_{0} \right\vert \tilde{U} = u\right] \cdot \mathbb{P}\left[\left. \tilde{Y}_{1} = y_{1} \right\vert \tilde{U} = u\right]$ for simplicity. Define $\mathbb{P}\left[\left. \tilde{Y}_{0} = 0 \right\vert \tilde{U} = u\right] = 1 - \mathbb{P}\left[\left. \tilde{Y}_{0} = 1 \right\vert \tilde{U} = u\right]$ and $\mathbb{P}\left[\left. \tilde{Y}_{1} = 0 \right\vert \tilde{U} = u\right] = 1 - \mathbb{P}\left[\left. \tilde{Y}_{1} = 1 \right\vert \tilde{U} = u\right]$. Consequently, I only have to define $\mathbb{P}\left[\left. \tilde{Y}_{0} = 1 \right\vert \tilde{U} = u\right]$ and $\mathbb{P}\left[\left. \tilde{Y}_{1} = 1 \right\vert \tilde{U} = u\right]$.

		\item Define the set $\mathcal{U} \coloneqq \left\lbrace \left. \tilde{u} \in \mathbb{R} \right\vert \tilde{u} = P_{\tilde{D}}\left(z\right) \text{ for some } z \in \mathcal{Z} \right\rbrace \subseteq \left[0, 1\right]$.

		\item If $u \notin \mathcal{U}$, $\mathbb{P}\left[\left. \tilde{Y}_{0} = 1 \right\vert \tilde{U} = u\right] = 0$ and $\mathbb{P}\left[\left. \tilde{Y}_{1} = 1 \right\vert \tilde{U} = u\right] = 0$.

		\item From now on, suppose that $u \in \mathcal{U}$. Define $$\mathbb{P}\left[\left. \tilde{Y}_{0} = 1 \right\vert \tilde{U} = u\right] = -\tilde{\theta}\left(P_{\tilde{D}}^{-1}\left(u\right)\right) \cdot \mathbf{1}\left\lbrace \tilde{\theta}\left(P_{\tilde{D}}^{-1}\left(u\right)\right) < 0 \right\rbrace$$ and $$\mathbb{P}\left[\left. \tilde{Y}_{1} = 1 \right\vert \tilde{U} = u\right] = \tilde{\theta}\left(P_{\tilde{D}}^{-1}\left(u\right)\right) \cdot \mathbf{1}\left\lbrace \tilde{\theta}\left(P_{\tilde{D}}^{-1}\left(u\right)\right) \geq 0 \right\rbrace.$$
	\end{enumerate}

	\item Note that $\tilde{Y}_{0}$ and $\tilde{Y}_{1}$ satisfy Assumption \ref{ASfinite} and Equation \eqref{EqSupport}.

	\item Observe that, for any $z \in \mathcal{Z}$,
	\begin{align*}
	\mathbb{E}\left[\left. \tilde{Y}_{1} - \tilde{Y}_{0} \right\vert U = P_{\tilde{D}}\left(z\right) \right] & = \mathbb{P}\left[\left. \tilde{Y}_{1} = 1 \right\vert \tilde{U} = P_{\tilde{D}}\left(z\right)\right] - \mathbb{P}\left[\left. \tilde{Y}_{0} = 1 \right\vert \tilde{U} = P_{\tilde{D}}\left(z\right)\right] \\
	& = \tilde{\theta}\left(P_{\tilde{D}}^{-1}\left(P_{\tilde{D}}\left(z\right)\right)\right) \cdot \mathbf{1}\left\lbrace \tilde{\theta}\left(P_{\tilde{D}}^{-1}\left(P_{\tilde{D}}\left(z\right)\right)\right) \geq 0 \right\rbrace \\
	& \hspace{20pt} - \left(-\tilde{\theta}\left(P_{\tilde{D}}^{-1}\left(P_{\tilde{D}}\left(z\right)\right)\right) \cdot \mathbf{1}\left\lbrace \tilde{\theta}\left(P_{\tilde{D}}^{-1}\left(P_{\tilde{D}}\left(z\right)\right)\right) < 0 \right\rbrace\right) \\
	& = \tilde{\theta}\left(z\right),
	\end{align*}
	ensuring that Equation \eqref{EqThetaTilde} holds.

	\item Note that, for any $z \in \mathcal{Z}$,
	\begin{align*}
	\dfrac{d\mathbb{E}\left[\left. \tilde{Y} \right\vert Z = z\right]}{dz} & = \dfrac{dP_{\tilde{D}}\left(z\right)}{dz} \cdot \tilde{\theta}\left(z\right) \\
	& = a\left(z\right) \cdot \dfrac{dP_{T}\left(z\right)}{dz} \cdot \tilde{\theta}\left(z\right) & \text{by \ref{StepPS}} \\
	& = a\left(z\right) \cdot \dfrac{dP_{T}\left(z\right)}{dz} \cdot \dfrac{1}{a\left(z\right)} \cdot f\left(z\right) & \text{by definition} \\
	& = \dfrac{d\mathbb{E}\left[\left. Y \right\vert Z = z\right]}{dz} & \text{by Equation \eqref{EqLIV}},
	\end{align*}
	implying that Equation \eqref{EqDataRestriction} holds.
\end{enumerate}

\newpage

\section{Estimation Details}\label{AppEstimation}
\setcounter{table}{0}
\renewcommand\thetable{H.\arabic{table}}

\setcounter{figure}{0}
\renewcommand\thefigure{H.\arabic{figure}}

\setcounter{equation}{0}
\renewcommand\theequation{H.\arabic{equation}}

\setcounter{theorem}{0}
\renewcommand\thetheorem{H.\arabic{theorem}}

\setcounter{proposition}{0}
\renewcommand\theproposition{H.\arabic{proposition}}

\setcounter{corollary}{0}
\renewcommand\thecorollary{H.\arabic{corollary}}

\setcounter{assumption}{0}
\renewcommand\theassumption{H.\arabic{assumption}}

In this Appendix, my main goal is to estimate the MTE function's sign (Corollary \ref{CorSign}) and its bounds (Proposition \ref{CorOuterMTE}).

To accomplish this task in my empirical application, I need to take covariates into account because trial judges are only randomly allocated to criminal cases after conditioning on the court district. Consequently, the propensity score becomes $P_{D}\left(z, x\right) = \mathbb{P}\left[\left. D = 1 \right\vert Z = z, X = x \right]$, and the target parameter (MTE function) is now given by $$\theta\left(z,x\right) = \mathbb{E}\left[\left. Y_{1} - Y_{0} \right\vert U = P_{D}\left(z, x\right), X = x \right]$$ for any value $z$ of the instrument and any value $x$ of the covariates, where the covariates encompass a full set of court district dummies.\footnote{When the model includes covariates, Assumption \ref{ASsharp} imposes that, for any value $x$ of the covariates, $P_{D}\left(\cdot, x\right)$ is invertible as a function of its first argument. In a separable model ($P_{D}\left(z, x\right) = g\left(z\right) + h\left(x\right)$) such as the ones used in this appendix, Assumption \ref{ASsharp} imposes that $g$ is invertible.}

To estimate the MTE function's sign and bounds, I need to estimate the LIV estimand $$f\left(z,x\right) = \dfrac{\sfrac{d\mathbb{E}\left[\left. Y \right\vert Z = z, X = x\right]}{dz}}{\sfrac{d\mathbb{E}\left[\left. T \right\vert Z = z, X = x\right]}{dz}}.$$ This object is also useful to understand the impact of ignoring misclassification of the treatment variable in my empirical application (Proposition \ref{PropLIV}).

Moreover, I am able to observe the correctly classified treatment variable $D$ (final sentence in each case) in my empirical application. Consequently, I can point identify the MTE function $\theta(\cdot,\cdot)$ and use it as a benchmark for the partial identification strategy described in Section \ref{Sbounds}. Estimating my target parameter is my second goal in this appendix. To do so, I need to estimate the correctly measured LIV estimand $$f^{*}\left(z,x\right) = \dfrac{\sfrac{d\mathbb{E}\left[\left. Y \right\vert Z = z, X = x\right]}{dz}}{\sfrac{d\mathbb{E}\left[\left. D \right\vert Z = z, X = x\right]}{dz}}$$ because $f^{*}\left(z,x\right) = \theta\left(z,x\right)$ as a direct consequence of Proposition \ref{PropLIV}.

I explain how to estimate the LIV estimand $f\left(\cdot,\cdot\right)$ and the correctly measured LIV estimand $f^{*}\left(\cdot,\cdot\right)$ in Subsections \ref{Sestmisclassified} and \ref{Sestcorrectly}, respectively. In both subsections, I focus on parametric estimators for brevity. In Subsection \ref{Sestbounds}, I also explain how to use the estimator of $f\left(\cdot, \cdot\right)$ to estimate the sign of $\theta\left(\cdot, \cdot\right)$ (Corollary \ref{CorSign}) and the bounds $\Theta_{1}$ (Proposition \ref{CorOuterMTE}). In Subsection \ref{Smontecarlo}, I implement a Monte Carlo exercise to illustrate the bias of the LIV estimand $f\left(\cdot,\cdot\right)$ and the finite sample behavior of the estimators described in Subsection \ref{Sestmisclassified} and \ref{Sestcorrectly}.

\subsection{Estimation with the Misclassified Treatment Variable}\label{Sestmisclassified}

In this subsection, I observe an independent and identically distributed sample $\left(Y_{i},Z_{i},X_{i},T_{i}\right)_{i=1}^{N}$, where $N$ is the sample size, index $i$ denotes a case-defendant pair, $Y_{i}$ indicates that defendant $i$ recidivated within two years of her case's final decision, $Z_{i}$ is the punishment rate of the trial judge who analyzed case $i$, $X_{i}$ is the court district where case $i$ was analyzed, $T_{i}$ is the trial judge's decision in case $i$.

Our goal is to estimate the LIV estimand $f\left(z,x\right) = \dfrac{\sfrac{d\mathbb{E}\left[\left. Y \right\vert Z = z, X = x\right]}{dz}}{\sfrac{d\mathbb{E}\left[\left. T \right\vert Z = z, X = x\right]}{dz}}$ for any value $z$ of the instrument and any value $x$ of the covariates.

To estimate the denominator, I use a polynomial model $$\mathbb{E}\left[\left. T_{i} \right\vert Z_{i} = z, X_{i} = x\right] = \gamma_{x} + \gamma_{1} \cdot z + \ldots + \gamma_{L} \cdot z^{L}$$ with court district fixed effects, where $L \in \mathbb{N}$. This model's coefficients can be estimated by ordinary least squares (OLS) and I denote its estimators by $\hat{\gamma}_{x}, \hat{\gamma}_{1}, \ldots, \hat{\gamma}_{L}$, implying that an estimator for $\sfrac{d\mathbb{E}\left[\left. T \right\vert Z = z, X = x\right]}{dz}$ is given by $\hat{\gamma}_{1} + \ldots + L \cdot \hat{\gamma}_{L} \cdot z^{L-1}$.

To estimate the numerator, I need to model the propensity score $\mathbb{P}\left[\left. D_{i} = 1 \right\vert Z_{i} = z, X_{i} = x \right]$ and the function $\mathbb{E}\left[\left. Y_{1,i} - Y_{0,i} \right\vert U_{i} = u, X_{i} = x \right]$. I use polynomial models with court district fixed effects for both  objects:
\begin{align}
P_{D}\left(z, x\right) & = \mathbb{P}\left[\left. D_{i} = 1 \right\vert Z_{i} = z, X_{i} = x \right] \nonumber \\
& = \mathbb{P}\left[\left. U \leq \gamma_{x}^{*} + \gamma_{1}^{*} \cdot z + \ldots + \gamma_{L^{*}}^{*} \cdot z^{L^{*}} \right\vert Z_{i} = z, X_{i} = x \right] \nonumber \\
& \label{EQcorrectPS} = \gamma_{x}^{*} + \gamma_{1}^{*} \cdot z + \ldots + \gamma_{L^{*}}^{*} \cdot z^{L^{*}}
\end{align}
and
\begin{equation}\label{EQmte}
\mathbb{E}\left[\left. Y_{1,i} - Y_{0,i} \right\vert U_{i} = u, X_{i} = x \right] = \beta_{x} + \beta_{1} \cdot u + \ldots + \beta_{K} \cdot u^{K},
\end{equation}
where $L^{*} \in \mathbb{N}$ and $K \in \mathbb{N}$. Based on Equations \eqref{EQcorrectPS} and \eqref{EQmte}, I have that
\begin{align}
\theta\left(z,x\right) & = \mathbb{E}\left[\left. Y_{1,i} - Y_{0,i} \right\vert U_{i} = P_{D}\left(z, x\right), X_{i} = x \right] \nonumber \\
& \label{EQtheta} = \beta_{x} + \beta_{1} \cdot \left(\gamma_{x}^{*} + \gamma_{1}^{*} \cdot z + \ldots + \gamma_{L^{*}}^{*} \cdot z^{L^{*}}\right) + \ldots + \beta_{K} \cdot \left(\gamma_{x}^{*} + \gamma_{1}^{*} \cdot z + \ldots + \gamma_{L^{*}}^{*} \cdot z^{L^{*}}\right)^{K}
\end{align}
and that
\begin{align}
\mathbb{E}\left[\left. Y_{i} \right\vert Z_{i} = z, X_{i} = x\right] & = \mathbb{E}\left[\left. Y_{0,i} \right\vert X_{i} = x\right] + \int_{0}^{P_{D}\left(z,x\right)} \mathbb{E}\left[\left. Y_{1,i} - Y_{0,i} \right\vert U_{i} = u, X_{i} = x \right] \, du \nonumber \\
& = \alpha_{x} + \int_{0}^{P_{D}\left(z,x\right)} \left(\beta_{x} + \beta_{1} \cdot u + \ldots + \beta_{K} \cdot u^{K}\right) \, du \nonumber \\
& = \alpha_{x} + \beta_{x} \cdot P_{D}\left(z,x\right) + \dfrac{\beta_{1}}{2} \cdot \left[P_{D}\left(z,x\right)\right]^{2} + \ldots + \dfrac{\beta_{K}}{K+1} \cdot \left[P_{D}\left(z,x\right)\right]^{K+1} \nonumber \\
& = \alpha_{x} + \beta_{x} \cdot \left[\gamma_{x}^{*} + \gamma_{1}^{*} \cdot z + \ldots + \gamma_{L^{*}}^{*} \cdot z^{L^{*}}\right] \nonumber \\
& \hspace{20pt} + \dfrac{\beta_{1}}{2} \cdot \left[\gamma_{x}^{*} + \gamma_{1}^{*} \cdot z + \ldots + \gamma_{L^{*}}^{*} \cdot z^{L^{*}}\right]^{2} \nonumber \\
& \hspace{20pt} + \ldots + \dfrac{\beta_{K}}{K+1} \cdot \left[\gamma_{x}^{*} + \gamma_{1}^{*} \cdot z + \ldots + \gamma_{L^{*}}^{*} \cdot z^{L^{*}}\right]^{K+1} \nonumber \\
& \label{EQlivnum} = \delta_{x} + \delta_{1,x} \cdot z + \ldots + \delta_{L^{*}\cdot\left(K+1\right),x} \cdot z^{L^{*}\cdot\left(K+1\right)}
\end{align}
where $\delta_{x}, \delta_{1,x}, \ldots, \delta_{L^{*}\cdot\left(K+1\right),x}$ are known functions of $\gamma_{x}^{*}, \gamma_{1}^{*}, \ldots,\gamma_{L^{*}}^{*},\beta_{x},\beta_{1},\ldots,\beta_{K}$ and $\alpha_{x} \coloneqq \mathbb{E}\left[\left. Y_{0,i} \right\vert X_{i} = x\right]$. The coefficients in Equation \eqref{EQlivnum} can be estimated by OLS and I denote its estimators by $\hat{\delta}_{x}, \hat{\delta}_{1,x}, \ldots, \hat{\delta}_{L^{*}\cdot\left(K+1\right),x}$, implying that an estimator for $\sfrac{d\mathbb{E}\left[\left. Y \right\vert Z = z, X = x\right]}{dz}$ is given by $\hat{\delta}_{1,x} + \ldots + \left(L^{*}\cdot\left(K+1\right)\right) \cdot \hat{\delta}_{L^{*}\cdot\left(K+1\right),x} \cdot z^{L^{*}\cdot\left(K+1\right)- 1}$.

Consequently, an estimator for the LIV estimand is given by $$\hat{f}\left(z,x\right) = \dfrac{\hat{\delta}_{1,x} + \ldots + \left(L^{*}\cdot\left(K+1\right)\right) \cdot \hat{\delta}_{L^{*}\cdot\left(K+1\right),x} \cdot z^{L^{*}\cdot\left(K+1\right)- 1}}{\hat{\gamma}_{1} + \ldots + L \cdot \hat{\gamma}_{L} \cdot z^{L-1}}.$$ In my empirical application, I impose that $L^{*} = L = 2$ and  $K = 1$, implying that
\begin{equation}\label{EQlivest}
\hat{f}\left(z,x\right) = \dfrac{\hat{\delta}_{1,x} + 2 \cdot \hat{\delta}_{2,x} \cdot z + 3 \cdot \hat{\delta}_{3} \cdot z^{2} + 4 \cdot \hat{\delta}_{4} \cdot z^{3}}{\hat{\gamma}_{1} + 2 \cdot \hat{\gamma}_{2} \cdot z}.
\end{equation}

Setting $L^{*} = L = 2$ and $K = 1$ imposes that the function $\mathbb{E}\left[\left. Y_{1,i} - Y_{0,i} \right\vert U_{i} = u, X_{i} = x \right]$ is linear and that the function $P_{D}\left(z, x\right)$ is quadratic. This type of parametric assumption is not only common in the empirical literature \citep{Cornelissen2018}, but also seems to be valid in my empirical application. In Appendix \ref{AppStandardMTE}, I estimate the function $\mathbb{E}\left[\left. Y_{1,i} - Y_{0,i} \right\vert U_{i} = u, X_{i} = x \right]$ semi-parametrically \citep{Robinson1988,Calonico2019} and find that this function is approximately constant. Moreover, my empirical results in Subsection \ref{Sresults} and in Appendix \ref{AppPSsemi} suggest that the propensity score function $P_{D}\left(z, x\right)$ is linear. Consequently, my parametric model seems sufficiently flexible for my empirical application.

Furthermore, this simple parametric model is sufficient to have a deep understanding of the misclassification bias. When $L^{*} = L = 2$, the ratio between the derivatives of the propensity score functions (Proposition \ref{PropLIV}) is given by $$\dfrac{\sfrac{dP_{D}\left(z,x\right)}{d z}}{\sfrac{dP_{T}\left(z,x\right)}{d z}} = \dfrac{\gamma_{1}^{*} + 2 \cdot \gamma_{2}^{*} \cdot z}{\gamma_{1} + 2 \cdot \gamma_{2} \cdot z}.$$ The right-hand side of this expression is a non-constant function of the instrument, allowing the scaling factor in Proposition \ref{PropLIV} to be greater than or less than 1 depending on the value of the instrument. Consequently, in this simple parametric model, the misclassification bias can either attenuate or enlarge the true MTE function depending on the value of the instrument.

\subsection{Estimating the Sign of $\theta\left(\cdot, \cdot\right)$ and the Bounds of Set $\Theta_{1}$}\label{Sestbounds}

In this subsection, I use the estimator $\hat{f}\left(\cdot,\cdot\right)$ to estimate the sign of $\theta\left(\cdot, \cdot\right)$ (Corollary \ref{CorSign}) and the upper and lower bounds of the set $\Theta_{1}$ (Proposition \ref{CorOuterMTE}).

To estimate the sign of $\theta\left(\cdot, \cdot\right)$, I first define the function $s\left(z,x\right) \coloneqq sign\left(\theta\left(z,x\right)\right)$ for any value $z$ of the instrument and any value $x$ of the covariates. Based on Corollary \ref{CorSign}, an estimator for $s\left(z,x\right)$ is given by
\begin{equation}\label{EQestsign}
\hat{s}\left(z,x\right) \coloneqq sign\left(\hat{f}\left(z,x\right)\right).
\end{equation}

To estimate the upper and lower bounds of set $\Theta_{1}$, I first define its upper bound $$\theta_{U}\left(z,x\right) \coloneqq \sup_{\tilde{\theta} \in \Theta_{1}} \tilde{\theta}\left(z,x\right)$$ and its lower bound $$\theta_{L}\left(z,x\right) \coloneqq \inf_{\tilde{\theta} \in \Theta_{1}} \tilde{\theta}\left(z,x\right),$$ where
\begin{equation*}
	\Theta_{1} \coloneqq \left\lbrace \tilde{\theta}\left(\cdot,\cdot\right) \left\vert \begin{array}{l}
		\text{For any any value } z \text{ of the instrument and any value } x \text{ of the covariates }, \\
		\tilde{\theta}\left(z,x\right) \in \left\lbrace \begin{array}{ll}
			\left[\dfrac{1}{c} \cdot f\left(z,x\right), c \cdot f\left(z\right) \right] & \text{if } f\left(z,x\right) \geq 0 \\
			\left[c \cdot f\left(z,x\right), \dfrac{1}{c} \cdot f\left(z,x\right) \right] & \text{if } f\left(z\right) < 0
		\end{array}  \right. .
	\end{array}   \right. \right\rbrace.
\end{equation*}
Based on Proposition \ref{CorOuterMTE}, estimators for $\theta_{U}\left(z,x\right)$ and $\theta_{L}\left(z,x\right)$ are given by
\begin{equation}\label{EQestupper}
\hat{\theta}_{U}\left(z,x\right) \coloneqq \left\lbrace \begin{array}{ll}
c \cdot \hat{f}\left(z,x\right) & \text{ if  } \hat{f}\left(z,x\right) \geq 0 \\
\dfrac{1}{c} \cdot \hat{f}\left(z,x\right) & \text{ if  } \hat{f}\left(z,x\right) < 0
\end{array} \right.
\end{equation}
and
\begin{equation}\label{EQestlower}
\hat{\theta}_{L}\left(z,x\right) \coloneqq \left\lbrace \begin{array}{ll}
\dfrac{1}{c} \cdot \hat{f}\left(z,x\right) & \text{ if  } \hat{f}\left(z,x\right) \geq 0 \\
c \cdot \hat{f}\left(z,x\right) & \text{ if  } \hat{f}\left(z,x\right) < 0
\end{array} \right.
\end{equation}
respectively.

For brevity, I do not write explicit formulas for the upper and lower bounds of $\Theta_{2}$ in Corollary \ref{CorOuterMTEsharp}. They are analogous to $\hat{\theta}_{U}\left(z,x\right)$ and $\hat{\theta}_{L}\left(z,x\right)$, where I replace the true LIV estimand $f\left(\cdot,\cdot\right)$ by its estimator $\hat{f}\left(\cdot,\cdot\right)$. Estimators for the bounds in Corollary \ref{CorBoundTE} can be written in a similar way.

\subsection{Estimation with the Correctly Classified Treatment Variable}\label{Sestcorrectly}

In this subsection, we observe an independent and identically distributed sample $\left(Y_{i},Z_{i},X_{i},D_{i}\right)_{i=1}^{N}$, where $N$ is the sample size, index $i$ denotes a case-defendant pair, $Y_{i}$ indicates that defendant $i$ recidivated within two years of her case's final decision, $Z_{i}$ is the punishment rate of the trial judge who analyzed case $i$, $X_{i}$ is the court district where case $i$ was analyzed, $D_{i}$ is the final decision in case $i$.

Our goal is to estimate the correctly measured LIV estimand $f^{*}\left(z,x\right) = \dfrac{\sfrac{d\mathbb{E}\left[\left. Y \right\vert Z = z, X = x\right]}{dz}}{\sfrac{d\mathbb{E}\left[\left. D \right\vert Z = z, X = x\right]}{dz}}$ for any value $z$ of the instrument and any value $x$ of the covariates. The only difference between this subsection and Subsection \ref{Sestmisclassified} is that, now, I can directly estimate the propensity score (Equation \eqref{EQcorrectPS}) using OLS and I denote its estimators by $\hat{\gamma}_{x}^{*},\hat{\gamma}_{1}^{*},\ldots,\hat{\gamma}_{L^{*}}^{*}$. Imposing that $L^{*} = L = 2$ and $K = 1$, an estimator for the correctly measured LIV estimand is given by
\begin{equation}\label{EQcorrectlivest}
	\hat{f}^{*}\left(z,x\right) = \dfrac{\hat{\delta}_{1,x} + 2 \cdot \hat{\delta}_{2,x} \cdot z + 3 \cdot \hat{\delta}_{3} \cdot z^{2} + 4 \cdot \hat{\delta}_{4} \cdot z^{3}}{\hat{\gamma}_{1}^{*} + 2 \cdot \hat{\gamma}_{2}^{*} \cdot z}.
\end{equation}
Moreover, this object is also an estimator for the true MTE function as proven by \cite{Chalak2017} in a nonparametric context.

\subsection{Monte Carlo Simulation}\label{Smontecarlo}

The data generating process (DGP) of this Monte Carlo Simulation is based on the discussion in Appendix \ref{Ebounded} and, consequently, satisfies Assumptions \ref{ASindependence}-\ref{ASfinite} and \ref{ASbounded}. The Monte Carlo's DGP is given by the following system of equations:
\begin{align*}
D & = \mathbf{1}\left\lbrace U \leq \gamma_{x}^{*} \cdot \mathbf{1}\left\lbrace X = x \right\rbrace + \gamma_{1}^{*} \cdot Z \right\rbrace \\
Y_{0} & = \alpha_{x} \cdot \mathbf{1}\left\lbrace X = x \right\rbrace + \epsilon_{0} \\
Y_{1} & = \beta_{x} \cdot \mathbf{1}\left\lbrace X = x \right\rbrace + \beta_{1} \cdot U + \epsilon_{1} \\
T & = \mathbf{1}\left\lbrace V \leq r \right\rbrace \cdot D + \mathbf{1}\left\lbrace V > r \right\rbrace \cdot \left(1 - D\right),
\end{align*}
where $U \sim Unif\left(0, 1\right)$, $X = 0$ with probability $p_{0}$, $X = 1$ with probability $p_{1}$, $X = 2$ with probability $p_{2} = 1 - p_{0} - p_{1}$,  $Z \sim Unif\left(0, 1\right)$, $\epsilon_{0} \sim N\left(0, \sigma^{2}\right)$, $\epsilon_{1} \sim N\left(0, \sigma^{2}\right)$, $V \sim Unif\left(0, 1\right)$, and $U$, $X$, $Z$, $\epsilon_{0}$, $\epsilon_{1}$ and $V$ are mutually independent.

Note that this DGP satisfies the parametric assumptions ($L^{*} = L = K = 1$) in Subsections \ref{Sestmisclassified} and \ref{Sestcorrectly} as shown by the following equations:
\begin{align*}
P_{D}\left(z,x\right) & = \gamma_{x}^{*} + \gamma_{1}^{*} \cdot z \\
\mathbb{E}\left[\left. Y_{1} - Y_{0} \right\vert U = u, X = x \right] & = \beta_{x} - \alpha_{x} + \beta_{1} \cdot u \\
\theta\left(z, x\right) & = \beta_{x} - \alpha_{x} + \beta_{1} \cdot \gamma_{x}^{*} + \beta_{1} \cdot \gamma_{1}^{*} \cdot z \\
\mathbb{E}\left[\left. T \right\vert Z = z, X = x \right] & = 1 - r + \left(2r - 1\right) \cdot \gamma_{x}^{*} + \left(2r - 1\right) \cdot \gamma_{1}^{*} \cdot z \\
f\left(z, x\right) & = \dfrac{\theta\left(z, x\right)}{2r - 1},
\end{align*}
implying that the misclassification bias of the LIV estimand f$\left(\cdot, \cdot\right)$ is given by
\begin{equation}\label{EQmcbias}
bias\left(z,x\right) = f\left(z, x\right) - \theta\left(z, x\right) = \theta\left(z, x\right) \cdot \left[\dfrac{2 - 2r}{2r - 1}\right].
\end{equation}

Observe that, for $r \in \left(0.5, 1\right]$, the misclassification bias is a decreasing function of $r$ and has the same sign of $\theta\left(z, x\right)$. As a consequence of the last property, note that the misclassification bias moves the point-estimates away from zero. Observe also that, if the misclassification rate is too large $\left(r \in \left[0,0.5\right)\right)$, Assumption \ref{ASsign} does not hold anymore, implying that misclassification bias is so intense that the LIV estimand $f$ does not capture the correct sign of the true MTE function $\theta$.

I choose the following set of parameters for this Monte Carlo Simulation: $\gamma_{x = 0}^{*} = 0.1$, $\gamma_{x = 1}^{*} = 0.2$, $\gamma_{x = 2}^{*} = 0.3$, $\gamma_{1}^{*} = 0.6$, $\alpha_{x = 0} = \alpha_{x = 1} = \alpha_{x = 2} = 0$, $\beta_{x = 0} = 0$, $\beta_{x = 1} = 0.5$, $\beta_{x = 2} = 1$, $\beta_{1} = -1$, $p_{0} = p_{1} = \sfrac{1}{3}$ and $\sigma^{2} = 0.5$. Moreover, I set $r = 0.95$ so that the amount of misclassification in this Monte Carlo simulation is similar to the share of cases whose trial judges' sentences are reversed in my empirical application.\footnote{See Subsection \ref{Sdescriptive}.} Moreover, sample size $N$ is equal to $3,000$ and the number of Monte Carlo repetitions $M$ is equal to $10,000$. Finally, in each Monte Carlo repetition, I estimate the bounds from Proposition \ref{CorOuterMTE} using four values of $c$: $\sfrac{10}{9}$, $\sfrac{12}{9}$, $\sfrac{14}{9}$, $\sfrac{16}{9}$. Note that, in this Monte Carlo's DGP, the smallest constant $c$ that satisfies Assumption \ref{ASbounded} is equal to $\sfrac{1}{\left(2 \cdot r - 1\right)} = \sfrac{10}{9}$.

This example illustrates the complexity of the misclassification bias and its possibly large magnitude. Given the chosen parameters, the misclassification bias is always negative for $x = 0$, always positive for $x = 2$, and changes its sign at $z = 0.5$ for $x = 1$, illustrating its complexity when estimating entire functions. Furthermore, the bias represents 11.1\% of the true MTE effect, illustrating that it can be large even when the misclassification rate is as small as 5\%. If the misclassification rate were as large as the share of reversed criminal cases in State Courts in the U.S. (12\%, \citealp{Waters2015}), the misclassification bias would represent 31.6\% of the true MTE effect.\footnote{A larger misclassification rate also impacts the smallest constant $c$ that satisfies Assumption \ref{ASbounded}. If the misclassification rate was as large as the share of reversed criminal cases in State Courts in the U.S. (12\%, \citealp{Waters2015}), the smallest constant $c$ that satisfies Assumption \ref{ASbounded} is equal to $\sfrac{25}{19}$ which is larger than the baseline value $\left(\sfrac{10}{9}\right)$ in the Monte Carlo Simulation. Consequently, the width of the bounds in Proposition \ref{CorOuterMTE} would have to increase to accommodate a larger misclassification rate.}

\begin{figure}[!htbp]
	\begin{center}
		\begin{subfigure}[b]{0.47\textwidth}
			\centering
			\includegraphics[width = \textwidth]{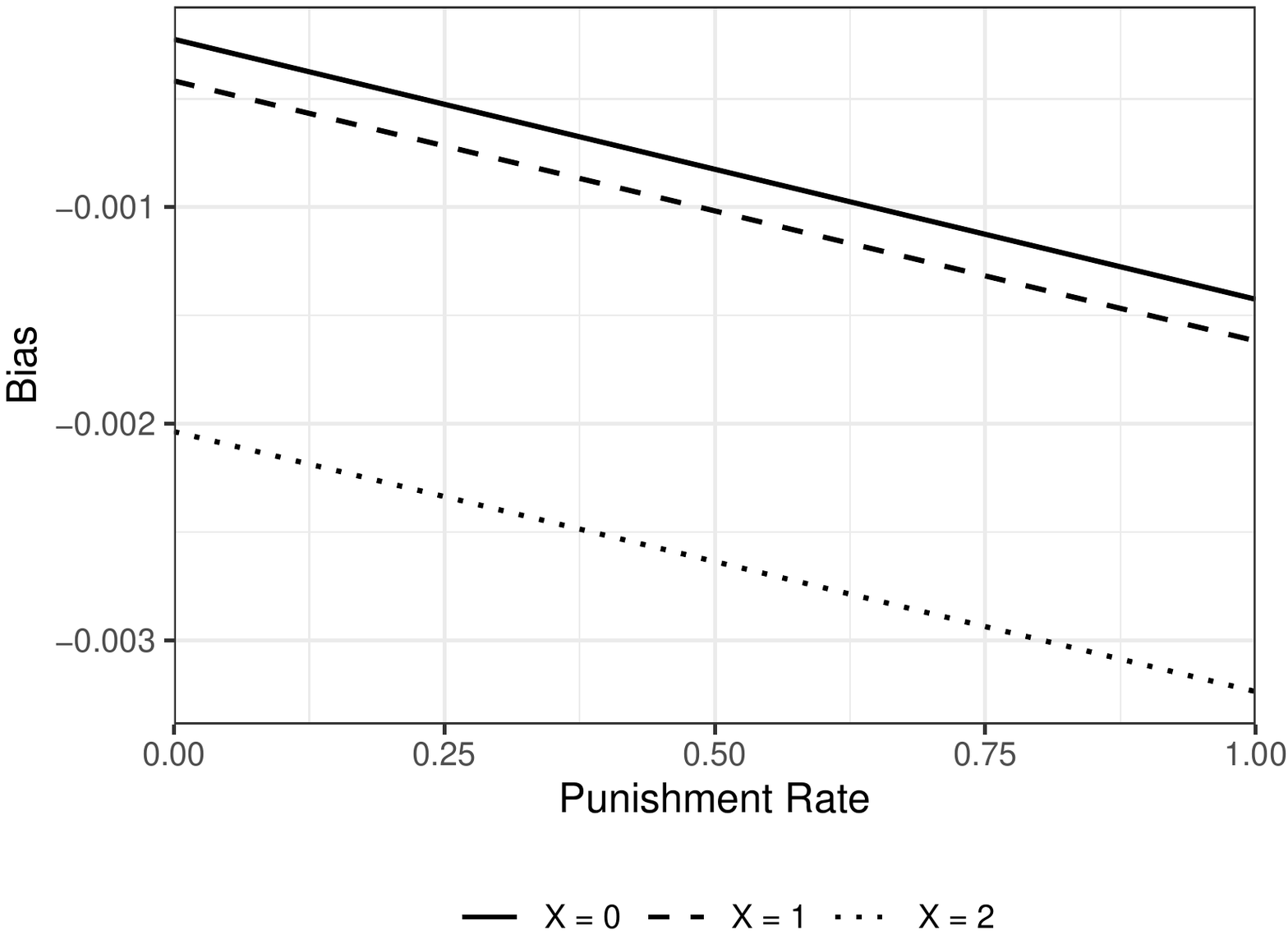}
			\caption{Correctly Measured LIV Estimator ($\hat{f}^{*}$)}
			\label{FigBiasMTE}
		\end{subfigure}
		\hfill
		\begin{subfigure}[b]{0.47\textwidth}
			\centering
			\includegraphics[width = \textwidth]{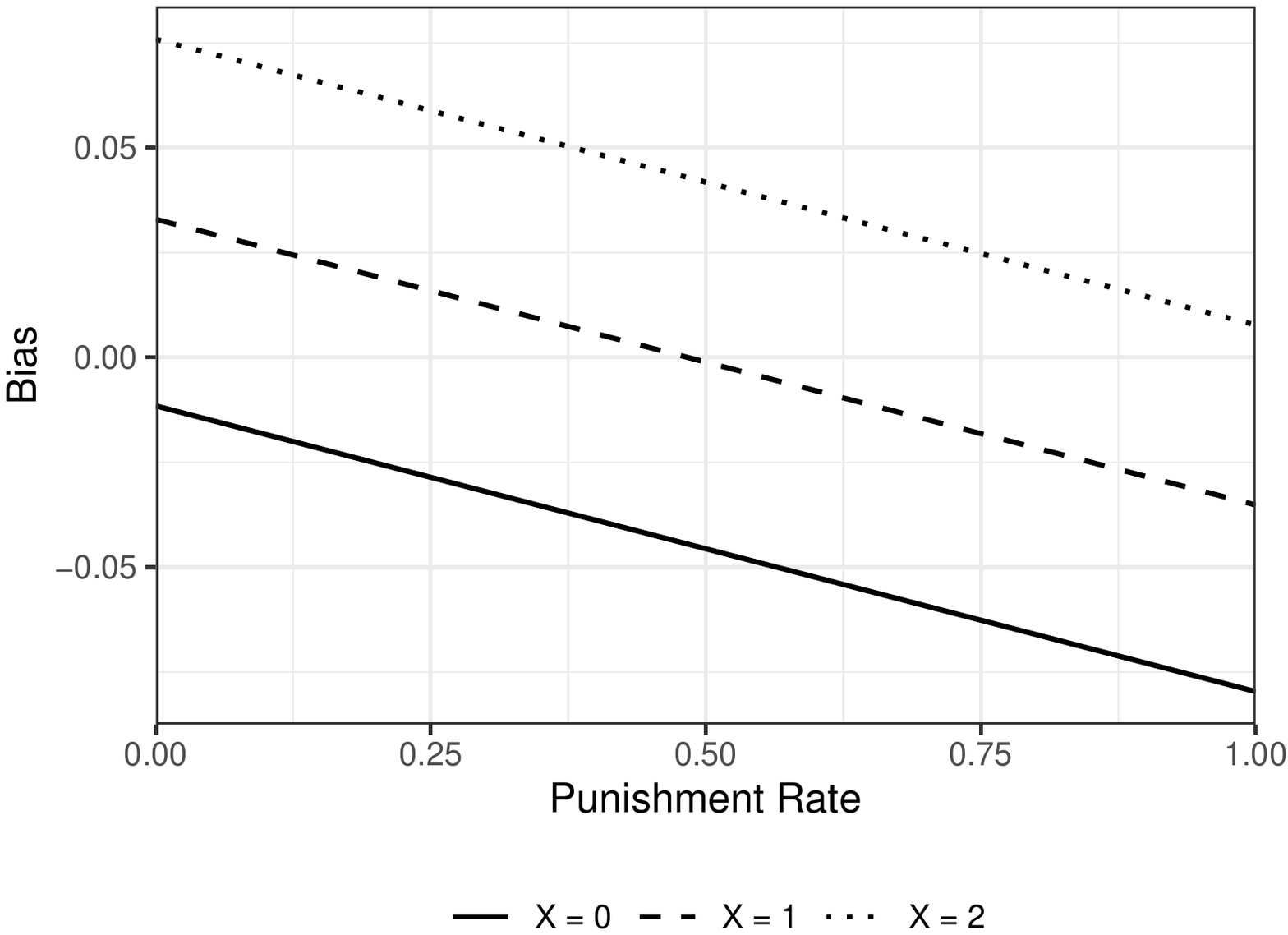}
			\caption{Mismeasured LIV Estimator ($\hat{f}$)}
			\label{FigBiasLIV}
		\end{subfigure}
		\caption[]{Bias of the LIV Estimators}
		\label{FigBias}
	\end{center}
	\footnotesize{Notes: The solid lines denote the bias of LIV estimators $\hat{f}^{*}\left(\cdot,0\right)$ and $\hat{f}\left(\cdot,0\right)$. The dashed lines denote the bias of LIV estimators $\hat{f}^{*}\left(\cdot,1\right)$ and $\hat{f}\left(\cdot,1\right)$. The dotted lines denote bias of LIV estimators $\hat{f}^{*}\left(\cdot,2\right)$ and $\hat{f}\left(\cdot,2\right)$.}
\end{figure}

Using this Monte Carlo exercise, I can compute the bias of the correctly measured LIV estimator $\hat{f}^{*}\left(\cdot,\cdot\right)$ and of the mismeasured LIV estimator  $\hat{f}\left(\cdot,\cdot\right)$. Subfigure \ref{FigBiasMTE} shows that the bias of $\hat{f}^{*}\left(\cdot,\cdot\right)$ is very small, reaching values that are no larger than 0.004 in magnitude. This result is expected because the parametric model is correctly specified in this simulation. Moreover, Subfigure \ref{FigBiasLIV} shows that the bias of $\hat{f}\left(\cdot,\cdot\right)$ is much larger than the bias of $\hat{f}^{*}\left(\cdot,\cdot\right)$, reaching values as large as 0.075 in magnitude.

\begin{figure}[!htbp]
	\begin{center}
		\begin{subfigure}[b]{0.3\textwidth}
			\centering
			\includegraphics[width = \textwidth]{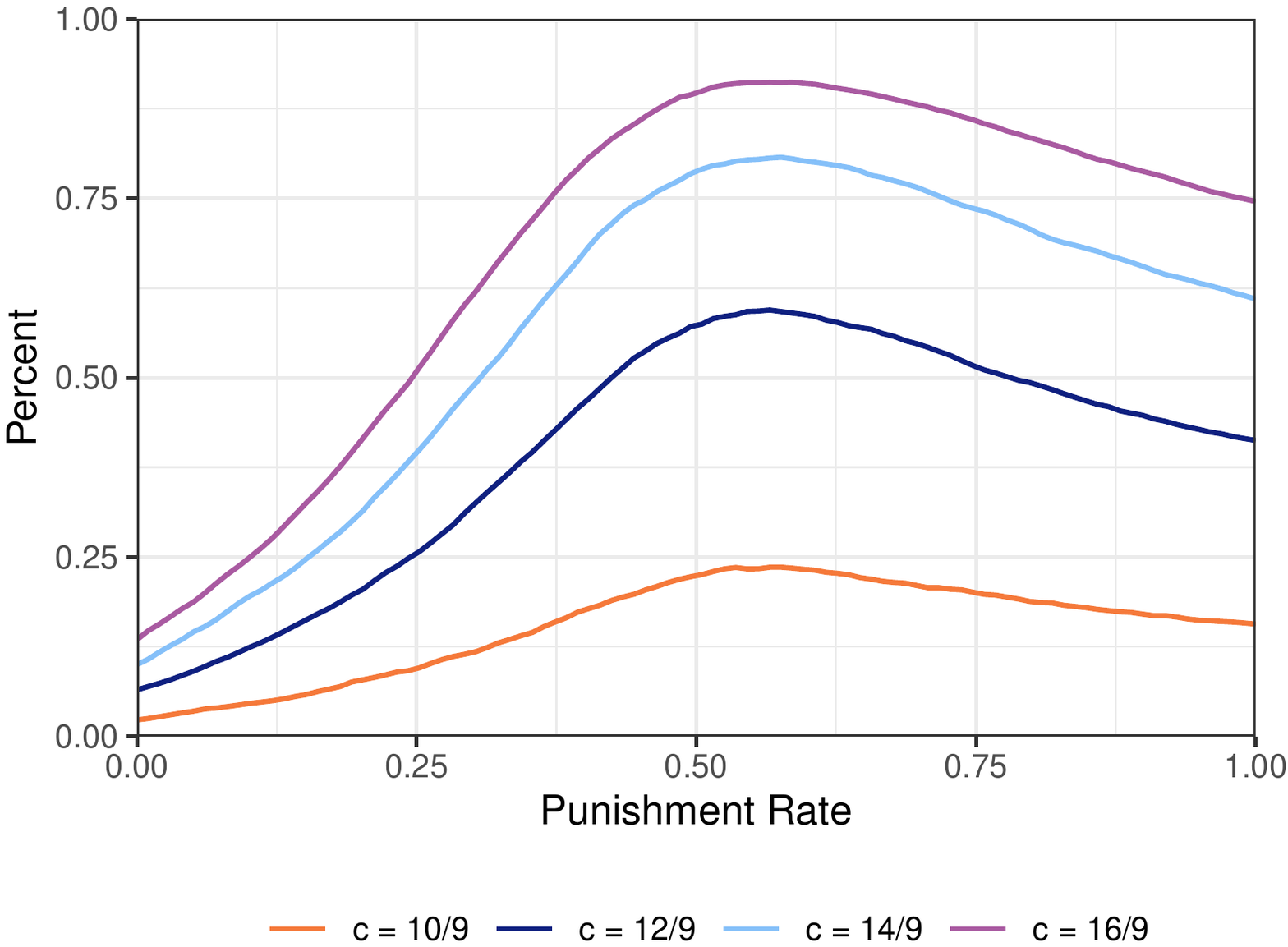}
			\caption{Coverage Rate: $x=0$}
			\label{FigCRX0}
		\end{subfigure}
		\hfill
		\begin{subfigure}[b]{0.3\textwidth}
			\centering
			\includegraphics[width = \textwidth]{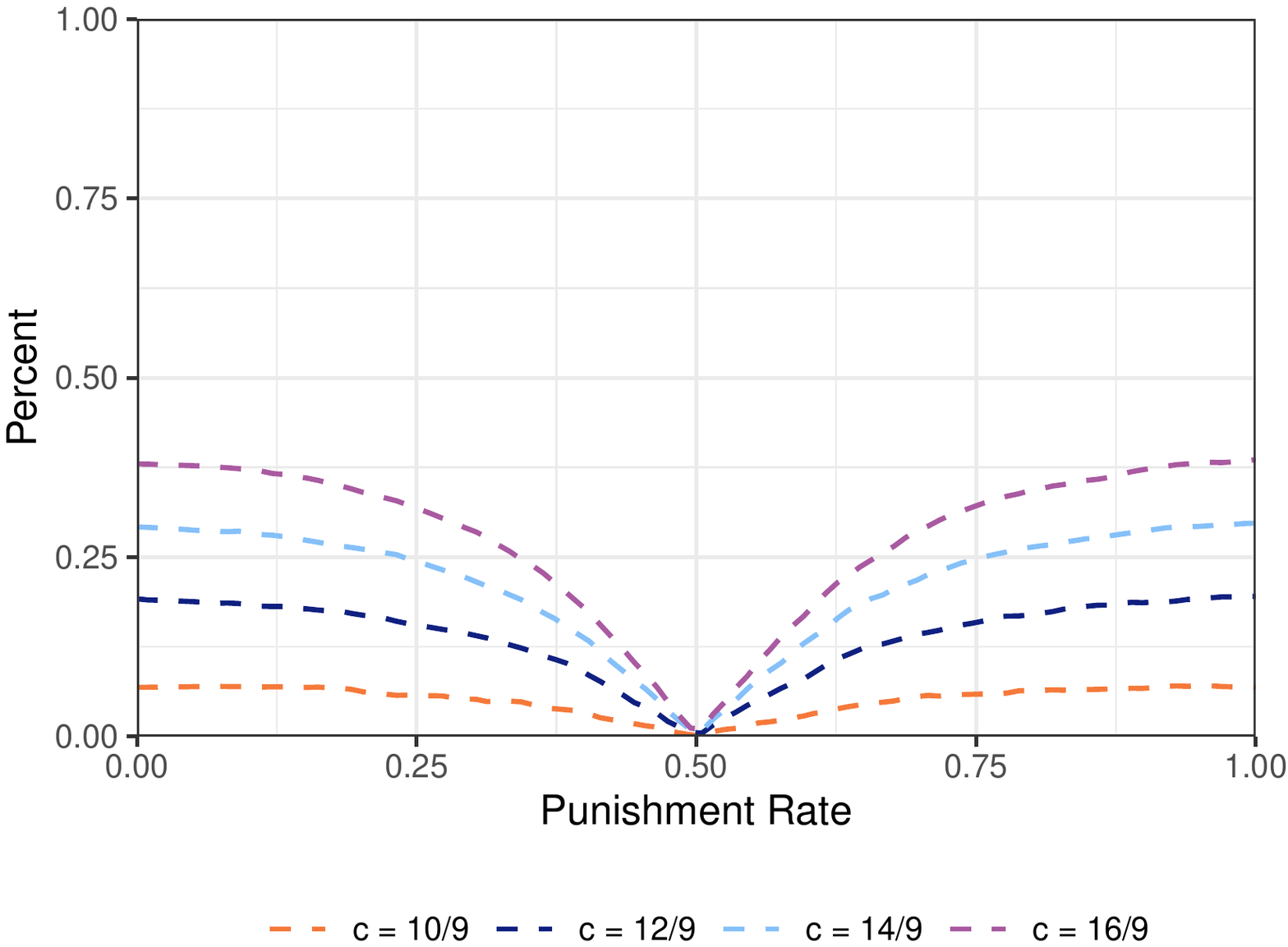}
			\caption{Coverage Rate: $x=1$}
			\label{FigCRX1}
		\end{subfigure}
		\hfill
		\begin{subfigure}[b]{0.3\textwidth}
			\centering
			\includegraphics[width = \textwidth]{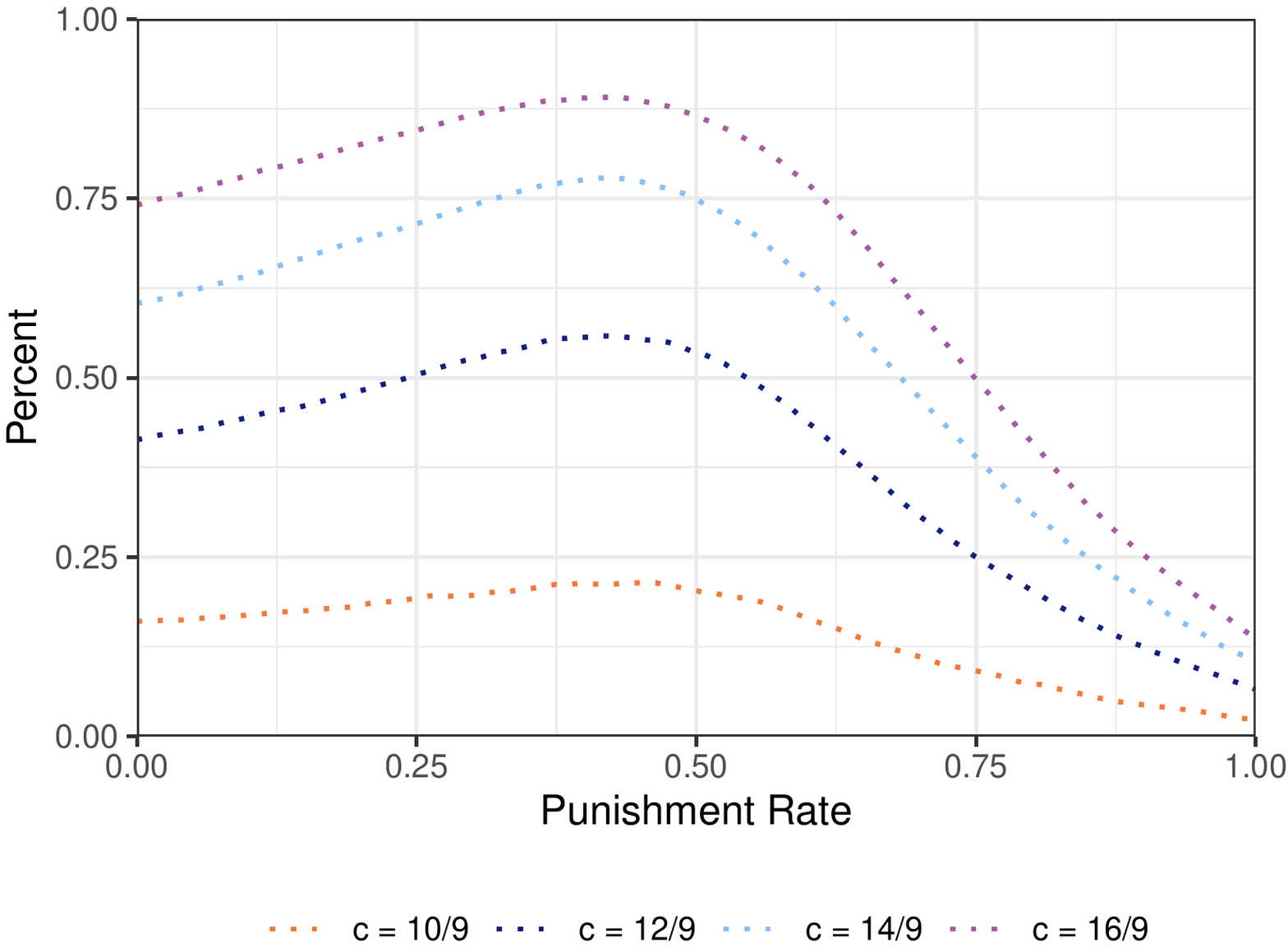}
			\caption{Coverage Rate: $x=2$}
			\label{FigCRX2}
		\end{subfigure}
		\caption[]{Coverage Rate of the Bounds in Proposition \ref{CorOuterMTE}}
		\label{FigCR}
	\end{center}
	\footnotesize{Notes: The solid lines are associated with bounds based on the LIV estimator $\hat{f}\left(\cdot,0\right)$. The dashed lines are associated with bounds based on the LIV estimator $\hat{f}\left(\cdot,1\right)$. The dotted lines are associated with bounds based on the LIV estimator $\hat{f}\left(\cdot,2\right)$. The orange lines are associated with bounds based on Assumption \ref{ASbounded} with c = \sfrac{10}{9}. The dark blue lines are associated with bounds based on Assumption \ref{ASbounded} with c = \sfrac{12}{9}. The light blue lines are associated with bounds based on Assumption \ref{ASbounded} with c = \sfrac{14}{9}. The purple lines are associated with bounds based on Assumption \ref{ASbounded} with c = \sfrac{16}{9}.}
\end{figure}

Using this Monte Carlo exercise, I can also compute the coverage rate of the estimated bounds $\hat{\theta}_{U}\left(\cdot,\cdot\right)$ and $\hat{\theta}_{L}\left(\cdot,\cdot\right)$. Formally, I can estimate $\mathbb{P}\left[\theta\left(z,x\right) \in \left[\hat{\theta}_{L}\left(z,x\right), \hat{\theta}_{U}\left(z,x\right)\right]\right]$ for the data generating process described above and each value $z$ of the instrument and each value $x$ of the covariates. Figure \ref{FigCR} shows the coverage rate for each covariate value $x$, separated in different subfigures and denoted by different line types, and each possible value of $c$, denoted by different line colors.

First, note that, when $c= \sfrac{10}{9}$ (orange line), the coverage rate is small regardless of the covariate value $x$. This result is expected because $\hat{\theta}_{L}\left(z,x\right) \overset{p}{\rightarrow} \theta\left(z,x\right)$ when $c= \sfrac{10}{9}$ and the estimated bounds do not account for sampling uncertainty. If I constructed bootstrapped confidence bands around the estimated bounds, the coverage rate would equal the nominal confidence level.

Moreover, when $c$ increases to $\sfrac{12}{9}$ (dark blue line), $\sfrac{14}{9}$ (light blue line) and $\sfrac{16}{9}$ (purple line), the coverage rate increases monotonically. This result is also expected because the true MTE function $\theta$ is in the interior of $\Theta_{1}$ when $c$ increases, reducing the importance of sampling uncertainty.

Second, note that the coverage rate is smaller on the left side of Subfigure \ref{FigCRX0}, on the middle of Subfigure \ref{FigCRX1} and on the right side of Subfigure \ref{FigCRX2}. This decrease in the coverage rate is caused by an MTE function $\theta\left(z,x\right)$ that is close to zero in those regions. When the true MTE function $\theta$ is close to zero, the LIV estimator $\hat{f}$ will incorrectly estimate the sign of $\theta$ due to sampling uncertainty. As a consequence of this mistake, the estimated bounds will not cover the true MTE function $\theta$ by construction.

In the extreme case when the true MTE function $\theta$ is exactly zero as it happens when $z = 0.5$ and $x = 1$ (Subfigure \ref{FigCRX1}), the coverage rate is exactly zero. This negative result is expected because the LIV estimator $\hat{f}$ is different from zero with probability one due to sampling uncertainty. Consequently, the estimated bounds will not cover zero by construction with probability one.

For a similar reason, the coverage rate is higher away from the left corner of Subfigure \ref{FigCRX0}, away from the center of Subfigure \ref{FigCRX1} and away from the right corner of Subfigure \ref{FigCRX2}. In those regions, the true MTE function $\theta$ is far away from zero and the LIV estimator $\hat{f}$ is more likely to estimate the sign of $\theta$ correctly. Consequently, the estimated bounds have a higher probability of covering the true MTE function when $\theta$ is far away from zero.

Despite this last mechanism, the coverage rate decreases when $z$ is very large in Subfigure \ref{FigCRX0} and when $z$ is very small in Subfigure \ref{FigCRX2}. Even though the true MTE function achieves its largest magnitude in those regions, the coverage rate is slightly smaller there than it is in the center of these two subfigures. The reason for this phenomenon is that the residual variance conditional on $Z$ is larger in those regions, increasing sampling uncertainty and decreasing the coverage rate of the estimated bounds. If I constructed bootstrapped confidence bands around the estimated bounds, this problem would disappear because the higher residual variance would generate wider confidence bands.

\clearpage

\section{Constructing the Dataset}\label{AppDataGeneral}
\setcounter{table}{0}
\renewcommand\thetable{I.\arabic{table}}

\setcounter{figure}{0}
\renewcommand\thefigure{I.\arabic{figure}}

\setcounter{equation}{0}
\renewcommand\theequation{I.\arabic{equation}}

\setcounter{theorem}{0}
\renewcommand\thetheorem{I.\arabic{theorem}}

\setcounter{proposition}{0}
\renewcommand\theproposition{I.\arabic{proposition}}

\setcounter{corollary}{0}
\renewcommand\thecorollary{I.\arabic{corollary}}

\setcounter{assumption}{0}
\renewcommand\theassumption{I.\arabic{assumption}}

\subsection{Detailed Descriptive Statistics}\label{AppDescriptive}

In this appendix, I complement Section \ref{Sdescriptive} with detailed descriptive statistics.

\begin{figure}[!htbp]
	\begin{center}
		\begin{subfigure}[b]{0.47\textwidth}
			\centering
			\includegraphics[width = \textwidth]{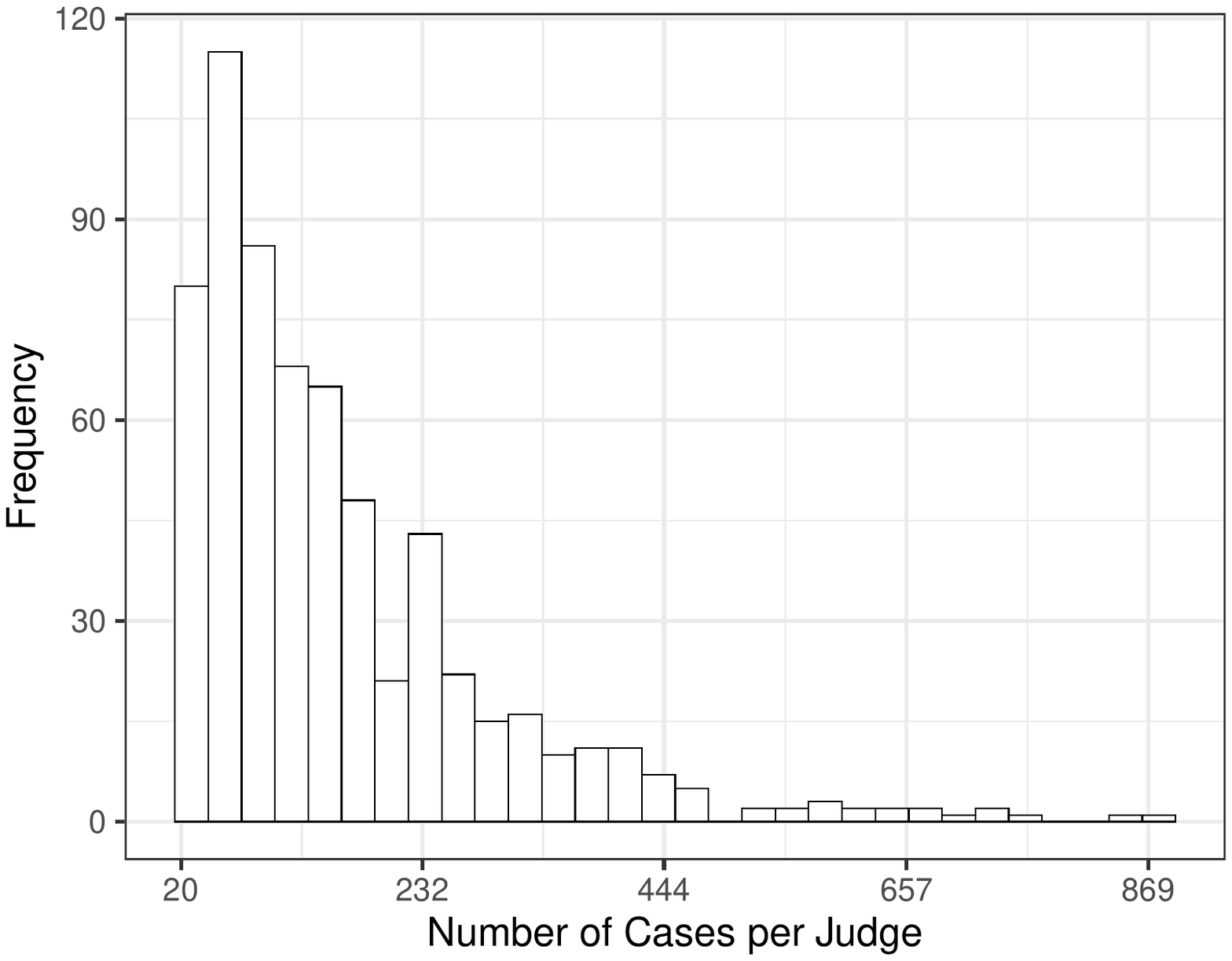}
			\caption{Number of Cases per Judge}
			\label{FigJudgeCases}
		\end{subfigure}
		\hfill
		\begin{subfigure}[b]{0.47\textwidth}
			\centering
			\includegraphics[width = \textwidth]{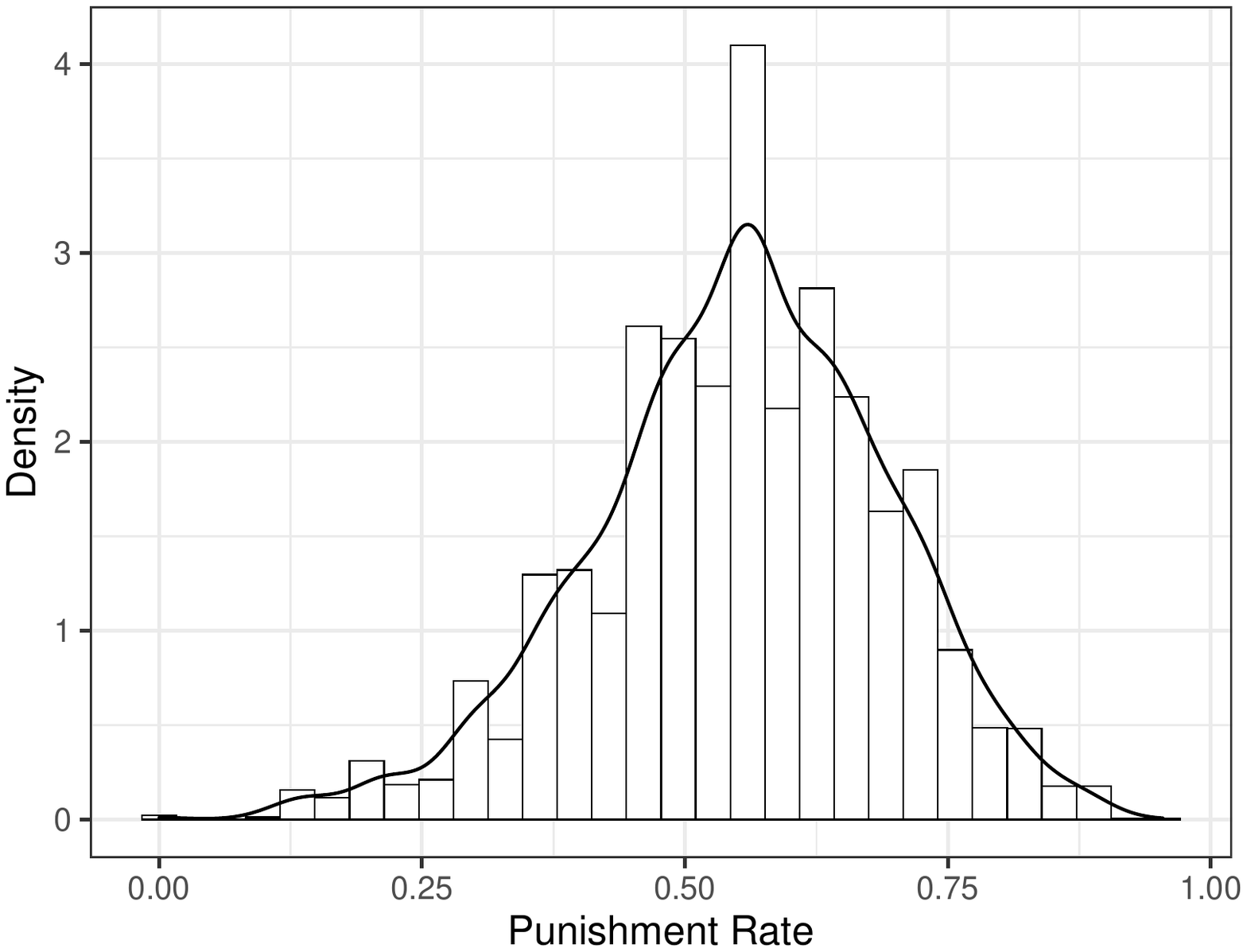}
			\caption{Trial Judge's Leniency Rates}
			\label{FigJudgeLeniency}
		\end{subfigure}
		\caption[]{Trial Judge's Descriptive Statistics}
		\label{FigJudgeDescriptive}
	\end{center}
	\footnotesize{Notes: The density function in Subfigure \ref{FigJudgeLeniency} is estimated using a Gaussian kernel.}
\end{figure}

Subfigure \ref{FigJudgeCases} shows the histogram of the number of case-defendant pairs analyzed by each judge in my sample. Note that it is not uncommon to find judges who analyzed more than 100 cases. For this reason, I treat the instrument $Z$ as the correctly measured leniency of the trial judges, ignoring any estimation error from this stage in my estimation and inference methods.

Subfigure \ref{FigJudgeLeniency} shows the histogram and the smoothed density of the trial judge's leniency rate. Observe that there are judges who are very lenient, punishing less than 25\% of their cases, and judges who are very strict, punishing more than 75\% of their cases. This large dispersion of leniency rates is useful for my empirical strategy because it strengthens my instrument.\footnote{When I semiparametrically estimate the propensity score functions (Appendix \ref{AppPSsemi}), I find that the correctly measured propensity score and the mismeasured propensity score are nontrivial functions of the instrument. Their steep inclines also suggest that my instrument is relevant.}

Table \ref{TabOutcome} shows the recidivism rate by punishment group. Overall, I find that 33.6\% of the case-defendant pairs in my sample recidivated within two years of the final sentence in their cases. Conditioning on treatment status, approximately 36\% of the punished defendants and 31\% of the non-punished defendants recidivate. Consequently, the comparison between these two groups suggests that receiving an alternative sentence increases the probability of committing crimes in the future if I ignore endogenous selection-into-treatment. As discussed in Subsection \ref{Sresults}, this naive conclusion is not supported by an analysis that takes endogeneity into account.

\begin{table}[!htb]
	\centering
	\caption{{Recidivism Rate by Punishment Group}} \label{TabOutcome}
	\begin{lrbox}{\tablebox}
		\begin{tabular}{ccclcc}
			\hline \hline
			\multirow{2}{*}{Overall} & \multicolumn{2}{c}{Trial Judge's Ruling} &  & \multicolumn{2}{c}{Final Ruling} \\ \cline{2-3} \cline{5-6}
			& Punished & Not Punished &  & Punished & Not Punished \\ \hline
			33.6\% & 36.2\% & 30.6\% &  & 36.1\% & 31.0\% \\ \hline
		\end{tabular}
	\end{lrbox}
	\usebox{\tablebox}\\
	\settowidth{\tableboxwidth}{\usebox{\tablebox}} \parbox{\tableboxwidth}{\footnotesize{Note: Recidivism is defined using a fuzzy matching algorithm based on the Jaro–Winkler similarity metric.}
	}
\end{table}

\begin{figure}[!htbp]
	\begin{center}
		\includegraphics[width = 0.47\textwidth]{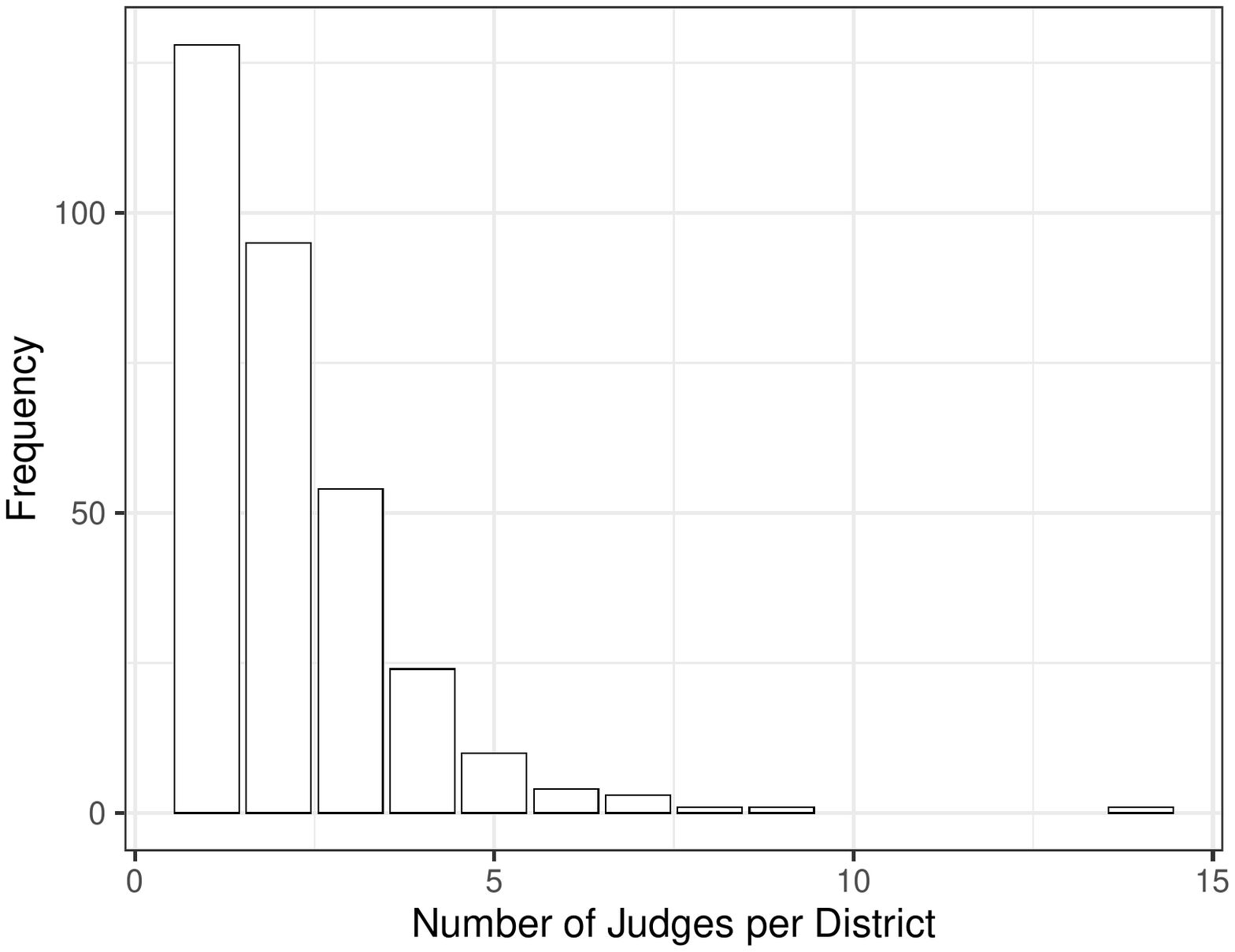}
		\caption{Number of Judges per Court District}
		\label{FigDistrict}
	\end{center}
\end{figure}

Figure \ref{FigDistrict} shows the histogram of the number of judges per court district. Note that most districts have at least two judges during my sampling period even though the modal district has only one judge. Since my identification strategy leverages the random allocation of judges to criminal cases, I only use the 193 districts with two or more judges.

Moreover, I can also find the defendant's sex based on the defendant's first name. This extra covariate is not used in my main analysis. Still, it can be used to check whether the trial judge's leniency rate is correlated with the defendant's sex given the court district. To do so, I use the \emph{R package} \texttt{genderBR} to define which names are typically male (37,304 defendants), female (4,936) or unisex (1,228) based on the Brazilian 2010 Census. Using only names that are typically male or female, I regress a sex dummy on the trial judge's leniency rate controlling for court district fixed effects and find a coefficient equal to 0.02 with a standard error equal to 0.03 when I cluster it at the court district level. This result indirectly suggests that trial judges are randomly allocated to criminal cases as mandated by law.

\subsection{Constructing the Dataset}\label{AppData}
In this appendix, I provide a detailed explanation of how I constructed the dataset used in my empirical application. I will explain the specific crime types included in my sample, the classification algorithms used to define which defendants were punished, and the fuzzy matching algorithm used to define which defendants recidivate.

The final dataset was created from four initial datasets.
\begin{enumerate}
	\item CPOPG (``Consulta de Processos de Primeiro Grau''): It contains information about all criminal cases in the Justice Court System in the State of São Paulo (TJ-SP) between 2010 and 2019. Its variables are described below.
	\begin{enumerate}
		\item \texttt{id}: An unique case identifier that can link cases across all datasets from TJ-SP.

		\item \texttt{status}: The case's status defines whether the trial judge has achieved her final decision in the trial or not, i.e., whether the case is open or not.

		\item \texttt{subject}: It denotes each case's crime type.

		\item \texttt{class}: The case's class defines whether the case's objective is to analyze whether a defendant is guilty or not. For example, some criminal cases aim to start a police investigation or to arrest a person before the trial judge's sentence.

		\item \texttt{assignment}: This variable contains information about the case's starting date and whether the case was randomly assigned to a judge within the case's court district or whether the case was assigned to a judge that was already analyzing a connected case.

		\item \texttt{trialjudge}: This variable contains the trial judge's full name.

		\item \texttt{parties}: This variable contains a list with the names of all the parties involved in the case, including defendants, prosecutors, defense attorneys and public defenders.
	\end{enumerate}

	\item CJPG (``Consulta de Julgados de Primeiro Grau''): It contains information about the last decision made by a trial judge in all criminal cases in TJ-SP between 2010 and 2019. Its variables are described below.
	\begin{enumerate}
		\item \texttt{id}: An unique case identifier that can link cases across all datasets from TJ-SP.

		\item \texttt{date}: It denotes the date of the last decision made by the case's trial judge.

		\item \texttt{courtdistrict}: It provide the court district's name.

		\item \texttt{sentence}: It provides the full text of the trial judge's final decision.
	\end{enumerate}

	\item CPOSG (``Consulta de Processos de Segundo Grau''): It contains information about all appealing criminal cases in TJ-SP between 2010 and 2019. Its variables are described below.
	\begin{enumerate}
		\item \texttt{id}: An unique case identifier that can link cases across all datasets from TJ-SP.

		\item \texttt{parties}: This variable contains a list with the names of all the parties involved in the case, including defendants, prosecutors, defense attorneys and public defenders.

		\item \texttt{composition}: It contains the name of the three Appeals judges who analyzed the appealing case, including their positions within the case (judge-rapporteur, revising judge, voting judge).

		\item \texttt{decision}: It contains the Appeals Court's final ruling.

		\item \texttt{date}: It denotes the date of the last decision made by the Appeals Court.
	\end{enumerate}

	\item Public Defenders' names: This dataset is a list with full names of all public defenders in the State of São Paulo from 2011 to 2019. It was constructed directly by the Public Defender's Office after a FOIA request. The Brazilian FOIA is known as ``Lei de Acesso à Informação'' and is regulated by Law n. 12.527/11.
\end{enumerate}

Starting from the CPOPG dataset, I implement the following steps.
\begin{enumerate}
	\item I only keep  cases that are currently in the Appeals Court (\texttt{status} equal to ``Em grau de recurso'' or ``Em grau de recurso $\vert$ (Tramitação prioritária)''), closed (\texttt{status} equal to ``Extinto'', ``Extinto $\vert$ (Tramitação prioritária)'' or ``Arquivado'') or whose status is empty. Those cases are already associated with a trial judge's sentence.

	\item I only keep cases whose crime types are associated with sentences that must be less than four years of incarceration. In particular, I keep cases whose \texttt{subject} is equal to ``Atentado Violento ao Pudor'' (sexual assault), ``Decorrente de Violência Doméstica'' (domestic violence), ``Violência Doméstica Contra a Mulher'' (domestic violence against a woman), ``Contravenções Penais'' (misdemeanors), ``Furto'' (theft), ``Furto (art. 155)'' (theft), ``Furto Privilegiado'' (qualified theft), ``Furto de coisa comum'' (theft of a common good), ``Desacato'' (contempt), ``Receptação'' (receiving stolen goods), ``Ameaça'' (threat), ``Violação de direito autoral'' (copyright violation), ``Crimes contra a Propriedade Intelectual'' (crimes against intellectual property), ``Posse de Drogas para Consumo Pessoal'' (drug consumption), ``Apropriação indébita'' (undue appropriation), ``Apropriação indébita (art. 168, caput)'' (undue appropriation), ``Quadrilha ou Bando'' (criminal conspiracy), ``Desobediência'' (disobedience), ``Resistência'' (resistence), ``Fato Atípico'' (atypical fact), ``Crimes de Abuso de Autoridade'' (abuse of authority), ``Crime Culposo'' (crime without criminal intent), ``Dano Qualificado'' (qualified harm), ``Violação de domicílio'' (trespassing), ``Favorecimento real'' (illegal favoring), ``Comunicação falsa de crime ou de contravenção'' (false criminal communication), ``Destruição / Subtração / Ocultação de Cadáver'' (destruction, subtraction or concealment of a corpse), ``Difamação'' (libel) and ``Injúria'' (insult).

	\item I only keep cases that aim to analyze whether a defendant is guilty or not. In particular, I drop all the cases whose \texttt{class} is equal to ``Execução da Pena'' (sentence execution), ``Habeas Corpus Criminal'' (habeas corpus), ``Execução Provisória'' (temporary sentence execution), ``Inquérito Policial'' (police investigation), ``Procedimento Investigatório Criminal (PIC-MP)'' (police investigation), ``Auto de Prisão em Flagrante'' (pre-trial arrest), ``Pedido de Prisão Preventiva'' (pre-trial arrest), ``Medidas Protetivas de urgência (Lei Maria da Penha) Criminal'' (urgent protective acts),``Relatório de Investigações'' (police report), ``Ação Penal de Competência do Júri'' (jury action), ``Mandado de Segurança Criminal'' (judicial mandate), ``Pedido de Busca e Apreensão Criminal'' (judicial mandate), ``Representação Criminal/Notícia de Crime'' (crime notification) and ``Termo Circunstanciado'' (report).

	\item I only keep cases that were randomly assigned to trial judges. In particular, I keep cases whose \texttt{assignment} contain the word ``Livre''.

	\item I only keep cases whose starting date is after January 1\textsuperscript{st}, 2010.
\end{enumerate}

After these steps, my dataset contains 98,552 cases. I, then, merged it with the CJPG dataset using cases' \texttt{id} codes. Since some cases do not have \texttt{id} codes, my dataset now contains 98,422 cases.

After this step, I randomly select 35 cases per year (2010-2019) for manual classification. I manually classify them into five categories: ``defendant died during the trial'', ``defendant is guilty'', ``defendant accepted a non-prosecution agreement'' (``transação penal'' in Portuguese), ``case was dismissed'' (``processo suspenso'' in Portuguese) and ``defendant was acquitted''. Since some sentences are missing or incomplete, I am able to manually classify only 325 sentences.

Now, I use those 325 manually classified cases to train a classification algorithm. To do so, I divide them into a training sample (216 cases) and a validation sample (109 sentences).

First, I design an algorithm to identify which defendants died during the trial. To do so, I check whether the sentence contains any reference to the first paragraph of Article 107 from the Brazilian Criminal Code. This specific part of the Brazilian Criminal Code states that a dead defendant cannot be punished in any way. In both my samples, I find that this deterministic algorithm perfectly classifies cases into the category ``defendant died during the trial''.

Second, I design an algorithm to identify which cases were dismissed. To do so, I check whether the sentence contains any reference to Article 89 in Law n. 9099/95. This specific law article defines the criteria for dismissing a case. In both my samples, I find that this deterministic algorithm correctly classifies 98\% of the cases into the category ``case was dismissed''. The few cases that were misclassified are cases that were initially dismissed but reopened because the defendant committed a second crime.

Third, I design an algorithm to identify which defendants accepted a non-prosecution agreement. To do so, I check whether the sentence contains any of the following expressions: ``cumprimento da transação penal'' (non-prosecution agreement was fulfilled), ``Homologo a proposta'' (I accept the proposition), ``homologo a transação'' (I accept the non-prosecution agreement), ``HOMOLOGO O ACORDO'' (I accept the agreement), ``proposta transação'' (A non-prosecution agreement was proposed), ``transação penal'' (non-prosecution agreement), ``Acolho a proposta'' (I accept the proposal) and ``aceitação da proposta'' (proposal acceptance). Those expressions were selected because, when manually classifying the sentences, I noticed that they were strong signals of a defendant who accepted a non-prosecution agreement. In both my samples, I find that this deterministic algorithm correctly classified almost all the cases into the category ``defendant accepted a non-prosecution agreement'', making only three mistakes. In the misclassified sentences, the judge mentioned that the prosecutor proposed a non-prosecution agreement, but the defendant missed the agreement's session.

Finally, I design an algorithm to classify the remaining cases into two categories: ``defendant is guilty'' and ``defendant was acquitted''. To do so, I define a bag of words that were found to be strong signals of acquittal and guilt when I manually classified the cases in my samples. This bag of words contains the following expressions: ``absolv'' (all words related to acquittal contain this expression in Portuguese), ``art. 107, inciso IV'' (the fourth paragraph of Article 107 from the Brazilian Criminal Code defines that a defendant cannot be punished if he or she is not guilty) and related expressions, ``extinta a punibilidade'' (it means that the defendant cannot be punished) and related expressions, ``improcedente'' (unfounded), ``prescrição'' (statute of limitations), ``conden'' (all words related to punishment contain this expression in Portuguese), ``pena'' (sentence), ``procedente'' (well-founded), ``cumprimento da pena'' (sentence is fulfilled) and related expressions, ``dosimetria'' (dosimetry of the penalties) and related expressions, and ``rol dos culpados'' (book of the guilty). I, then, count how many times each one of those expressions appear in each sentence and I normalize those counts to be between 0 and 1.

Using the normalized counts, I train six algorithms using my training sample: k-Nearest Neighbors, Random Forest, L2-Regularized Logistic Regression, L1-Regularized Logistic Regression, Naive Bayes and xgboost. I, then, validate those algorithms using my validation sample and find that the k-Nearest Neighbors algorithm correctly classifies 95.3\% of the cases, the Random Forest algorithm correctly classifies 96.5\% of the cases, the L2-Regularized Logistic Regression algorithm correctly classifies 97.7\% of the cases, the L1-Regularized Logistic Regression algorithm correctly classifies 98.8\% of the cases, the Naive Bayes algorithm correctly classifies 84.9\% of the cases and the xgboost algorithm correctly classifies 91.9\% of the cases. Given these success rates, I use the L1-Regularized Logistic Regression algorithm to define the treatment variable in my full sample.

Having designed the above algorithms, I use them to define the misclassified treatment variable $T$ in the full sample. First, I find which defendants died during their trials and drop them from my sample. I, then, use the second and third algorithms to define which cases were dismissed and which cases are associated with a non-prosecution agreement. Moreover, I use the trained L1-regularized Logistic Regression algorithm to classify the remaining cases into the categories ``defendant is guilty'' and ``defendant was acquitted''. Finally, I combine the categories ``defendant was acquitted'' and ``case was dismissed'' into the untreated group (``not punished'', $T = 0$) and the categories ``defendant accepted a non-prosecution agreement'' and ``defendant is guilty'' into the treated group (``punished'', $T = 1$). At the end, my dataset contains 96,225 cases.

Now, I merge my current dataset with the CPOSG dataset using each case's \texttt{id} code. When merging these datasets, I create an indicator variable that denotes which cases went to the Appeals Court, i.e., which cases were matched. I, then,  randomly select 50 cases per year for manual classification (2010-2019) and divide them into three categories: ``cases that went to the Appeals Court, but were immediately returned due to bureaucratic errors'', ``cases whose trial judge's sentences were affirmed'' and ``cases whose trial judge's sentences were reversed''.

Now, I use those 500 manually classified cases to train a classification algorithm. To do so, I divide them into a training sample (300 cases) and a validation sample (200 sentences).

First, I design an algorithm to identify which cases went to the Appeals Court but were immediately returned. To do so, I simply check whether Appeals Court's \texttt{decision} is empty.

Finally, I design an algorithm to classify the non-empty cases into two categories: ``cases whose trial judge's sentences were affirmed'' and ``cases whose trial judge's sentences were reversed''. To do so, I define a bag of words that were found to be strong signals of sentence reversal when I manually classified the cases in my sample. This bag of words contains the following expressions: ``absolv'' (all words related to acquittal contain this expression in Portuguese), ``art. 107, inciso IV'' (the fourth paragraph of Article 107 from the Brazilian Criminal Code defines that a defendant cannot be punished if he or she is not guilty) and related expressions, ``extinta a punibilidade'' (it means that the defendant cannot be punished) and related expressions, ``prescrição'' (statute of limitations), ``negaram provimento'' (it means that the Appeals Court affirmed the trial judge's sentence) and related expressions, ``deram provimento'' (it means that the Appeals Court reversed the trial judge's sentence) and related expressions, and ``parcial provimento'' (it means that the Appeals Court reduced the penalty established by the trial judge) and related words. I, then, count how many times each one of those expressions appear in each sentence and I normalize those counts to be between 0 and 1.

Using the normalized counts, I train six algorithms using my training sample: k-Nearest Neighbors, Random Forest, L2-Regularized Logistic Regression, L1-Regularized Logistic Regression, Naive Bayes and xgboost. I, then, validate those algorithm using my validation sample and find that the k-Nearest Neighbors algorithm correctly classifies 97.8\% of the cases, the Random Forest algorithm correctly classifies 97.8\% of the cases, the L2-Regularized Logistic Regression algorithm correctly classifies 97.2\% of the cases, the L1-Regularized Logistic Regression algorithm correctly classifies 96.2\% of the cases, the Naive Bayes algorithm correctly classifies 95.6\% of the cases and the xgboost algorithm correctly classifies 96.1\% of the cases. Given the success rates in this dataset and the dataset that focuses on trial judges' sentences, I use the L1-Regularized Logistic Regression to define the treatment variable in my full sample.

Having designed the above algorithms, I use them to define the correctly classified treatment variable $D$ in the full sample. First, I set $D = T$ if a case did not go to the Appeals Court or if a case went to the Appeals Court, but was immediately returned. Second, I use the trained L1-Regularized Logistic Regression algorithm to classify the remaining cases into the categories ``reversed trial judge's sentence'' and ``affirmed trial judge's sentence''. I, then, set $D = T$ if the trial judge's sentence was affirmed and $D = 1 - T$ if the trial judge's sentence was reversed. Moreover, I also drop the cases whose dates (starting date, trial judge's sentence date and Appeal Court's decision date) are not appropriately ordered. At the end, my dataset contains 95,119 cases.

Now, my goal is to find the defendants' names in each case. To do so, I use the variable \texttt{parties} from the CPOPG dataset and search for names listed as ``réu'', ``ré'', ``indiciado'', ``indiciada'', ``denunciado'', ``denunciada'', ``coré'', ``coréu'', ``investigado'', ``infrator'', ``acusado'', ``autordofato'', ``autoradofato'', ``averiguada'', ``averiguado'', ``infrator'', ``querelado'', ``querelada'', ``representado'', ``reqdo'' and ``reqda''.\footnote{All these expression are associated with the word ``defendant'' in Portuguese.} Moreover, I use the variable \texttt{parties} from the CPOSG dataset and search for names listed as ``apelante'', ``recorrente'', ``requerente'', ``apelado'', ``corréu'', ``recorrido'', ``apelada'', ``réu'', ``corré'' and ``querelado''.\footnote{All these expression are associated with the word ``defendant'' or ``appealing party'' in Portuguese.} Furthermore, I analyze the full sentences from the CJPG dataset and search for names listed as  ``Réu:'', ``Ré:'', ``RÉU:'', ``RÉ:'', ``Réu'', ``Ré'', ``Autor do Fato:'', ``Autora do Fato:'', ``Autor(a) do Fato:'', ``Indiciado:'', ``Indiciada:'',  ``Sentenciado:'', ``Sentenciada:'', ``Sentenciado(a):'', ``Querelado:'', ``Querelada:'', ``Averiguado:'', ``Averiguada:'',  ``Sujeito Passivo:'', ``Denunciada:'', ``Denunciado:'', ``Requerido:'' and ``Requerida:''.\footnote{All these expressions are associated with the word ``defendant'' in Portuguese.} Finally, I delete names that are not a person's name --- such as district attorney offices, public defender offices and ``unknown author'' --- or names that are listed in the Public Defenders dataset. My sample now contains 103,423 case-defendant pairs.

Furthermore, I repeat the steps in the last paragraph to find defendants' names in a dataset that contains all cases from the CPOPG dataset, including cases that are still open and cases with severe crimes. To build this dataset, I followed the steps described above but did not subset my sample based on the variables \texttt{status} and \texttt{subject}. Moreover, when subsetting my sample based on the variable \texttt{class}, I only dropped the cases whose \texttt{class} was equal to ``Execução da Pena'', ``Habeas Corpus Criminal'', ``Execução Provisória'', ``Pedido de Busca e Apreensão Criminal'' and ``Termo Circunstanciado''. At the end, this dataset contains 1,027,120 case-defendants pairs.

Now, I use these two datasets to define my outcome variable ($Y = $``recidivism within 2 years of the final sentence''). A defendant $i$ in a case $j$ in the smaller dataset recidivated ($Y_{ij} = 1$) if and only if defendant $i$'s full name appears in a case $\bar{j}$ in the larger dataset and if case $\bar{j}$'s starting date is within 2 years after case $j$'s final sentence's date. To match defendants' names across cases, I use the Jaro–Winkler similarity metric and I define a match if the similarity between full names in two different cases is greater than or equal to 0.95.

At the end, I delete the case-defendant pairs whose cases started in 2018 and 2019 because their outcome variable is not properly defined due to right-censoring. Consequently, my dataset contains 51,731 case-defendants pairs. This dataset is used in my empirical analysis as described in Subsection \ref{Sdescriptive}.

\clearpage

\section{Additional Empirical Results}\label{AppAddEmpirical}
\setcounter{table}{0}
\renewcommand\thetable{J.\arabic{table}}

\setcounter{figure}{0}
\renewcommand\thefigure{J.\arabic{figure}}

\setcounter{equation}{0}
\renewcommand\theequation{J.\arabic{equation}}

\setcounter{theorem}{0}
\renewcommand\thetheorem{J.\arabic{theorem}}

\setcounter{proposition}{0}
\renewcommand\theproposition{J.\arabic{proposition}}

\setcounter{corollary}{0}
\renewcommand\thecorollary{J.\arabic{corollary}}

\setcounter{assumption}{0}
\renewcommand\theassumption{J.\arabic{assumption}}

\subsection{Tables}\label{AppTables}

\begin{table}[h]
	\centering
	\caption{{Joint Distribution: Trial Judge's Ruling v. Final Ruling}} \label{TabMisclassification}
	\begin{lrbox}{\tablebox}
		\begin{tabular}{lccc}
			\hline \hline
			&  & \multicolumn{2}{c}{Final Ruling} \\ \cline{3-4}
			&  & Not Punished & Punished \\ \hline
			\multirow{2}{*}{Trial Judge's Ruling} & \multicolumn{1}{l}{Not Punished} & 45.8\% & 0.9\% \\
			& \multicolumn{1}{l}{Punished} & 3.5\% & 49.8\% \\ \hline
		\end{tabular}
	\end{lrbox}
	\usebox{\tablebox}\\
	\settowidth{\tableboxwidth}{\usebox{\tablebox}} \parbox{\tableboxwidth}{\footnotesize{Note: Most cases (67.3\%) do not go the Appeals Courts. In those cases, the trial judge's ruling and the final ruling must be equal by construction.}
	}
\end{table} \clearpage

\begin{table}[p]
	\centering
	\caption{{First Stage Results}} \label{TabFirstStage}
	\begin{lrbox}{\tablebox}
		\begin{tabular}{llllll}
			\hline \hline
			& \multicolumn{2}{c}{Final Ruling} & \multicolumn{1}{c}{} & \multicolumn{2}{c}{Trial Judge's Ruling} \\ \cline{2-3} \cline{5-6}
			& \multicolumn{1}{c}{(1)} & \multicolumn{1}{c}{(2)} & \multicolumn{1}{c}{} & \multicolumn{1}{c}{(3)} & \multicolumn{1}{c}{(4)} \\ \hline
			Linear & 0.765*** & 0.688*** &  & 0.777*** & 0.848*** \\
			& (0.027) & (0.132) &  & (0.026) & (0.131) \\
			\\
			Quadratic &  & 0.070 & & & -0.065  \\
			&  & (0.117) &   & & (0.116)  \\
			 \hline
			District FE & \multicolumn{1}{c}{$\checkmark$} & \multicolumn{1}{c}{$\checkmark$} & & \multicolumn{1}{c}{$\checkmark$} & \multicolumn{1}{c}{$\checkmark$} \\
			F-statistic & \multicolumn{1}{c}{827} & \multicolumn{1}{c}{827} &  & \multicolumn{1}{c}{868} & \multicolumn{1}{c}{868} \\
			Sample Size & \multicolumn{2}{c}{43,468} & & \multicolumn{2}{c}{43,461} \\ \hline
		\end{tabular}
	\end{lrbox}
	\usebox{\tablebox}\\
	\settowidth{\tableboxwidth}{\usebox{\tablebox}} \parbox{\tableboxwidth}{\footnotesize{Note: In Columns (1) and (2), I regress the final punishment in each case on a polynomial series of the trial judge's leave-one-out punishment rate and court district fixed effects. In Columns (3) and (4), I regress the trial judge's ruling in each case on the trial judge's leave-one-out punishment rate  and court district fixed effects. The F-statistic is the test statistic of a $\chi^{2}$-test whose null hypothesis is that all the coefficients in the associated column are equal to zero.  Heteroskedasticity-robust standard errors (HC3) are reported in parenthesis. Significance levels are indicated by *$p \leq 0.10$, **$p \leq 0.05$ and ***$p \leq 0.01$.}
	}
\end{table} \clearpage

	\begin{table}
		\centering
		\caption{LIV Estimand's Bias --- Summary Statistics} \label{TabBias}
		\begin{lrbox}{\tablebox}
			\begin{tabular}{lrcrrrrcrr}
				\hline \hline
				& \multicolumn{4}{c}{Panel A: Raw Estimates} & \multicolumn{1}{c}{} & \multicolumn{4}{c}{Panel B: Trimmed Estimates} \\ \cline{2-5} \cline{7-10}
				& \multicolumn{1}{c}{Mean} & \multicolumn{1}{c}{Std. Dev.} & \multicolumn{1}{c}{Min.} & \multicolumn{1}{c}{Max} & \multicolumn{1}{c}{} & \multicolumn{1}{c}{Mean} & \multicolumn{1}{c}{Std. Dev.} & \multicolumn{1}{c}{Min.} & \multicolumn{1}{c}{Max} \\ \hline
				$Z = .3$ & 2.98 & 24.94 & -51.47 & 309.08 &  & 0.00 & 0.02 & -0.11 & 0.09 \\
				$Z = .4$ & 1.40 & 12.56 & -40.98 & 155.31 &  & 0.00 & 0.02 & -0.07 & 0.07 \\
				$Z = .5$ & 0.39 & 4.35 & -24.66 & 51.61 &  & 0.00 & 0.01 & -0.03 & 0.03 \\
				$Z = .6$ & -0.01 & 0.26 & -2.46 & 2.52 &  & 0.00 & 0.00 & -0.01 & 0.01 \\
				$Z = .7$ & 0.13 & 3.39 & -10.35 & 40.51 &  & 0.00 & 0.01 & -0.05 & 0.04 \\
				\hline
			\end{tabular}
		\end{lrbox}
		\usebox{\tablebox}\\
		\settowidth{\tableboxwidth}{\usebox{\tablebox}} \parbox{\tableboxwidth}{\footnotesize{Note: Panel A reports summary statistics for the raw difference between the estimated LIV estimand and the estimated MTE function for five values of the instrument $\left(z \in \left\lbrace .3, .4, .5, .6, .7 \right\rbrace\right)$ across 192 court districts. Panel B reports the same summary statistics, but, before taking the difference between the estimated LIV estimand and the estimated MTE function, it trims these estimates to be between the minimum possible treatment effect (-1) and the maximum possible treatment effect (1).}
		}
	\end{table} \clearpage

\subsection{Estimating the Sign of the MTE function}\label{AppSign}
\begin{figure}[!htb]
	\begin{center}
		\includegraphics[width = 0.47\textwidth]{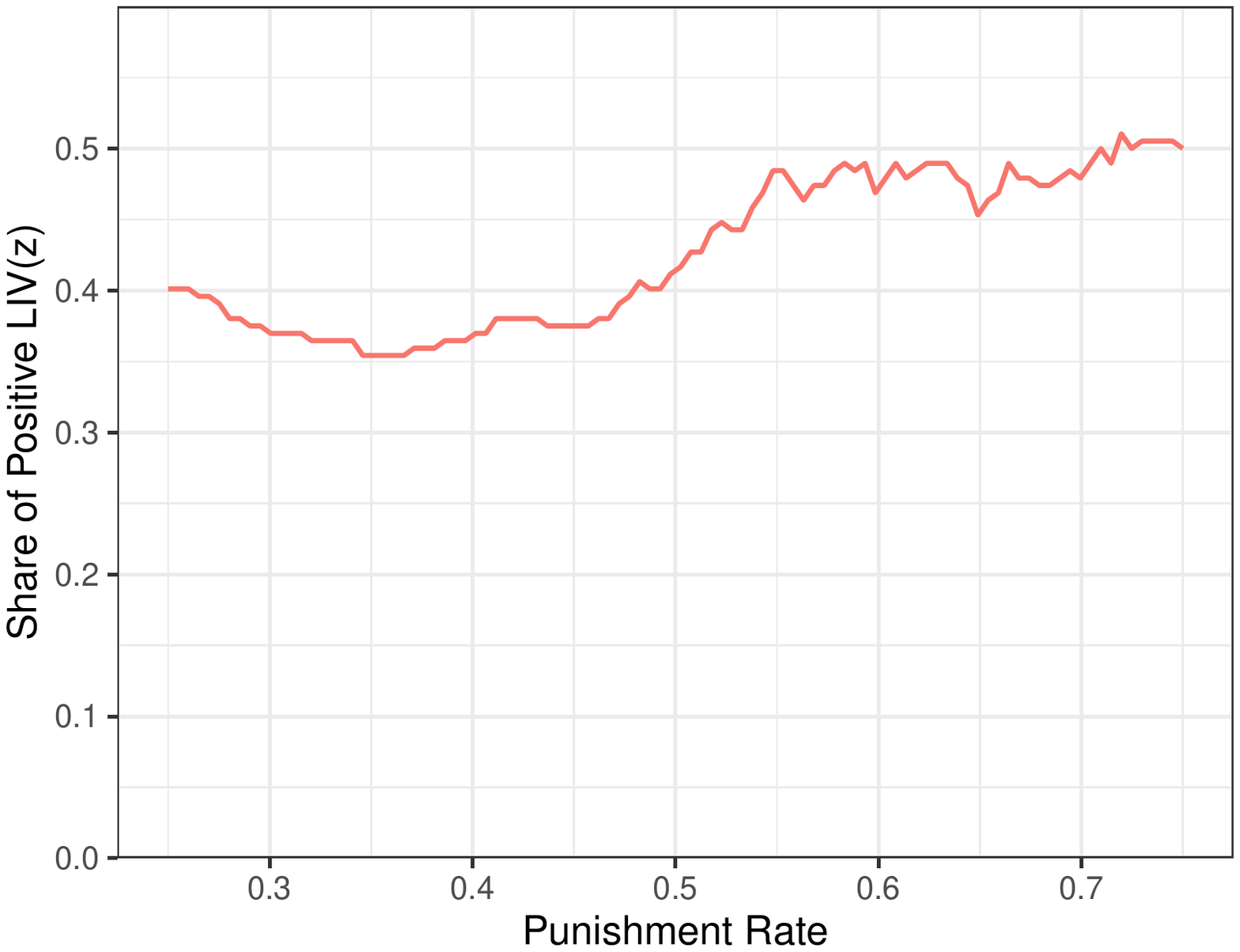}
		\caption{Sign of the MTE function $\theta\left(\cdot, \cdot \right)$ --- Corollary \ref{CorSign}}
		\label{FigMTEsign}
	\end{center}
	\footnotesize{Notes: The orange line reports the share of points $\theta\left(z, \cdot \right)$ that are estimated to be positive (Corollary \ref{CorSign} and Equation \eqref{EQestsign}) across court districts for each value of the instrument.}
\end{figure}

In this appendix, I discuss the estimated signs of the MTE function (Corollary \ref{CorSign} and Equation \eqref{EQestsign}). Figure \ref{FigMTEsign} plots the share of points $\theta\left(z, \cdot \right)$ that are estimated to be positive across court districts for each value of the instrument. For example, for a punishment rate equal to 0.5, approximately 40\% of the court districts are estimated to have a positive MTE function.

Figure \ref{FigMTEsign} implies that alternative sentences may increase or decrease recidivism depending on cities' contexts. This result illustrates the importance of accounting for geographic heterogeneity when discussing the effect of alternative sentences on recidivism.

Moreover, I can also estimate the sign of the true MTE function (Appendix \ref{Sestcorrectly}) because I observe the correctly classified treatment variable $D$ (``each case's final ruling) in my dataset. When I compare the estimated sign of the MTE function against the sign of the correctly estimated MTE function, I find that my method reaches the right conclusion for all evaluated points.

\subsection{Taking Sample Uncertainty into Account}\label{AppCI}

\begin{figure}[!htbp]
	\begin{center}
		\begin{subfigure}[b]{0.47\textwidth}
			\centering
			\includegraphics[width = \textwidth]{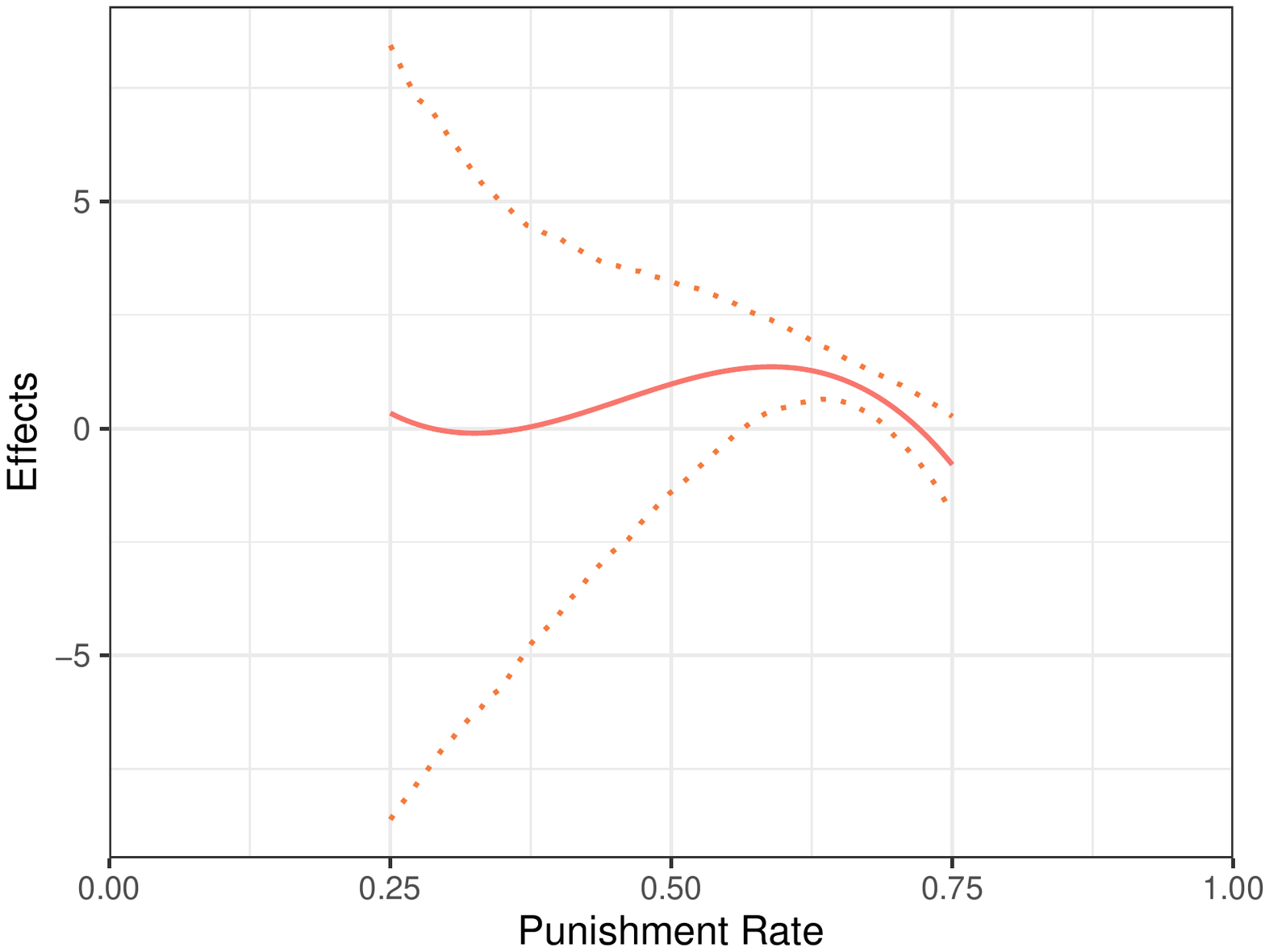}
			\caption{Confidence Bands around the MTE function $\theta\left(\cdot, \cdot \right)$}
			\label{FigMTECI}
		\end{subfigure}
		\hfill
		\begin{subfigure}[b]{0.47\textwidth}
			\centering
			\includegraphics[width = \textwidth]{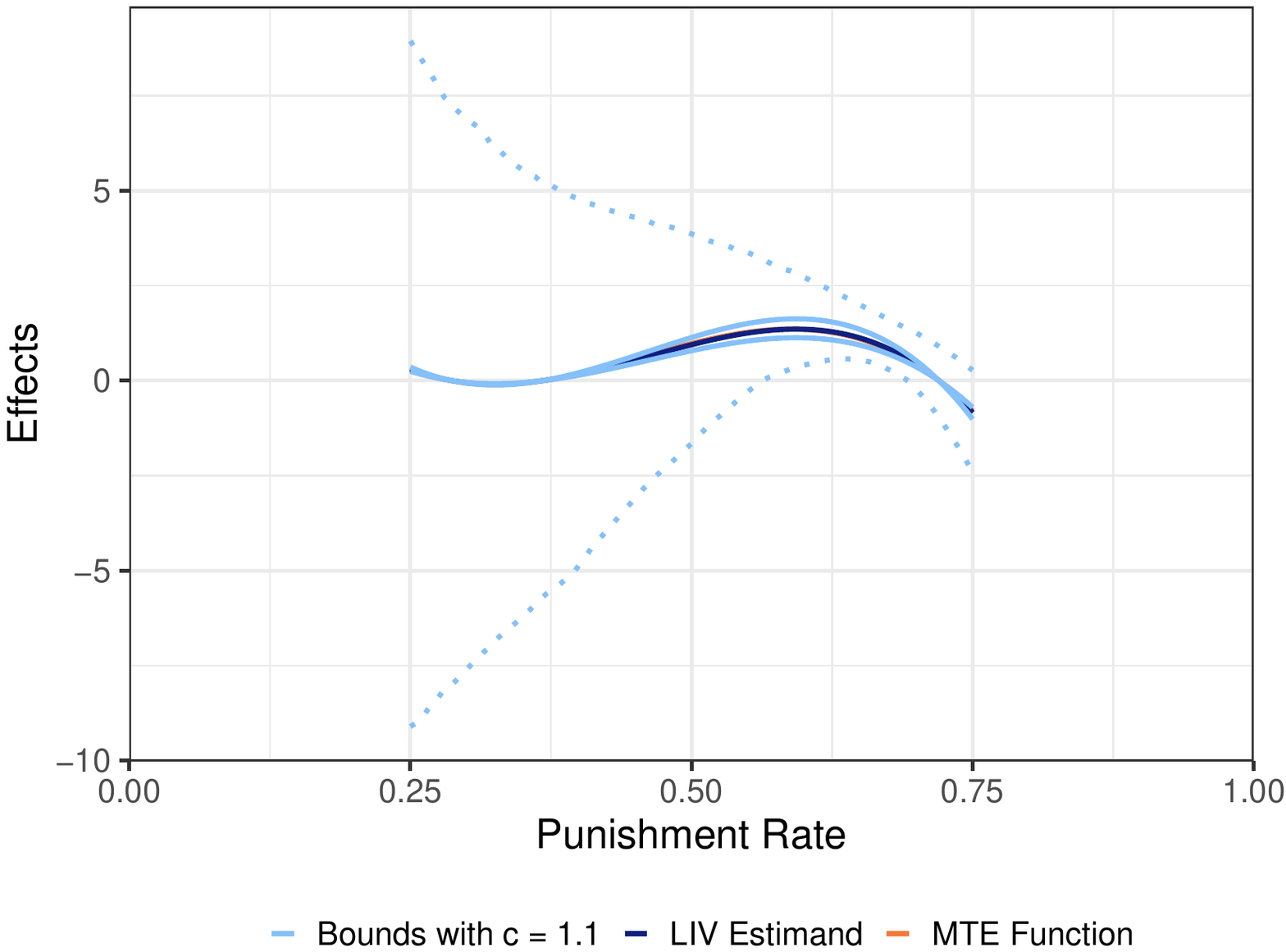}
			\caption{Confidence Bands around the set $\Theta_{1}$ --- Proposition \ref{CorOuterMTE}}
			\label{FigMTERibeiraoCI}
		\end{subfigure}
		\caption[]{Confidence Bands for Ribeirão Preto}
		\label{FigCI}
	\end{center}
	\footnotesize{Notes: The orange lines are the estimated MTE functions $\theta\left(\cdot, \cdot \right)$ based on the LIV estimator $\hat{f}^{*}$ that uses the correctly classified treatment variable $D$ (Equation \eqref{EQcorrectlivest}). The dark blue lines are the point-estimates of the LIV estimator $\hat{f}$ that uses the misclassified treatment variable $T$ (Equation \eqref{EQlivest}). The light blue solid lines are the estimated upper and lower bounds $\hat{\theta}_{U}$ and $\hat{\theta}_{L}$ based on a constant $c = 1.2$ (Equations \eqref{EQestupper} and \eqref{EQestlower}). The dotted lines are bootstrapped 90\%-confidence bands (1,000 repetitions) around the functions with the same color.}
\end{figure}

See Figure \ref{FigCI}.

\subsection{Bounds around the MTE function $\theta\left(\cdot, \cdot \right)$ --- Sensitivity Analysis}\label{AppSensitivity}

\begin{figure}[!htbp]
	\begin{center}
		\includegraphics[width = \textwidth]{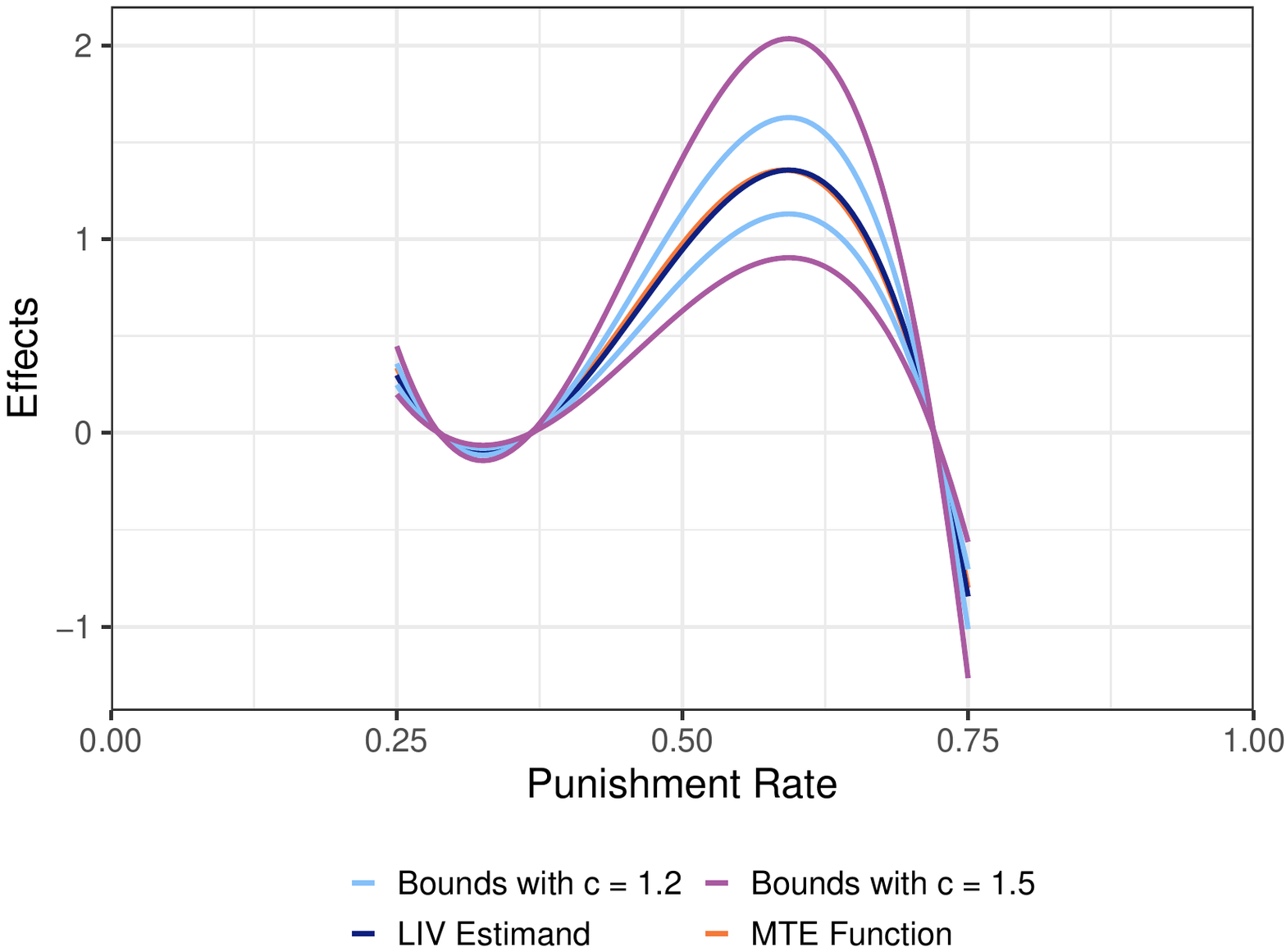}
		\caption{Sensitivity Analysis for Ribeirão Preto}
		\label{FigMTESensitivity}
	\end{center}
	\footnotesize{Notes: The orange line is the estimated MTE function $\theta\left(\cdot, \cdot \right)$ based on the LIV estimator $\hat{f}^{*}$ that uses the correctly classified treatment variable $D$ (Equation \eqref{EQcorrectlivest}). The dark blue line is the point-estimate of the LIV estimator $\hat{f}$ that uses the misclassified treatment variable $T$ (Equation \eqref{EQlivest}). The light blue lines and the purple lines are the estimated upper and lower bounds $\hat{\theta}_{U}$ and $\hat{\theta}_{L}$ based on constants $c = 1.2$ and $c = 1.5$ (Equations \eqref{EQestupper} and \eqref{EQestlower}), respectively. }
\end{figure}

See Figure \ref{FigMTESensitivity}.

\subsection{Results based on 2SLS Estimation}\label{App2SLS}

In this appendix, I evaluate the effect of alternative sentences on recidivism using standard 2SLS with fixed effects. The outcome variable denotes whether the defendant recidivates within 2 years after the case's final ruling. The treatment variable is either whether the defendant was punished according to the trial judge's ruling or whether the defendant was punished according to the final ruling. The instrument is either a full set of trial judges' dummies (``Judge FE'') or the leave-one-out punishment rate.

All regressions in this appendix control for a full set of court district fixed effects. Differently from the main text, standard errors are clusterized at the court district level.

Table \ref{Tab2SLS} shows the results of the four regressions described above. When I regress recidivism on the cases' final rulings using the trial judge's punishment rate as an instrument (Column (4) in Table \ref{Tab2SLS}), I find a point-estimate equal to .054 and marginally significant at the 10\%-significance level. When I use a full set of trial judge's dummies as my instrumental variables (Column (3) in Table \ref{Tab2SLS}), I find an even smaller point-estimate (.048) based on a weak set of instruments. Since 2SLS estimands may assign negative weights to interpretable treatment effect parameters \citep{Heckman2006} and the instruments in Column (3) are weak, the significance of those results should be considered cautiously in light of the null effects that were found when analyzing the MTE function in Appendix \ref{AppStandardMTE}.

\begin{table}[!htbp]
	\centering
	\caption{{Effect of Alternative Sentences on Recidivism: 2SLS Estimation}} \label{Tab2SLS}
	\begin{lrbox}{\tablebox}
		\begin{tabular}{ccclcc}
			\hline \hline
			\multirow{2}{*}{Overall} & \multicolumn{2}{c}{Trial Judge's Ruling} &  & \multicolumn{2}{c}{Final Ruling} \\ \cline{2-3} \cline{5-6}
			& Judge FE & Punishment Rate &  & Judge FE & Punishment Rate \\
			& (1) & (2) &  & (3) & (4) \\ \hline
			2SLS Estimate & .049** & .053* &  & .048** & .054* \\
			Cluster S.E. & (.020) & (.031) &  & (.021) & (.032) \\
			District FE & $\checkmark$ & $\checkmark$ &  & $\checkmark$ & $\checkmark$ \\
			1\textsuperscript{st} Stage's F-Stat & 1.76 & 246 &  & .568 & 275 \\ \hline
		\end{tabular}
	\end{lrbox}
	\usebox{\tablebox}\\
	\settowidth{\tableboxwidth}{\usebox{\tablebox}} \parbox{\tableboxwidth}{\footnotesize{Note: The table reports the estimated effect of alternative sentences on recidivism based on standard two-stage least squares regressions. The treatment variable is described by the first heading in the columns, while the instrumental variables are described by the second heading in the columns. The first stage F-statistic is the test statistic of a $\chi^{2}$-test whose null hypothesis is that all the coefficients associated with the instrumental variables are equal to zero in the first stage regression.  Robust standard errors clusterized at the court district level are reported in parenthesis. Significance levels are indicated by *$p \leq 0.10$, **$p \leq 0.05$ and ***$p \leq 0.01$.}
	}
\end{table}

Table \ref{Tab2SLS} also illustrates the danger of ignoring misclassification bias in a 2SLS regression. When I use the possibly misclassified trial judge's ruling as my treatment variable, I find point-estimates equal to 0.053 and .049 depending on the chosen instruments. Consequently, the misclassification bias may be equal to 1.3\% or 2.2\% of the estimated 2SLS estimand depending on the chosen instruments.

\subsection{Semiparametric Estimation of the Propensity Score Functions}\label{AppPSsemi}

In this appendix, I estimate the functions $P_{D}$ and $P_{T}$ semiparametrically. To do so, I use the semi-parametric estimator proposed by \cite{Robinson1988} combined with the local linear estimator proposed by \cite{Calonico2019}.

\begin{figure}[!htb]
	\begin{center}
		\includegraphics[width = 0.47\textwidth]{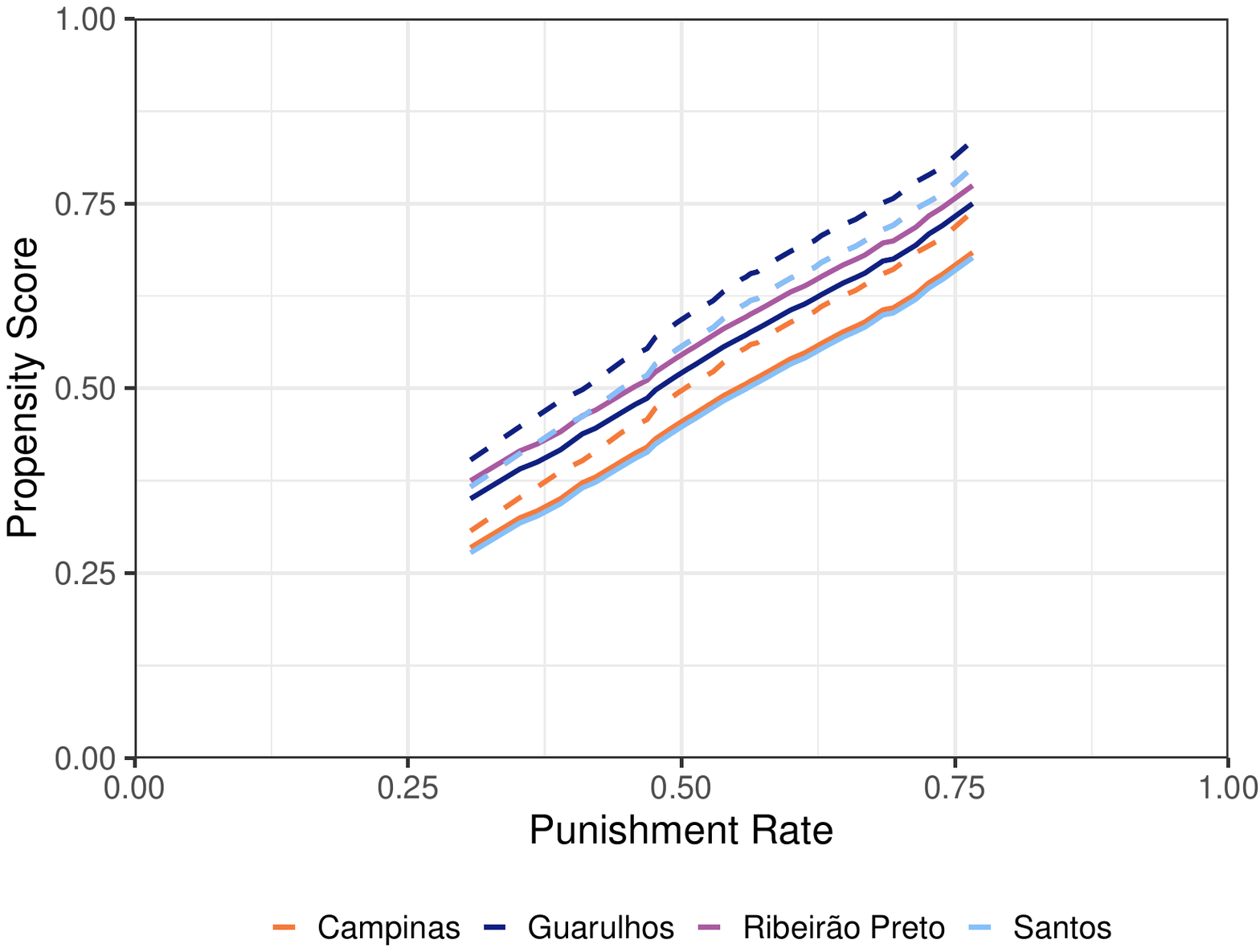}
		\caption{Semiparametric Estimates of the Propensity Score Functions}
		\label{FigPS}
	\end{center}
	\footnotesize{Notes: The solid lines are the semiparametrically estimated $P_{D}$ functions using the semi-parametric estimator proposed by \cite{Robinson1988}. The dashed lines are the semiparametrically estimated $P_{T}$ functions using the semi-parametric estimator proposed by \cite{Robinson1988}. Colors indicate the court district associated to each curve.}
\end{figure}

Figure \ref{FigPS} presents the results of four court districts (Campinas, Guarulhos, Ribeirão Preto and Santos) for illustrative purposes.\footnote{These districts are the four largest districts with respect to the number of judges in my sample.} The solid lines represent the estimated correctly measured propensity score functions while dashed lines represent the mismeasured propensity score functions. Each court district  are represented by a different color.

\subsection{Standard Approach to MTE Estimation}\label{AppStandardMTE}

In this appendix, I estimate the function $\mathbb{E}\left[\left. Y_{1} - Y_{0} \right\vert U = u, X = x \right]$ defined by \cite{Heckman2006}. To do so, I use a semi-parametric estimator \citep{Robinson1988,Calonico2019}.

\begin{figure}[!htb]
	\begin{center}
		\begin{subfigure}[b]{0.47\textwidth}
			\centering
			\includegraphics[width = \textwidth]{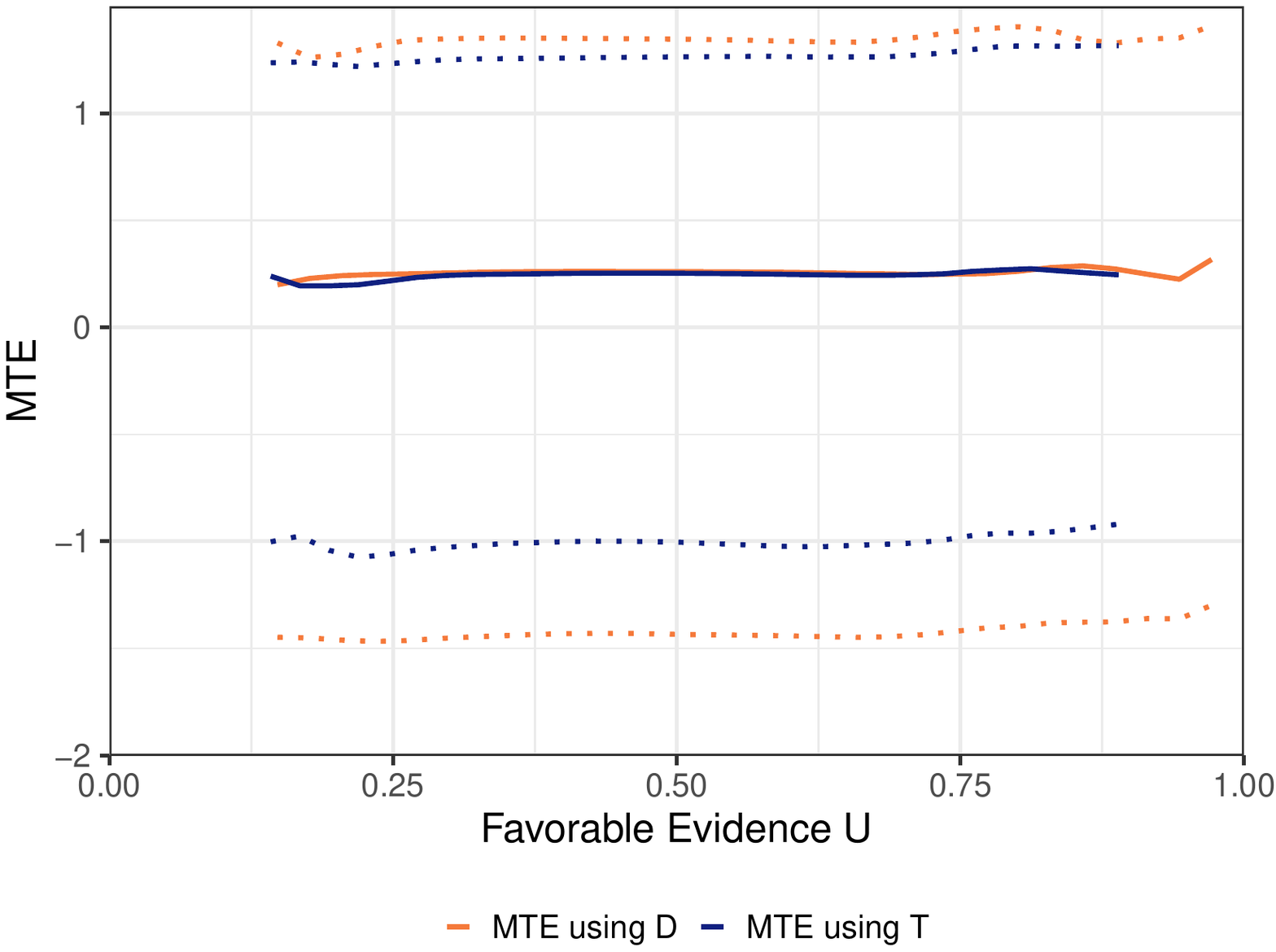}
			\caption{Ribeirão Preto}
			\label{FigMTERibeiraoStandard}
		\end{subfigure}
		\hfill
		\begin{subfigure}[b]{0.47\textwidth}
			\centering
			\includegraphics[width = \textwidth]{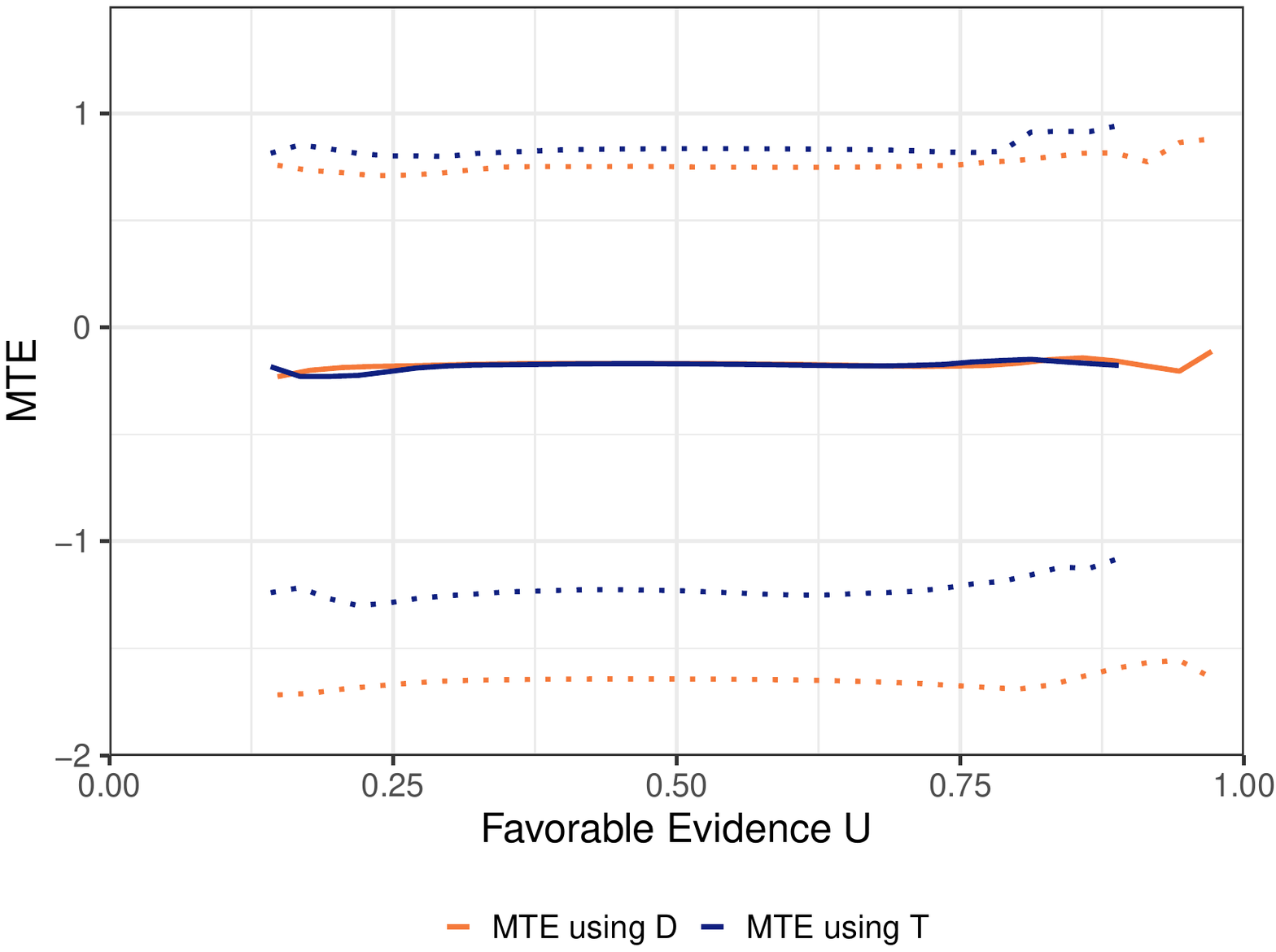}
			\caption{Campinas}
			\label{FigMTECampinasStandard}
		\end{subfigure}
		\begin{subfigure}[b]{0.47\textwidth}
			\centering
			\includegraphics[width = \textwidth]{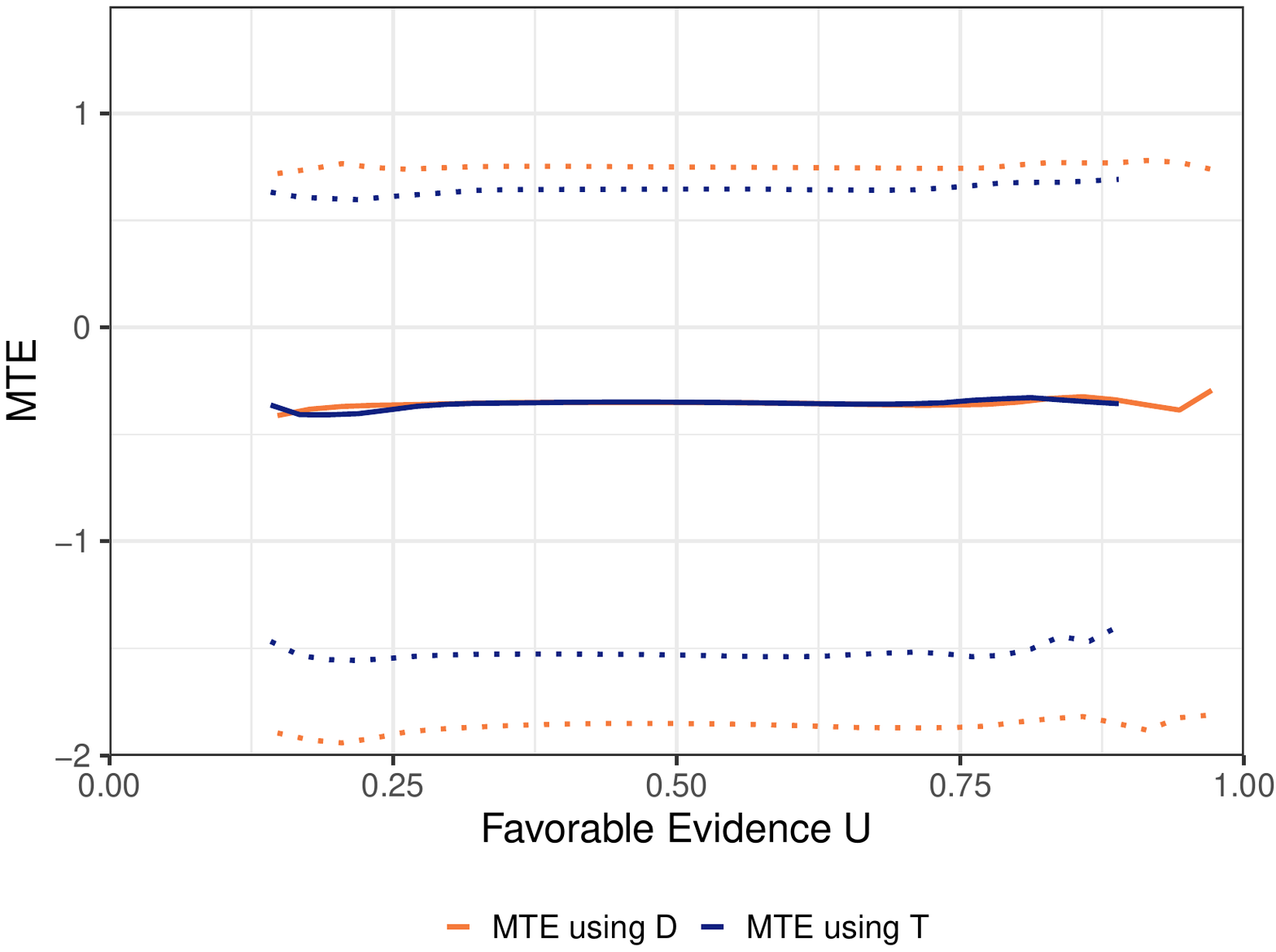}
			\caption{Santos}
			\label{FigMTESantosStandard}
		\end{subfigure}
		\hfill
		\begin{subfigure}[b]{0.47\textwidth}
			\centering
			\includegraphics[width = \textwidth]{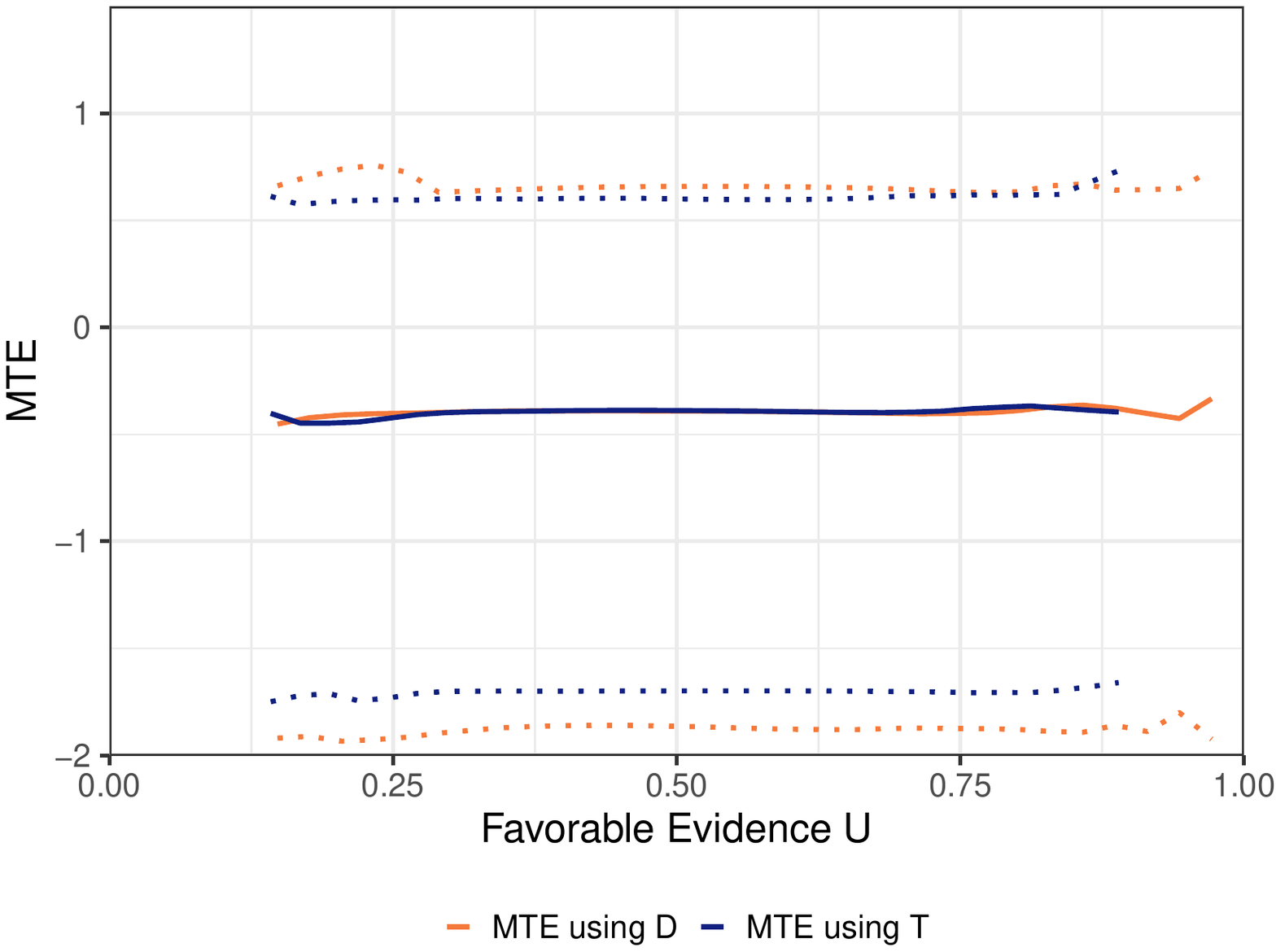}
			\caption{Guarulhos}
			\label{FigMTEGuarulhosStandard}
		\end{subfigure}
		\caption[]{Standard MTE Functions ($\mathbb{E}\left[\left. Y_{1} - Y_{0} \right\vert U = u, X = x \right]$) --- Semi-parameteric Estimation}
		\label{FigMTEStandard}
	\end{center}
	\footnotesize{Notes: The orange solid lines are the estimated MTE functions using a semi-parametric estimator \citep{Robinson1988,Calonico2019} and the correctly classified treatment variable $D$. The dark blue solid lines are the estimated MTE functions using a semi-parametric estimator \citep{Robinson1988,Calonico2019} and the misclassified treatment variable $T$. The dotted lines are bootstrapped 90\%-confidence bands (100 repetitions) around the estimated MTE functions and follow the same color scheme as the solid lines.}
\end{figure}

Figure \ref{FigMTEStandard} presents the results of four court districts (Campinas, Guarulhos, Ribeirão Preto and Santos) for illustrative purposes.\footnote{These districts are the four largest districts with respect to the number of judges in my sample.} The orange line is the estimated MTE function based on the correctly measured propensity score $P_{D}$, while the dark blue line is the estimated LIV estimand $\tilde{f}$ based on the mismeasured propensity score $P_{T}$. The dotted lines follow the same color scheme and show bootstrapped 90\%-confidence intervals.

Since the MTE functions seem constant, these results suggest that the unobserved heterogeneity is possibly unimportant when discussing the effect of alternative sentences on recidivism. Furthermore, all eight confidence bands contain the zero function, indicating that the effect of alternative sentences is likely small.

\clearpage

\section{Comparison between Proposition \ref{CorOuterMTE} and Appendix G by \cite{Acerenza2021}}\label{AppComparison}
\setcounter{table}{0}
\renewcommand\thetable{K.\arabic{table}}

\setcounter{figure}{0}
\renewcommand\thefigure{K.\arabic{figure}}

\setcounter{equation}{0}
\renewcommand\theequation{K.\arabic{equation}}

\setcounter{theorem}{0}
\renewcommand\thetheorem{K.\arabic{theorem}}

\setcounter{proposition}{0}
\renewcommand\theproposition{K.\arabic{proposition}}

\setcounter{corollary}{0}
\renewcommand\thecorollary{K.\arabic{corollary}}

\setcounter{assumption}{0}
\renewcommand\theassumption{K.\arabic{assumption}}

In this appendix, I analytically compare the bounds in Proposition \ref{CorOuterMTE} against the bounds proposed by \citet[Appendix G]{Acerenza2021}.

Without imposing that the instrument is independent of the misreporting decision, \citet[Appendix G]{Acerenza2021} are able to bound the MTE function using $$\min \left\lbrace \lim_{z^{\prime} \rightarrow z} \dfrac{\mathbb{E}\left[\left. Y \right\vert Z = z\right] - \mathbb{E}\left[\left. Y \right\vert Z = z^{\prime}\right]}{TV_{Y}\left(z^{\prime}, z\right)}, 0 \right\rbrace \leq \theta\left(z\right) \leq \max \left\lbrace \lim_{z^{\prime} \rightarrow z} \dfrac{\mathbb{E}\left[\left. Y \right\vert Z = z\right] - \mathbb{E}\left[\left. Y \right\vert Z = z^{\prime}\right]}{TV_{Y}\left(z^{\prime}, z\right)}, 0 \right\rbrace,$$
where $TV_{Y}\left(z^{\prime}, z\right) \coloneqq \dfrac{1}{2} \bigints \left\vert f_{\left. Y \right\vert Z} \left(\left. y \right\vert z\right) - f_{\left. Y \right\vert Z} \left(\left. y \right\vert z^{\prime}\right) \right\vert \, dF_{Y}\left(y\right)$, $f_{\left. Y \right\vert Z}$ is the conditional density of $Y$ given $Z$ and $F_{Y}$ is the cumulative distribution function of $Y$.

To simplify the exposition, I impose that $\lim_{z^{\prime} \rightarrow z} \dfrac{TV_{Y}\left(z^{\prime}, z\right)}{z - z^{\prime}}$ exists and is strictly positive, that $\dfrac{d P_{T}\left(z\right)}{d z} > 0$ and that $\dfrac{d \mathbb{E}\left[\left. Y \right\vert Z = z\right]}{dz} > 0$.

Under these assumptions, the upper bound proposed by \citet[Appendix G]{Acerenza2021} is given by $$UB_{ABK}\left(z\right) \coloneqq \dfrac{\sfrac{d \mathbb{E}\left[\left. Y \right\vert Z = z\right]}{dz}}{\lim_{z^{\prime} \rightarrow z} \dfrac{TV_{Y}\left(z^{\prime}, z\right)}{z - z^{\prime}}}$$ and the lower bound is given $$LB_{ABK}\left(z\right) \coloneqq 0.$$

In this case, the upper bound in Proposition \ref{CorOuterMTE} is given by $$\theta_{U}\left(z\right) \coloneqq c \cdot \dfrac{\sfrac{d \mathbb{E}\left[\left. Y \right\vert Z = z\right]}{dz}}{\sfrac{d P_{T}\left(z\right)}{d z}}$$ and the lower bound is given by $$\theta_{L}\left(z\right) \coloneqq \dfrac{1}{c} \cdot \dfrac{\sfrac{d \mathbb{E}\left[\left. Y \right\vert Z = z\right]}{dz}}{\sfrac{d P_{T}\left(z\right)}{d z}}.$$

Consequently, the upper bound in Proposition \ref{CorOuterMTE} is less than the bound proposed by \citet[Appendix G]{Acerenza2021} $\left( \theta_{U}\left(z\right) \leq UB_{ABK}\left(z\right) \right)$ if and only if $1 \leq c \leq c_{1}$ where $$c_{1} \coloneqq \dfrac{\sfrac{d P_{T}\left(z\right)}{d z}}{\lim_{z^{\prime} \rightarrow z} \dfrac{TV_{Y}\left(z^{\prime}, z\right)}{z - z^{\prime}}},$$ which is a testable restriction. Note that $c_{1} \geq 1$ is a necessary condition for $\theta_{U}\left(z\right) \leq UB_{ABK}\left(z\right)$ while this appendix's assumptions only imply that $c_{1} \geq 0$. Hence, there may exist cases when $UB_{ABK}\left(z\right)$ must be less than $\theta_{U}\left(z\right)$.

Moreover, the lower bound proposed by \citet[Appendix G]{Acerenza2021} is equal to zero while the lower bound Proposition \ref{CorOuterMTE} is strictly greater than zero for any choice of $c \geq 1$.

Additionally, the length of the bounds in Proposition \ref{CorOuterMTE} is less than the length of the bounds proposed by \citet[Appendix G]{Acerenza2021} $\left( \theta_{U}\left(z\right) - \theta_{L}\left(z\right) \leq UB_{ABK}\left(z\right) - LB_{ABK}\left(z\right) \right)$ if and only if $c \leq c_{2}$ where $c_{2} \coloneqq \dfrac{c_{1} + \sqrt{\left(c_{1}\right)^{2} + 4}}{2}$.\footnote{Observe that $\dfrac{c_{1} - \sqrt{\left(c_{1}\right)^{2} + 4}}{2} \leq 1$ for any value of $c_{1} \in \mathbb{R}$.} Note that $c_{2} \geq 1$ for any value of $c_{1} \geq 0$. Since this appendix's assumptions imply that $c_{1} \geq 0$, there must exist a $\tilde{c}$ that ensures that $\theta_{U}\left(z\right) - \theta_{L}\left(z\right) \leq UB_{ABK}\left(z\right) - LB_{ABK}\left(z\right)$.\footnote{I thank an anonymous referee for suggesting this result.}

This last result illustrates the identifying power of Assumption \ref{ASbounded} in comparison with the worst-case bounds derived by \citet[Appendix G]{Acerenza2021}.

\clearpage

\section{Differential Measurement Error}\label{AppDifferential}
\setcounter{table}{0}
\renewcommand\thetable{L.\arabic{table}}

\setcounter{figure}{0}
\renewcommand\thefigure{L.\arabic{figure}}

\setcounter{equation}{0}
\renewcommand\theequation{L.\arabic{equation}}

\setcounter{theorem}{0}
\renewcommand\thetheorem{L.\arabic{theorem}}

\setcounter{proposition}{0}
\renewcommand\theproposition{L.\arabic{proposition}}

\setcounter{corollary}{0}
\renewcommand\thecorollary{L.\arabic{corollary}}

\setcounter{assumption}{0}
\renewcommand\theassumption{L.\arabic{assumption}}

In the measurement error literature, there are two types of misclassification problems. The first and simplest type is known as non-differential measurement error and imposes that the potential misclassified treatment variables are independent of the potential outcomes, i.e., $\left(T_{0}, T_{1}\right) \independent \left(Y_{0}, Y_{1}\right)$ \citep{Calvi2019}. The second type of misclassification problem is known as differential measurement error and allows for the measured treatment to depend on the outcome conditional on the true treatment \citep{Ura2018, Tommasi2020, Acerenza2021}.

Similarly to \citet{Ura2018}, \citet{Tommasi2020} and \citet{Acerenza2021}, my framework is compatible with differential measurement error. The easiest way to verify this compatibility is through a simple numerical example where the data-generating process has differential measurement error and satisfies Assumptions \ref{ASindependence}-\ref{ASfinite} and \ref{ASbounded}.

Let $Y_{0} \sim N\left(0,1\right)$, $Y_{1} \sim N\left(0,1\right)$ and $Z \sim Unif\left(0, 1\right)$ be mutually independent. Define $V \coloneqq Y_{1} - Y_{0} \sim N\left(0, 2\right)$ and $U \coloneqq F_{V}\left(V\right) \sim Unif\left(0, 1\right)$, where $F_{V}$ is the cumulative distribution function of $V$. Moreover, let $D \coloneqq \mathbf{1}\left\lbrace U \leq P_{D}\left(Z\right) \right\rbrace$ and $T \coloneqq \mathbf{1}\left\lbrace U \leq P_{T}\left(Z\right) \right\rbrace$, where $P_{D}\colon \left[0, 1\right] \rightarrow \left[0, 1\right]$ and $P_{T}\colon \left[0, 1\right] \rightarrow \left[0, 1\right]$ are given by $P_{D}\left(z\right) = \alpha_{D} + \beta_{D} \cdot z$ and $P_{T}\left(z\right) = \alpha_{T} + \beta_{T} \cdot z$ with $\alpha_{D} = 0$, $\beta_{D} = 1$, $\alpha_{T} = \sfrac{1}{4}$ and $\beta_{T} = \sfrac{1}{2}$.

Note that Assumptions \ref{ASindependence}-\ref{ASfinite} hold by construction. Moreover, Assumption \ref{ASbounded} holds with $c = 2$.

In this example, there is differential measurement error. To visualize this phenomenon, we can numerically compute $\mathbb{P}\left[\left. T = 1 \right\vert Y = y, D = 1\right]$ and $\mathbb{P}\left[\left. T = 1 \right\vert Y = y, D = 0\right]$ for a set of outcome values $y$ and verify that these two objects depend on the value $y$. I do so using numerical integration with a sample size equal to 100,000 and the nonparametric estimator proposed by \cite{Calonico2019}.

\begin{figure}[!htb]
	\begin{center}
		\includegraphics[width = 0.47\textwidth]{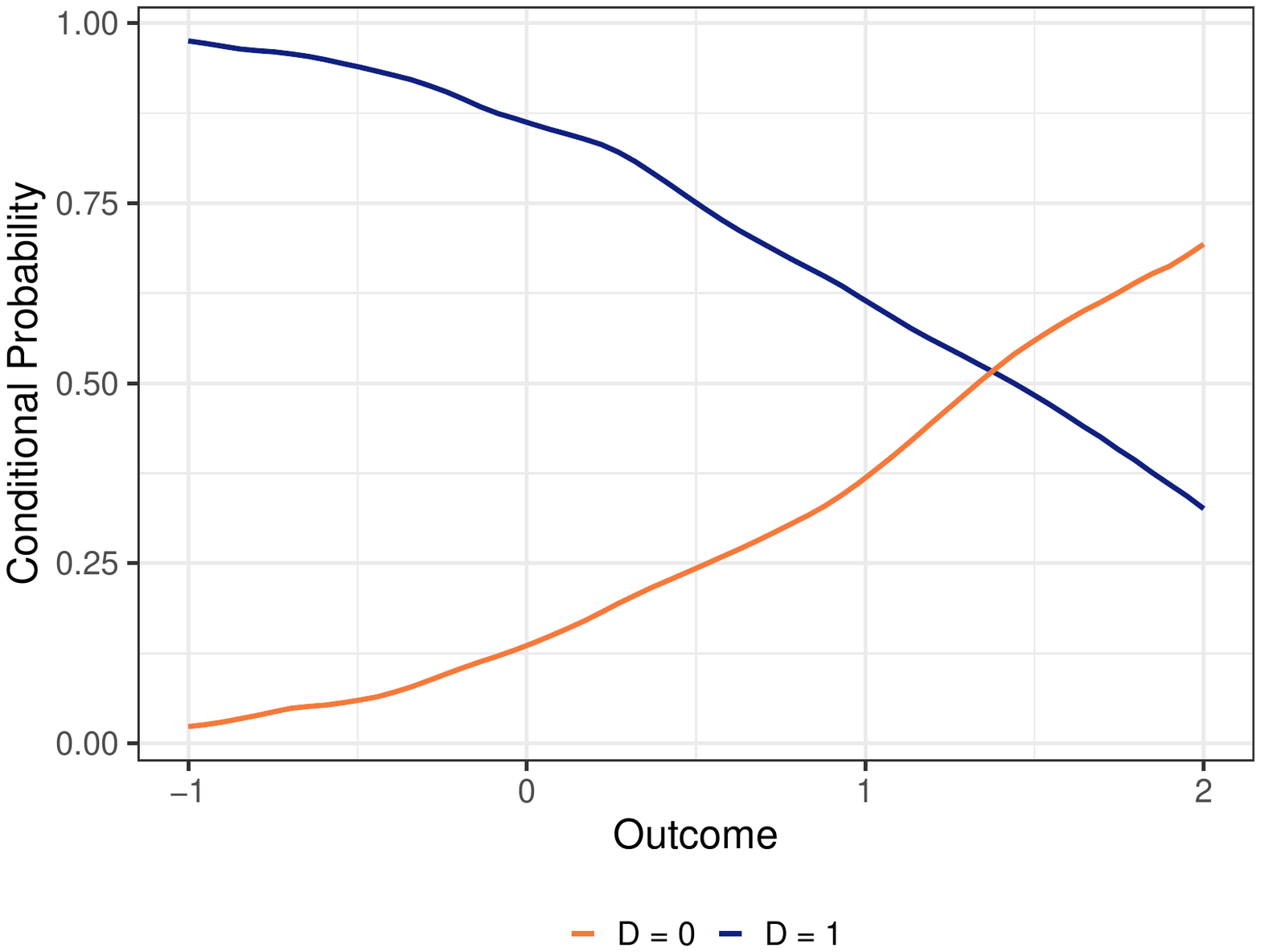}
		\caption{Differential Measurement Error --- Functions $\mathbb{P}\left[\left. T = 1 \right\vert Y = y, D = d\right]$}
		\label{FigDifferential}
	\end{center}
	\footnotesize{Notes: Function $\mathbb{P}\left[\left. T = 1 \right\vert Y = y, D = 0\right]$ is the orange line and function $\mathbb{P}\left[\left. T = 1 \right\vert Y = y, D = 1\right]$ is the dark blue line. Both functions were computed using numerical integration with a sample size equal to 100,000 and the nonparametric estimator proposed by \cite{Calonico2019}.}
\end{figure}

Figure \ref{FigDifferential} plots function $\mathbb{P}\left[\left. T = 1 \right\vert Y = y, D = 0\right]$ in orange and function $\mathbb{P}\left[\left. T = 1 \right\vert Y = y, D = 1\right]$ in dark blue. Both functions clearly depend on the value $y$ of the outcome variable. Hence, there is a data-generating process that has differential measurement error and satisfies Assumptions \ref{ASindependence}-\ref{ASfinite} and \ref{ASbounded}.

\end{document}